%% file: main.tex
\documentclass{article}

\usepackage{a4wide}
\usepackage{hyperref}

\usepackage{proof}
\usepackage{amsfonts}

\usepackage{amssymb}
\usepackage{amsmath}

\usepackage{url}
\usepackage[utf8]{inputenc}
\usepackage[capitalise]{cleveref}
\usepackage[all]{xy}

\newtheorem{lemma}{Lemma}[section]
\newtheorem{theorem}{Theorem}[section]

\newtheorem{example}{Example}[section]
\newtheorem{definition}{Definition}[section]
\newtheorem{proposition}{Proposition}[section]
\newtheorem{corollary}{Corollary}[section]

\input{macros}

\begin{document}

\title{Coinductive proof search for polarized logic with applications to full intuitionistic propositional logic}

\author{Jos\'{e} Esp\'{\i}rito Santo, Ralph Matthes, Lu\'{\i}s Pinto}
\date{December 4, 2025}

\maketitle
\begin{abstract}
  The approach to proof search dubbed ``coinductive proof search'' (CoIPS), and previously developed by the authors for implicational intuitionistic logic, is in this paper extended to $\PIL$, a focused sequent-calculus presentation of polarized intuitionistic logic, including an array of positive and negative connectives. As before, this includes developing a coinductive description of the search space generated by a sequent, an equivalent inductive syntax describing the same space, and decision procedures for inhabitation problems in the form of predicates defined by recursion on the inductive syntax. Inhabitation is taken in a generalized sense, because we also consider when a sequent has a solution, that is a (possibly infinite) run of bottom-up proof search which never fails to apply another inference rule. In this view, proofs are just the finite solutions, and finiteness of a sequent may refer to finiteness of the number of proofs, or to finiteness of all solutions (two problems we show to be decidable). In fact, we provide a very general scheme whose instances are decision problems about $\LJP$ having algorithms through the inductive syntax.
  Moreover, polarized logic and $\PIL$ can be used as a platform from which proof search for other logics is understood. We illustrate the technique with the well-known proof systems $\LJT$ and $\LJQ$ for intuitionistic logic, both equipped with all the connectives. For that we work out respectively a negative and a positive interpretation into $\PIL$, which map formulas of the source logic into formulas in $\PIL$ of the said polarity; and this is done even at the level of the coinductive versions of the three involved proof systems. The interpretations are proved to be, not only faithful, but actually full embeddings, establishing a bijection between the solutions (resp.~proofs) of an intuitionistic sequent and the solutions (resp.~proofs) of its polarized interpretation. This allows the inheritance to the source systems of the decidability and other results previously obtained for $\PIL$, thereby vastly generalizing the previous results of CoIPS, which were confined to $\LJT$ and implicational intuitionistic logic.  
\end{abstract}


\input{intro}
\input{background}
\input{proof-search-PIPL}
\input{analysis-PIPL}
\input{decision-problems-PIPL}
\input{applications-IPL}
\input{final}


\bibliographystyle{plain}


\appendix
\section{Appendix with some more technical details}
\input{app-analysis-PIPL}
\input{appendix-negative-embedding}
\input{appendix-positive-embedding}
\end{document}

%% file: macros.tex
\newcommand{\LJQ}{\mathsf{LJQ}}
\newcommand{\LJQouru}{\LJQ^{u}} 
\newcommand{\LJTour}{\LJT^{e}} 

\newcommand{\crc}[1]{\langle#1\rangle} 

\newcommand{\finfinsymb}{\mathsf{finfin}}
\newcommand{\finfin}[1]{\finfinsymb(#1)}
\newcommand{\inffinsymb}{\mathsf{inffin}}
\newcommand{\inffin}[1]{\inffinsymb(#1)}
\newcommand{\finfinmemsymb}{\mathsf{finfinext}}
\newcommand{\finfinmem}[1]{\finfinmemsymb(#1)}
\newcommand{\finfinmemi}[2]{\finfinmemsymb_{#1}(#2)}
\newcommand{\inffinmemsymb}{\mathsf{inffinext}}
\newcommand{\inffinmem}[1]{\inffinmemsymb(#1)}
\newcommand{\inffinmemisymb}[1]{\inffinmemsymb_{#1}}
\newcommand{\inffinmemi}[2]{\inffinmemisymb{#1}(#2)}

\newcommand{\nofinmemsymb}{\mathsf{nofinext}}
\newcommand{\nofinmem}[1]{\nofinmemsymb(#1)}
\newcommand{\nofinmemisymb}[1]{\nofinmemsymb_{#1}}
\newcommand{\nofinmemi}[2]{\nofinmemisymb{#1}(#2)}
\newcommand{\exfinmemsymb}{\mathsf{exfinext}}
\newcommand{\exfinmem}[1]{\exfinmemsymb(#1)}
\newcommand{\exfinmemi}[2]{\exfinmemsymb_{#1}(#2)}

\newcommand{\memsymb}{{\sf mem}}
\newcommand{\mem}[2]{\memsymb(#1,#2)}

\newcommand{\extsymb}{{\mathcal{E}}}
\newcommand{\ext}[1]{\extsymb(#1)}

\newcommand{\finextsymb}{{\mathcal{E}_\mathrm{fin}}} 
\newcommand{\finext}[1]{\finextsymb(#1)}

\newcommand{\solletter}{\mathcal S}
\newcommand{\solfunction}[1]{{\solletter}(#1)}

\newcommand{\vect}[1]{\overrightarrow{#1}} 

\newcommand{\falsity}{\mathop{\perp}}
\newcommand{\impl}{\supset}
\newcommand{\cosign}{\mathit{co}}
\newcommand{\gfpsymb}{\mathsf{gfp}}
\newcommand{\gfp}{\gfpsymb\kern0.1em}  
\newcommand{\oo}{\mathbb{O}}
\newcommand{\lb}{\lambda}

\newcommand{\s}[2]{\sum\limits_{#1}^{}{#2}} 

\newcommand{\interps}[1]{[\![#1]\!]}

\newcommand{\finrepsymb}{{\mathcal F}}
\newcommand{\finrep}[2]{\finrepsymb(#1;#2)}
\newcommand{\finrepempty}[1]{\finrepsymb(#1)}

\newcommand{\nofinsymb}{{\sf nofin}}
\newcommand{\nofin}[1]{\nofinsymb(#1)}

\newcommand{\exfinsymb}{{\sf exfin}}
\newcommand{\exfin}[1]{\exfinsymb(#1)}

\newcommand{\nosolsymb}{{\sf nosol}}
\newcommand{\exsolsymb}{{\sf exsol}}
\newcommand{\exsolisymb}[1]{\exsolsymb_{#1}}
\newcommand{\exsoli}[2]{\exsolisymb{#1}(#2)}
\newcommand{\allfinsymb}{{\sf allfin}}
\newcommand{\exinfsymb}{{\sf exinf}}
\newcommand{\nosolmemsymb}{\nosolsymb{\sf ext}}
\newcommand{\exsolmemsymb}{\exsolsymb{\sf ext}}
\newcommand{\allfinmemsymb}{\allfinsymb{\sf ext}}
\newcommand{\exinfmemsymb}{\exinfsymb{\sf ext}}
\newcommand{\nosolmem}[1]{\nosolmemsymb(#1)}
\newcommand{\exsolmem}[1]{\exsolmemsymb(#1)}
\newcommand{\allfinmem}[1]{\allfinmemsymb(#1)}
\newcommand{\exinfmem}[1]{\exinfmemsymb(#1)}
\newcommand{\nosol}[1]{\nosolsymb(#1)}
\newcommand{\exsol}[1]{\exsolsymb(#1)}
\newcommand{\allfin}[1]{\allfinsymb(#1)}
\newcommand{\exinf}[1]{\exinfsymb(#1)}
\newcommand{\exinfisymb}[1]{\exinfsymb_{#1}}
\newcommand{\exinfi}[2]{\exinfisymb{#1}(#2)}
\newcommand{\exinfmemisymb}[1]{\exinfmemsymb_{#1}}
\newcommand{\exinfmemi}[2]{\exinfmemisymb{#1}(#2)}

\newcommand{\ipredsymb}{\mathbb{P}}
\newcommand{\cpredsymb}{\overline\ipredsymb}
\newcommand{\ipredd}[1]{\ipredsymb_{#1}}
\newcommand{\cpredd}[1]{\cpredsymb_{#1}}
\newcommand{\ipred}[2]{\ipredsymb_{#1}(#2)}
\newcommand{\cpred}[2]{\cpredsymb_{#1}(#2)}
\newcommand{\ipredsymbf}{\mathsf{F}\mathbb{P}}
\newcommand{\cpredsymbf}{\overline{\mathsf{F}\ipredsymb}}
\newcommand{\ipredfd}[1]{\ipredsymbf_{#1}}
\newcommand{\cpredfd}[1]{\cpredsymbf_{#1}}
\newcommand{\ipredf}[2]{\ipredsymbf_{#1}(#2)}
\newcommand{\cpredf}[2]{\cpredsymbf_{#1}(#2)}
\newcommand{\cpredisymb}[1]{\cpredsymb_{#1}}
\newcommand{\cpredi}[2]{\cpredisymb{#1}(#2)}

\newcommand{\truepred}{\mathbb{T}}
\newcommand{\falsepred}{\mathbb{F}}

\newcommand{\binfsymb}{\mbox{${\sf NEF}$}}
\newcommand{\binfpsymb}[1]{\binfsymb\kern-0.152em_{#1}} 
\newcommand{\binfp}[2]{\binfpsymb{#1} (#2)}
\newcommand{\binfcanindex}{*}  
\newcommand{\binfcansymb}{\binfpsymb\binfcanindex}
\newcommand{\binfcan}[1]{\binfp\binfcanindex{#1}}

\newcommand{\Pcansymb}{P_\binfcanindex}
\newcommand{\Pcan}[1]{P_\binfcanindex(#1)}

\newcommand{\nbinfsymb}{\mathsf{EF}}    
\newcommand{\nbinfpsymb}[1]{\nbinfsymb\kern-0.152em_{#1}} 
\newcommand{\nbinfp}[2]{\nbinfpsymb{#1} (#2)}
\newcommand{\nbinfcansymb}{\nbinfpsymb\binfcanindex}
\newcommand{\nbinfcan}[1]{\nbinfp\binfcanindex{#1}}

\newcommand{\stucksymb}{{\sf NES}}
\newcommand{\nstucksymb}{{\sf ES}}

\newcommand{\stuckpsymb}[1]{\stucksymb_{#1}}
\newcommand{\stuckp}[2]{{\stuckpsymb{#1}}(#2)}
\newcommand{\stuckcansymb}{\stuckpsymb\binfcanindex}

\newcommand{\nstuckpsymb}[1]{\nstucksymb_{#1}}
\newcommand{\nstuckp}[2]{{\nstuckpsymb{#1}}(#2)}
\newcommand{\nstuckcansymb}{\nstuckpsymb\binfcanindex}
\newcommand{\nstuckcan}[1]{\nstuckp\binfcanindex{#1}}

\newcommand{\exfsymb}{\mathsf{ex}}
\newcommand{\exf}[1]{\exfsymb(#1)}
\newcommand{\exfvarsymb}{\widetilde{\mathsf{ex}}}
\newcommand{\exfvar}[1]{\exfvarsymb(#1)}

\newcommand{\ipl}{IPL}
\newcommand{\pipl}{PIPL}

\newcommand{\lbpol}{\lambda_G^\pm}

\newcommand{\PIL}{\mathsf{LJP}}
\newcommand{\pil}{\mathsf{LJP}}

\newcommand{\inj}[3]{\mathsf{in}_{#1}^{#2}(#3)} 
\newcommand{\thunksymb}{\mathsf{thunk}}
\newcommand{\thunk}[1]{\thunksymb(#1)}
\newcommand{\thunkbig}[1]{\thunksymb\bigl(#1\bigr)}
\newcommand{\ea}[1]{\ulcorner#1\urcorner} 
\newcommand{\ep}[1]{\lceil#1\rceil} 
\newcommand{\nil}{\mathsf{nil}}
\newcommand{\cothunk}[1]{\mathsf{cothunk}(#1)}
\newcommand{\abortt}[1]{\mathsf{abort}^{#1}}
\newcommand{\abort}{\mathsf{abort}}
\newcommand{\dlvsymb}{\mathsf{dlv}}
\newcommand{\dlv}[1]{\dlvsymb(#1)}
\newcommand{\dlvbig}[1]{\dlvsymb\bigl(#1\bigr)}
\newcommand{\coretsymb}{\mathsf{coret}}
\newcommand{\coret}[2]{\coretsymb(#1,#2)}
\newcommand{\coretR}[3]{\coretsymb(#1,#2)^{#3}} 
\newcommand{\retsymb}{\mathsf{ret}}
\newcommand{\ret}[1]{\retsymb(#1)}
\newcommand{\retbig}[1]{\retsymb\bigl(#1\bigr)}

\newcommand{\Idt}{{\sf ID}}
\newcommand{\Churcht}{{\sf CHURCH}}
\newcommand{\Inft}{{\sf INFTY}}

\newcommand{\heightsymb}{\mathsf{h}}
\newcommand{\height}[1]{\heightsymb (#1)}

\newcommand{\omegatwo}{x(y,\_ .\cdot)^\omega}

\newcommand{\omegafour}{{{x(y,\_ .\cdot)}^{\omega}_Q}}

\newcommand{\copil}{\PIL^\cosign}
\newcommand{\copils}{\PIL^\cosign_{\Sigma}}
\newcommand{\valuesco}{v^\cosign}
\newcommand{\termsco}{t^\cosign}
\newcommand{\covaluesco}{s^\cosign}
\newcommand{\cotermsco}{p^\cosign}
\newcommand{\expsco}{e^\cosign}
\newcommand{\Valuesco}{v^\cosign_{\Sigma}}
\newcommand{\Termsco}{t^\cosign_{\Sigma}}
\newcommand{\Covaluesco}{s^\cosign_{\Sigma}}
\newcommand{\Cotermsco}{p^\cosign_{\Sigma}}
\newcommand{\Expsco}{e^\cosign_{\Sigma}}

\newcommand{\pils}{\PIL^\gfpsymb_{\Sigma}}

\newcommand{\focl}[3]{(#1#2)^{#3} }

\newcommand{\FPV}{\mathit{FPV}}

\newcommand{\jay}{\sigma}    

\newcommand{\fgt}[1]{|#1|} 
\newcommand{\fgtp}[2]{|#1|_{#2}} 
\newcommand{\fgtQ}[1]{\fgtp{#1}{\qsymb}}

\newcommand{\LJT}{\mathsf{LJT}}

\newcommand{\coLJT}{\LJT^{co}} 
\newcommand{\coLJQ}{\LJQ^{co}} 

\newcommand{\finrepsymbL}{{\mathcal F}}

\newcommand{\finrepemptyL}[1]{\finrepsymbL(#1)}

\newcommand{\EFsymbL}{\mathsf{EF}^L}    
\newcommand{\EFpsymbL}[1]{\EFsymbL\kern-0.152em_{#1}} 

\newcommand{\EFsymb}{\nbinfsymb}    
\newcommand{\EFpsymb}[1]{\EFsymb\kern-0.152em_{#1}} 

\newcommand{\nofinisymb}[1]{\nofinsymb_{#1}}
\newcommand{\nofini}[2]{\nofinisymb{#1}(#2)}
\newcommand{\bigasssymb}{\mathcal{A}}
\newcommand{\bigass}[2]{\bigasssymb_{#1}(#2)}

\newcommand{\FFsymb}{{\sf FF}}

\newcommand{\FFpsymb}[1]{\FFsymb\kern-0.152em_{#1}}
\newcommand{\FFp}[2]{\FFpsymb{#1}(#2)}

\newcommand{\NFFsymb}{\mbox{${\sf NFF}$}}

\newcommand{\NFFpsymb}[1]{\NFFsymb\kern-0.152em_{#1}}
\newcommand{\NFFp}[2]{\NFFpsymb{#1}(#2)}

\newcommand{\inffinisymb}[1]{\inffinsymb_{#1}}
\newcommand{\inffini}[2]{\inffinisymb{#1}(#2)}

\newcommand{\qsymb}{\mathsf{Q}}

\newcommand{\lbforqunomacro}[2]{\thunk{\lb(#1.\dlv{\ep{#2}})}}
\newcommand{\pairforqunomacro}[2]{{\thunk{\langle \ep{#1},\ep{#2}\rangle}}}

\newcommand{\ntermsymb}{\untsymb\mathsf{t}} 
\newcommand{\nexpsymb}{\untesymb\mathsf{e}} 
\newcommand{\nspinesymb}{\untsymb\mathsf{s}} 
\newcommand{\nterm}[1]{\ntermsymb(#1)}
\newcommand{\nexp}[1]{\nexpsymb(#1)}
\newcommand{\nspine}[1]{\nspinesymb(#1)}

\newcommand{\pexpsymb}{\uptsymb\mathsf{e}} 
\newcommand{\vvalsymb}{\urntvsymb\mathsf{v}} 
\newcommand{\pexp}[1]{\pexpsymb(#1)}
\newcommand{\vval}[1]{\vvalsymb(#1)}

\newcommand{\upshift}{\mathop{\uparrow}}
\newcommand{\downshift}{\mathop{\downarrow}}
\newcommand{\Downsymb}{\mathop{\Downarrow}}
\newcommand{\Down}[1]{\Downsymb #1}
\newcommand{\Upsymb}{\mathop{\Uparrow}}
\newcommand{\Up}[1]{\Upsymb #1}
\newcommand{\horseshoe}{\mathbin{\supset}}

\newcommand{\pairljq}[2]{\langle#1,#2\rangle}

\newcommand{\pairljt}[2]{\langle#1,#2\rangle}

\newcommand{\abortljqt}[2]{{\sf abort}(#1)^{#2}}
\newcommand{\conjleftljqt}[6]{#1(#5,#2^{#3}.#4)^{#6}}
\newcommand{\disjleftljqt}[8]{#1(#2^{#3}.#4, #5^{#6}.#7)^{#8}}
\newcommand{\impleftljqt}[6]{#1(#2,#3^{#4}.#5)^{#6}}

\newcommand{\lbforqut}[2]{\lb_\qsymb#1.#2} 
\newcommand{\pairforqut}[2]{{\langle #1,#2\rangle}_\qsymb} 

\newcommand{\elimimpforqt}[5]{#1(#2,#3.#4)^{#5}_\qsymb} 
\newcommand{\elimandforqt}[5]{#1(#4,#2.#3)^{#5}_\qsymb} 
\newcommand{\elimorforqt}[6]{#1(#2.#3\,,\,#4.#5)^{#6}_\qsymb} 
\newcommand{\elimabsforqt}[2]{\abort(#1)^{#2}_{\qsymb}} 

\newcommand{\elimimpforqnomacrot}[5]{\coretR #1{{#2}::\cothunk{#3.#4}}{#5}}
\newcommand{\elimandforqnomacrot}[5]{\coretR#1{{#4}::\cothunk{#2.#3}}{#5} }
\newcommand{\elimorforqnomacrot}[6]{\coretR #1{\cothunk{[#2.#3\,,\,#4.#5]}}{#6}}
\newcommand{\elimabsforqnomacrot}[2]{\coretR #1{\cothunk{\abort^#2}}#2}

\newcommand{\LJP}{\PIL}

\newcommand{\untsymb}{{\mathfrak{n}}} 
\newcommand{\uptsymb}{{\mathfrak{p}}} 
\newcommand{\ultsymb}{{\mathfrak{l}}} 
\newcommand{\urtsymb}{{\mathfrak{r}}} 
\newcommand{\unt}[1]{#1^\untsymb} 
\newcommand{\upt}[1]{#1^\uptsymb} 
\newcommand{\ult}[1]{#1^\ultsymb} 
\newcommand{\urt}[1]{#1^\urtsymb} 

\newcommand{\untesymb}{\dagger} 
\newcommand{\urntvsymb}{\ddagger} 
\newcommand{\unte}[1]{#1^\untesymb} 
\newcommand{\urntv}[1]{#1^\urntvsymb} 

\newcommand{\nfasymb}{{\mathcal P}} 

\newcommand{\lopvar}{\circledcirc} 
\newcommand{\lopvarbinop}{\mathop{\,\lopvar\,}} 
\newcommand{\ovllopvarbinop}{\mathop{\,\overline\lopvar\,}} 
\newcommand{\lopvaralt}{\circledast} 

\newcommand{\AFsymb}{{\sf AF}}
\newcommand{\AFpsymb}[1]{\AFsymb\kern-0.152em_{#1}}

\newcommand{\NAFsymb}{\mbox{${\sf NAF}$}}
\newcommand{\NAFpsymb}[1]{\NAFsymb\kern-0.152em_{#1}}
\newcommand{\NAFp}[2]{\NAFpsymb{#1}(#2)}


%% file: intro.tex
\section{Introduction and Motivation}

The authors developed for the implicational fragment of intuitionistic logic a ``coinductive approach'' to proof search (CoIPS)~\cite{EspiritoSantoMatthesPintoInhabitation,JESRMLP-FI2019,JESRMLP-APAL2021}. Proof search is understood as the process of bottom-up application of the inference rules of the sequent calculus under consideration. One guiding idea of the approach is the emphasis on the generative aspect of the search process, which entails considering on an equal foot the outcomes of all successful runs, and the preoccupation with representing the entire search space and the set of all those outcomes. Such outcomes we call \emph{solutions} (to a proof search problem determined by a given sequent) and they are the possibly infinite trees generated by the search process, when all the branches are only required to never lead to failure -- and failure here is a sequent from where no inference rule is applicable (bottom-up). In this view, proofs are just the finite solutions.

Another guiding idea of the approach is to represent the entire search space for a given sequent as a single proof term.
This requires extending the concept of proof term in two directions: first, since as just said, naive proof search can run into cycles, hence non-terminating computation generating infinite branches, we adopt a coinductive interpretation of proof terms, so that they may represent non-wellfounded trees of locally correct applications of proof rules;
second, choice points are added to represent choices found in the search process in the application of proof rules.
The obtained expressions serve for a mathematical specification of the search space of a logical sequent. Such expressions can then have immediate use in the study of meta-theoretic properties of the proof system, but also in the precise formulation of decision problems related to proof search.
Algorithms for these decision problems are written in an alternative, equivalent, inductively defined syntax, where cycles are represented by formal fixed-point operators. Both the coinductive syntax for specification and the inductive syntax for the algorithms are ways of extending, to proof search, the Curry-Howard paradigm of representation of proofs (by typed $\lambda$-terms).

Our previous work targeted implicational intuitionistic logic and the sequent calculus $\LJT$~\cite{HerbelinCSL94}, and applications were concentrated on inhabitation problems for simple types. The viability of CoIPS requires the enlargement of its scope to a logic language with a full repertoire of connectives. In the conference paper~\cite{JESRMLP-TYPES20} we started this move in a somewhat radical way: we moved our attention to polarized intuitionistic logic~\cite{LiangMillerTCS09}, which is a language not only with a more expressive set of logical operators, but also with the capability of interpreting a range of sequent calculi for intuitionistic logic. The ambition is that, through the study of the polarized logic, we study indirectly and simultaneously that range of other systems. The present paper is the full accomplishment of that initial move reported in~\cite{JESRMLP-TYPES20}.

We start by developing the CoIPS of the proof system $\PIL$ for polarized intuitionistic logic already used in ~\cite{JESRMLP-TYPES20}, which is a minor variant  of the cut-free, focused sequent calculus $\lbpol$, developed by the first author in \cite{JESENTCS17}. In this logic, the connectives are classified as negative (resp.~positive) if their right (resp.~left) introduction rule(s) is (are) invertible. This imposes a partition of formulas according to their \emph{polarity}, which is extended to atoms. Explicit operators to shift the polarity of a formula are included. $\PIL$ is organized to impose in proof search the alternation of inversion and focusing phases typical of the focusing discipline \cite{LiangMillerTCS09,SimmonsTOCL2014,JESENTCS17}.

As mentioned above, we extend $\PIL$ in two steps, taking a coinductive view of the syntax of proof terms ($\copil$) and adding sums to represents choice points ($\copils$). A coinductive representation $\solfunction{\sigma}\in\copils$ of the solution space determined by a given sequent $\sigma$ is developed. Next, we introduce the alternative, inductively generated syntax $\pils$, the semantics $\interps{\cdot}$ of the latter into $\copils$, together with the finitary representation $\finrepempty{\sigma}$ of the solution space. The correctness of this alternative representation is illustrated as the commutation of the ``triangle'' with solid arrows in the left half of \cref{fig:roadmap}, starting with $\sigma\in\PIL$.

The main application of this infrastructure is in decision problems related to proof search in $\PIL$. With a sequent $\sigma$ as input, we show the decidability of the predicates ``$\sigma$ is inhabited'', ``$\sigma$ is solvable'' (where being solvable means to have a solution), ``$\sigma$ is finite'' (where being finite means to have finitely many inhabitants), and ``$\sigma$ has an infinite solution'' (which amounts to $\sigma$ having a non-terminating run of proof search). Each of such predicates is equivalent to $P(\solfunction{\sigma})$, for some predicate $P$ over the coinductive terms of $\copils$. This $P$ is initially given simply in terms of the extension (the set of members) of the coinductive term, but an alternative, inductive or coinductive characterization is later provided. Next, a corresponding predicate $\mathsf{F}P$, over the terms of $\pils$ is identified, enjoying a syntax-directed, recursive definition, being therefore computable, and such that $P(\solfunction{\sigma})=\mathsf{F}P(\finrepemptyL{\sigma})$. Since $\finrepsymb$ is computable, so is $\mathsf{F}P\circ\finrepsymb$. Hence $P\circ\solletter$ is computable and the original predicate on $\sigma$ decidable. The algorithm is a two-stage process: first calculate $\finrepemptyL{\sigma}$, next recursively traverse this term to decide $\mathsf{F}P$.

\begin{figure}\caption{Roadmaps}\label{fig:roadmap}
\begin{center}
\begin{tabular}{ccc}
\xymatrix{
	\copils\\
	\copil \ar@{--}[u]\\
	\pil \ar@{--}[u] \ar[r]_\finrepsymb \ar@/^3pc/[uu]^\solletter& \pils \ar[luu]_{\interps{\cdot}}
}
&\qquad\qquad&
\xymatrix{
& \copil\\
\coLJT \ar[ur]^{\unt{(\cdot)}}&&\coLJQ \ar[ul]_{\upt{(\cdot)}}\\
&\pil \ar@{--}[uu]\\
\LJT \ar[ur]^{\unt{(\cdot)}} \ar@{--}[uu]&&\LJQ \ar@{--}[uu] \ar[ul]_{\upt{(\cdot)}}
}
\end{tabular}
\end{center}
\end{figure}

As said, one advantage of studying proof search in $\PIL$ (and in polarized logics in general \cite{LiangMillerTCS09}) is that, indirectly and simultaneously, we may study proof search of other proof systems though their interpretation in $\PIL$. As soon as the interpretation of $\mathsf{S}$ in $\PIL$ is \emph{faithful}, it allows the reduction of provability in $\mathsf{S}$ to provability in $\PIL$. Here we seek a property stronger than faithfulness, namely the property of being a \emph{full embedding}, which means that the proof terms inhabiting $\sigma\in\mathsf{S}$ are in bijective correspondence with the proof terms inhabiting the interpretation of $\sigma$ in $\PIL$. This allows ``proof relevant'' reductions of decision problems, for instance: ``Is the number of inhabitants of $\sigma\in\mathsf{S}$ finite'' is reduced to the similar decision for the interpretation of $\sigma$.

This possibility has already been explored  in \cite{JESRMLP-TYPES20}  for (a minor variant of) the focused sequent calculus $\LJT$ \cite{HerbelinPhD}, related to proof search by \emph{backward chaining} (stressed for example in \cite{LiangMillerTCS09}). The faithful interpretation of $\LJT$ into $\PIL$ provided in \cite{JESRMLP-TYPES20}
 is based on a negative polarization $\unt{(\_)}$ of the formulas of intuitionistic logic 
(in other words, $\unt A$ (called $A^*$ in \cite{JESRMLP-TYPES20}) is a negative $\PIL$ formula for any intuitionistic formula $A$). 
Similar translations were developed in various contexts \cite{ZeilbergerPOPL2008,LiangMillerTCS09,CurienMunchMaccagnoni10}.
A new case study offered here is system $\LJQ$ \cite{DyckhoffLengrand2006,DyckhoffLengrand2007}, a well-known focused sequent calculus with a long history in proof theory, following the \emph{forward chaining} strategy (as emphasized in \cite{LiangMillerTCS09}), and connected to {\em call-by-value} functional programming.
Our new proposed interpretation of $\LJQ$ into $\PIL$ is through a positive polarization $\upt{(\_)}$, which we prove to be a full embedding. The algorithms obtained for $\LJT$ or $\LJQ$ are now three-staged, since the polarized interpretation is pre-composed to the two-staged algorithms obtained before for decision problems about $\LJP$.

In fact, the reduction of decision problems through the negative/positive interpretations just described is a corollary of what we actually prove. For $\mathsf{S}$ either $\LJT$ or $\LJQ$, we take its coinductive extension $\mathsf{S}^{co}$, with its own notion of coinductive proof term and solution, and define an interpretation, either negative or positive, of $\mathsf{S}^{co}$ into $\copil$ and prove it to be a full embedding. This allows the inheritance from $\LJP$ of decidability results also for solution-related problems. For instance, solvability in $\mathsf{S}^{co}$ is reduced to solvability in $\copil$, of which the above-mentioned reduction between provability is just a consequence. The situation is depicted in the right half of \cref{fig:roadmap}. This diagram is an illustration of the role of $\LJP$ as a framework: by developing CoIPS in it, we are dispensed of developing CoIPS in the well-known systems $\LJT$ and $\LJQ$, but inherit for them the same benefits as if we had done such separate developments.

\textbf{Comparison with our previous \cite{JESRMLP-TYPES20}.} The consideration of a full language of propositional logic was initiated in \cite{JESRMLP-TYPES20}, where $\LJP$ was studied and the negative interpretation of $\LJT$ introduced.  To illustrate the generalizations developed here, notice that in \cite{JESRMLP-TYPES20} we just obtained reduction between decision problems along the arrow $\LJT\to\LJP$ in the right half of \cref{fig:roadmap}. So here we opened the side relative to $\LJQ$ and lifted the analysis to the coinductive level.

In addition, even if we did not mention this before, the use of $\LJP$ as a framework is developed here for the inheritance, not only of decidability results, but also of some other meta-theoretic results like the disjunction property, and another property which, when it holds, restricts finiteness to the case of unprovability.

The present paper also makes significant investment into meta-syntax in the favour of conciseness, which is very much needed for the luxuriant syntax of $\PIL$ (typical of focused systems, rich in various forms of judgments). The meta-syntax not only concerns the (co-)proof terms, but we also develop a specific syntax for predicates on \emph{forests}, which are our proof-search expressions. They in particular use placeholders for the multi-ary conjunction and disjunction connectives and thus not only avoid writing out several instances of inference rules but, notably, bring uniformity into the proof of the technical statements in the appendix that underly our four main decision procedures (for each of the considered proof systems).

\smallskip
\noindent
\textbf{Plan of the paper.} The sequent-calculus presentation of polarized logic from \cite{JESENTCS17} is reviewed in \cref{sec:background}, including the notational device concerning its five forms of sequents and a number of concrete examples of proof terms.

Coinductive proof search (CoIPS) for $\PIL$ first comes with coinductive syntax in \cref{sec:proof-search-PIPL}.
This starts with the definition of $\copil$ (a parity condition is applied instead of full coinductive reading of the raw syntax) and examples (\cref{sec:copil}), then has the extension $\copils$ with choice points  (\cref{sec:copils}) which gives the forest representation, allowing to associate a forest to each logical sequent (given with examples) that contains all inhabitants and even all solutions, in a precise sense, in \cref{sec:solfunction}.
In \cref{sec:predsforests}, we then introduce the notational device for properties on forests and exhibit the four main examples that are intimately related to the decision procedures that come later in the paper. We illustrate the usefulness of the developed notions with two meta-theoretic properties of $\pil$ in \cref{sec:firstmetathms}: the disjunction property under hypotheses, adapted to polarization, and a sufficient condition for having either none or infinitely many inhabitants.

\Cref{sec:proof-search-PIPL-partII} brings in \cref{sec:pils} the inductive syntax $\pils$
of the finitary forests, and its interpretation into forests, while \cref{sec:finrep} defines the finitary forests associated to logical sequents that represent again the entire search spaces (\cref{thm:equiv-PIPL-simpl}). Another notation system for properties of finitary forests is introduced in \cref{sec:analysis-PIPL}. 
The inductive and the coinductive systems fit well together, as seen in \cref{prop:genfinchar} (whose proof needs inductively defined ``slices'' of the coinductive predicates, corresponding to observations up to a given depth -- this technical is part delegated to the appendix), and through this device, the decision algorithms are obtained in \cref{sec:decision-problems-PIPL}.

Applications to full intuitionistic logic are extracted in Section \ref{sec:applications-IPL}.
As said above, the results concern two rather different and well-known proof systems, $\LJT$ and $\LJQ$, but they are developed for their coinductive extensions, and the decision algorithms also respond to solution-related questions. This is developed in \cref{subsec:LJT-LJTco} and \cref{subsec:LJQ-LJQco}, respectively.

\Cref{sec:final} concludes, with pointers to related work beyond the comparisons made throughout the main text.

There is also an appendix of approximately 10 pages with technical material that the authors consider as an obstacle to a normal flow of reading. The appendix is not just complementary material. It provides technical notions that are used in proofs of the main results of the paper, proofs that are also only carried out in the appendix.
\Cref{sec:appendix-definednessofS} introduces a weight for logical sequents that allows to justify well-definedness of the forest of solutions associated with a logical sequent, more specifically: that the forest satisfies the parity condition. The same weight serves in \cref{sec:appendix-terminationofF} as element of justification of termination of the finitary representation function (for the solutions associated with a logical sequent). In \cref{sec:appendix-exfinnofin}, we introduce ``sliced'' versions of the predicates on forests mentioned above, and we give their properties, their relation with the non-sliced predicates and identify the slices of some of our example predicates. \Cref{sec:appendix-decontraction} is an interjection on decontraction, which is a forest transformation in the situation of inessential extensions of contexts. In \cref{sec:appendix-mainpropgencoindanalysis}, the crucial \cref{prop:genfinchar} is proven based on the material of \cref{sec:appendix-exfinnofin}. \Cref{app:appendix-negative-embedding} and \cref{app:appendix-positive-embedding} give technical complements on the forgetful maps back from subsystems of $\copil$ into $\coLJT$ and $\coLJQ$, respectively.


%% file: background.tex
\section{Background on the system $\PIL$ of polarized propositional logic}\label{sec:background}

We describe the formulas of polarized intuitionistic propositional logic (\pipl) and then
introduce the sequent calculus $\PIL$ for \pipl. $\PIL$ is a variant of the cut-free fragment of $\lbpol$ \cite{JESENTCS17}, and it corresponds to the system with the same name presented in \cite[Section 2]{JESRMLP-TYPES20}, up to minor differences explained below.

\smallskip
\noindent
\textbf{Formulas.} Let us start with the formulas of intuitionistic propositional logic (\ipl), with a presentation that makes the later extension to polarized formulas more visible.
The 
formulas we consider for \ipl\ are made from atoms, from absurdity and are constructed with implication and binary conjunction and disjunction. We subdivide them as follows:
\[
  \begin{array}[4]{rrcl}
    \textrm{(intuitionistic formulas)}&A,B&::= & N\mid R\\
    \textrm{(negative intuitionistic formulas)}&N&::= & A\impl B\mid A\wedge B\\
    \textrm{(positive intuitionistic formulas)}&P&::= &\falsity\mid A\vee B\\
    \textrm{(right intuitionistic formulas)}&R &::= &a\mid P\\[0.5ex]
    \textrm{(left intuitionistic formulas)}&L &::= &a\mid N
\end{array}
\]
where $a$ ranges over atoms, of which an infinite supply is assumed, and the symbols $\perp$, $\wedge$ and $\vee$ obviously stand for falsity, conjunction and disjunction, and $\impl$ stands for implication. A connective is negative (resp.~positive) if its right (resp.~ left) introduction rules are invertible. The classes of negative and positive formulas are auxiliary notions that just classify them according to the outermost connective. The distinction between all (intuitionistic) formulas and right formulas will be crucial in the proof systems for \ipl\ studied in \cref{sec:applications-IPL}. The left formulas do not enter the other productions, so this class is here for comparison with the polarized notion to be described next.

The formulas of \pipl\ are those of \ipl\ enriched with polarity. All formulas, including atoms, are assigned a unique polarity (positive or negative): this is either the polarity of the outermost connective, or is explicitly indicated as $a^+$ and $a^-$, for positive and negative polarity, respectively, in the case of an atomic formula.
There are special unary connectives to switch polarity, so a negative formula can always be converted (or disguised) as a positive formula or vice-versa. This allows us to impose constraints on the polarity of constituent formulas without loss of expressiveness, e.\,g.,the constituents of a disjunction are required to be positive.

More formally, \emph{formulas} of \pipl\ are as follows:
\[
\begin{array}[4]{rrclrrcl}
(\textrm{formulas})
&A&::=&N\mid P  \\
(\textrm{negative})
&N,M&::=&a^- \mid C \\
(\textrm{composite negative})
&C&::=&\upshift P \mid P\impl N\mid N\wedge M\\
  (\textrm{positive})
&P,Q&::=&a^+ \mid D \\
  (\textrm{composite positive})
&D&::=&\downshift N\mid\falsity\mid P\vee Q
\end{array}
\]
Here, as for \ipl, we assume a supply of names of atoms, denoted typically by $a$, and the markers $-$ and $+$ for polarity are added to the atom name as superscripts, giving rise to negative resp.\ positive atoms (for a given name $a$, $a^-$ and $a^+$ are distinct atoms). The symbols $\uparrow$ and $\downarrow$ are polarity shifts (as they are commonly denoted in the literature).

We introduce the auxiliary categories of left and right formulas:
\[
\begin{array}{rrcl}
(\textrm{left/$L$}) &L&::=&a^+\mid N\\
 (\textrm{right/$R$}) &R&::=&a^-\mid P
\end{array}
\]
These taken together comprise all formulas - but the two categories are not disjoint since both categories contain all atoms. The set of formulas is partitioned in three ways: into negative $N$ and positive $P$ formulas; into composite negative $C$ and right formulas $R$; and into composite positive $D$ and left formulas $L$. 

We define syntactic operations on left and right formulas: 
\[
\begin{array}{rclcrcl}
	\Down N&:=&\downshift N && \Down{a^+} &:=& a^+\\ 
	\Up P &:=& \upshift P &&\Up{a^-} &:=&a^-
\end{array}
\] indicating the application of polarity shift only where necessary to obtain the positive formula $\Down L$ and the negative formula $\Up R$, respectively. Thus, the negative formulas are also partitioned as $\Up R\mid P\impl N\mid N\wedge M$, and the positive formulas are also partitioned as $\Down L\mid\falsity\mid P\vee Q$.

There is an obvious forgetful map $\fgt{\_}$ from \pipl\ formulas to \ipl\ formulas (assuming the same supply of atom names), where the polarity shifts and the polarity annotations at the atoms are removed. This mapping only works globally for the set of all formulas, not for the identified subclasses, e.\,g., the composite negative formula $\upshift\falsity$ is mapped to the positive formula $\falsity$, so in particular, negativity is not preserved by the mapping.

In \cref{sec:applications-IPL}, we will use a number of translations $f$ from \ipl\ formulas into \pipl\ formulas that are all right inverses to $\fgt{\_}$, in other words, they are \emph{sections} of that forgetful map, in symbols: $\fgt{f A}=A$ for intuitionistic formulas $A$. Intuitively, this means that they all give a way of exploiting the richer syntax of formulas of \pipl\ w.\,r.\,t.\ \ipl\ in ``decorating'' the formula trees with extra elements present in \pipl. In particular, these translations $f$ are injective (as right inverses). Inspired by the use of that notion for individual formulas in \cite{LiangMiller2024}, we call such a function $f$ a \emph{polarization} of \ipl\ formulas, and we may call $\fgt A$ the \emph{depolarized} version of $A$, hereby directly following \cite{LiangMiller2024}.

\smallskip
\noindent
\textbf{Proof terms.} The proof system we will introduce employs \emph{proof terms}, which are organized in five syntactic categories as follows:\\[-1ex]
\[
\begin{array}{lcrcl}
\textrm{(values)} &  & v& ::= & x\mid{\thunk t}\mid\inj i {P}{v}\\
\textrm{(terms)} &  & t& ::= & \ep{e}\mid\ea{e}\mid\lb p\mid\langle t_1,t_2\rangle\\
\textrm{(co-values/spines)} &  & s & ::= & \nil\mid\cothunk p\mid v::s\mid i::s\\
\textrm{(co-terms)} &  & p& ::= & 
x^L.e\mid \abortt A\mid [p_1,p_2]\\
\textrm{(stable expressions)} &  & e & ::= & \dlv{t}\mid\ret{v}\mid\coretR xsR
\end{array}
\]
where $i\in\{1,2\}$, and 
$x$ ranges over a countable set of variables.

The syntactic categories will correspond to the different forms of sequents handled by the proof system. At first sight, these proof terms are far removed from any familiar sort of $\lb$-terms; and the fact that cut-elimination does not belong to this paper means that no reduction semantics will be given here to help grasping what they are. Very roughly, a term is either a $\lb$-abstraction, a pair of terms, or an expression which can be a returned value, or an ``applicative'' expression, comprising a head variable $x$ and its spine $s$. The values stacked in the spine can be terms-turned-into-values by means of the operator $\thunk{\cdot}$; and the spine, instead of ending with $\nil$, can continue with a $\cothunk{\cdot}$. Thus we can write the ``applicative'' expression $\coret{x}{\thunk{t}::\cothunk{y^L.\ret{\thunk{t'}}}}$, which we can recognize as a kind of generalized application $x(t,y.t')$, if we filter all the fine-grained tagging.

As detailed in \cite{JESENTCS17}, this language refines call-by-push-value \cite{Levy06}, with the positive/negative distinction being related to the value/computation distinction. Example \ref{ex:types} at the end of the current section shows variant forms of the identity combinator or the Church numerals allowed in this syntax; and in Section \ref{sec:applications-IPL} the translation of the more familiar proof terms from $\LJT$ into these proof terms gives additional insight. Bear in mind proof terms are the cornerstone of coinductive proof search, as both the coinductive and the finitary representations of search spaces are based on them.

Now some technical comments. Notice that we restrict the upper index in $\inj i {P}{v}$ to positive formulas $P$ already in the syntax, not only later through the typing rules. Likewise for the restriction of the upper index in $x^L.e$ to left formulas $L$ and for the restriction of the upper index in $\coretR xsR$ to right formulas $R$. Unlike $\PIL$ in our previous presentation \cite{JESRMLP-TYPES20}, which considers disjoint countable sets of \emph{positive variables} and \emph{negative variables} (ranged over by $z$ and $x$, respectively), here we consider only one class of  (arbitrary) variables, and amalgamate the two binding constructions into $x^L.e$, with $L$ a left formula. Consequently,  values now comprise arbitrary and not only positive variables, and the head $x$ of a stable expression $\coretR xsR$ is no longer limited to negative variables. However, the typing rules (to be introduced below) have these polarity conditions baked in, so our present system is no more permissive than before. In other words, as proof systems, the former and the present version of $\PIL$ are isomorphic. Notice also that the upper index in $\coretR xsR$ is not present in the former version of $\PIL$; it serves as a preparation for the coinductive variant $\copil$, introduced in \cref{sec:copil}, where this data is needed to exclude a striking counterexample concerning uniqueness of types.
We use this annotation also for qualifying a stable expression as \emph{atomic} iff it is of the form $\coretR xs{a^-}$, i.\,e., it must be in the last of the three cases in the grammar, and even with $R=a^-$.

We use the typical letters for denoting elements of the syntactic
categories as sorts: let $S:=\{v,t,s,p,e\}$ be the set of sorts, and use
letter $\tau$ to denote any element of $S$.

Often we refer to all proof terms of $\PIL$ as \emph{expressions}, and use letter $T$ to range over expressions in this wide sense ($T$ being reminiscent of terms, but not confined to the syntactic category $t$). To shorten notation, we communicate $\langle t_1,t_2\rangle$ and $[p_1,p_2]$ as $\langle t_i\rangle_i$ and $[p_i]_i$, respectively.

\smallskip
\noindent
\textbf{Proof system.} We are ready to present proof system $\PIL$, a focused sequent calculus for reasoning with \pipl\ formulas that slightly deviates from the system introduced in \cite{JESRMLP-TYPES20}, which in turn corresponds to a variant of the cut-free fragment of $\lbpol$ \cite{JESENTCS17}, as already mentioned.

As in the presentation of $\PIL$ in \cite{JESRMLP-TYPES20} (and following  \cite{JESENTCS17}), \emph{contexts} $\Gamma$  are made of associations of variables with left formulas. Note that, despite having given up on the separation of variables into positive variables and negative variables,  when reasoning in a context $\Gamma$ consisting of left formulas, we always know which variables are associated to positive atoms and which to negative formulas.

There are five forms of \emph{sequents}, one for each syntactic category $\tau$ of proof terms:
\[
\begin{array}[4]{rclrcl}
(\textrm{focus negative left})&& \Gamma[s:N]\vdash R & (\textrm{focus positive right})&& \Gamma\vdash[v:P]\\
(\textrm{invert positive left})&& \Gamma\mid p:P\Longrightarrow A & (\textrm{invert negative right})&& \Gamma\Longrightarrow t:N\\[1ex]
(\textrm{stable})&& \Gamma\vdash e:A
\end{array}
\]
The sequents can be uniquely decomposed into a proof term and a \emph{logical sequent}: the latter are sequents without proof-term annotations, i.\,e., 
$$\Gamma[N]\vdash R \qquad \Gamma\mid P\Longrightarrow A \qquad\Gamma\vdash A \qquad \Gamma\vdash[P]\qquad\Gamma\Longrightarrow N  \enspace.$$

The inference/typing rules of $\PIL$ are given in \cref{fig:typingpipl}.
\begin{figure}[tb]\caption{Typing rules of $\PIL$}\label{fig:typingpipl}
	\[
	\begin{array}{c}
		\infer[]{\Gamma,x:a^+\vdash[x: a^+]}{}\quad\infer[]{\Gamma\vdash[\thunk t:\downshift N]}{\Gamma\Longrightarrow t:N}\quad\infer[i\in\{1,2\}]{\Gamma\vdash[\inj i{P_{3-i}}v:P_1\vee P_2]}{\Gamma\vdash[v:P_i]}\\[2.5ex]
		\infer[]{\Gamma\Longrightarrow \ea e:a^-}{\Gamma\vdash e:a^-}\quad\infer[]{\Gamma\Longrightarrow \ep e:\upshift P}{\Gamma\vdash e:P}\quad\infer[]{\Gamma\Longrightarrow\lb p:P\impl N}{\Gamma\mid p:P\Longrightarrow N}\quad\infer[]{\Gamma\Longrightarrow\langle t_i\rangle_i: N_1\wedge N_2}{\Gamma\Longrightarrow t_i: N_i\quad\textrm{for $i=1,2$}}\\[2.5ex]
		\infer[]{\Gamma[\nil:a^-]\vdash a^-}{}\quad\!\infer[]{\Gamma[\cothunk p:\upshift P]\vdash R}{\Gamma\mid p:P\Longrightarrow R}\\[2.5ex]
		\quad\!	\infer[]{\Gamma[v::s:P\impl N]\vdash R}{\Gamma\vdash[v:P]\quad\!\Gamma[s:N]\vdash R}\quad 
		\infer[i\in\{1,2\}]{\Gamma[i::s:N_1\wedge N_2]\vdash R}{\Gamma[s:N_i]\vdash R} \\[2.5ex]
		\infer[]{\Gamma\mid x^L.e:\Down L\Longrightarrow A}{\Gamma,x:L\vdash e:A}\quad
		\quad\!	\infer[]{\Gamma\mid\abortt A:\falsity\Longrightarrow A}{}\quad\!\infer[]{\Gamma\mid[p_i]_i:P_1\vee P_2\Longrightarrow A}{\Gamma\mid p_1:P_1\Longrightarrow A\quad\Gamma\mid p_2:P_2\Longrightarrow A}\\[2.5ex]
		\infer[]{\Gamma\vdash\dlv t:C}{\Gamma\Longrightarrow t:C}\quad\infer[]{\Gamma\vdash\ret v:P}{\Gamma\vdash[v:P]}\quad\infer[]{\Gamma,x:N\vdash\coretR xsR:R}{\Gamma,x:N[s:N]\vdash R}
	\end{array}   
	\]
\end{figure}
These rules serve to derive sequents. When a sequent $\sigma$ can be derived by building a finite tree of rule applications in the usual manner, we say that $\sigma$ is \emph{valid}. The rules can be seen as typing rules assigning logical sequents $\sigma$ to proof terms $T$;
or they can be seen as inference rules, determining the theorems of the logic together with the proof terms which inhabit them.

The forms of the sequents relate to different instructions in the process of bottom-up proof search as organized by the focusing discipline. Two of them correspond to the instruction of keeping the focus, whenever possible, on a positive formula $P$ in the r.\,h.\,s.~or on a negative formula $N$ in the l.\,h.\,s.~of the sequent; two of them correspond to the inversion either of $N$ in the r.\,h.\,s.~or $P$ in the l.\,h.\,s.~of the sequent; and one -- the stable sequent -- asks for a decision as to which formula to focus on. The inference rules for the polarity shifts, when read bottom-up, signal the passage from the focusing phase to the inversion phase, or the ending of an inversion phase and the return to a stable sequent. We cannot further invert a positive formula $\Down L$ in the l.\,h.\,s.~and similarly we cannot further invert a negative formula $\Up R$ in the r.\,h.\,s. This is a rationale for the terminology ``left''/``right'' formula. More details in \cite{JESENTCS17} \footnote{The rule witnessed by the construction $\dlv{\cdot}$ is an exception to the nice symmetries of the focusing discipline. It concludes a stable sequent, but, when read bottom-up, is not about deciding a focus. Typically the rule is used in the middle of the process of inverting $P\impl N$. First we invert $P$ in the l.\,h.\,s.~, and in most cases we return to a stable sequent with $N$ as its succedent formula. If $N$ is composite, we want to jump immediately to its inversion, and this passage is the role of the $\dlv{\cdot}$ rule.}.

Some further notation and terminology:
\begin{itemize}
	\item We will systematically use the following notational device: If $\jay$ is any logical sequent and $T$ a proof term of the suitable syntactic category, let $\jay(T)$ denote the sequent obtained by placing ``$T:$'' properly into $\jay$, e.\,g., if $\jay=(\Gamma\mid P\Longrightarrow A)$, then $\jay(p)=(\Gamma\mid p:P\Longrightarrow A)$ (the parentheses around sequents are often used for better parsing of the text).
	\item If the syntactic category $\tau$ is the category of proof terms corresponding to a certain form of logical sequents, then we may use $\jay^\tau$ to range over logical sequents of that particular form, e.\,g., an arbitrary logical sequent $\Gamma\vdash A$ is indicated by $\jay^e$.
	\item The letters $\rho, \rho'$ etc.~will range over the logical sequents of the form $\Gamma\vdash R$, with an R-formula on the right-hand side. Those will be called \emph{R-stable sequents} (omitting ``logical''). R-stable sequents do not type proof terms of the form $\dlv{t}$.
	\item $\Gamma\leq\Gamma'$ iff $\Gamma\subseteq\Gamma'$ and $|\Gamma|=|\Gamma'|$, with $|\Delta|:=\{L\mid \exists \mathit{y},\, (\mathit{y}:L)\in\Delta\}$ for an arbitrary context $\Delta$ (where we write $\mathit{y}$ for an arbitrary variable). That is, $\Gamma\leq\Gamma'$ if $\Gamma'$ only has extra bindings w.\,r.\,t.\ $\Gamma$ that come with types that are already present in $\Gamma$. If $\Gamma\leq\Gamma'$, we say $\Gamma'$ is an \emph{inessential extension} of $\Gamma$. 
	\item $\rho\leq\rho'$ iff  for some $\Gamma\leq\Gamma'$ and for some right formula $R$, $\rho=(\Gamma\vdash R)$ and $\rho'=({\Gamma'}\vdash R)$. If $\rho\leq\rho'$, we say $\rho'$ is an \emph{inessential extension} of $\rho$. (The definitions of $\leq$ are an immediate adaptation of the corresponding definition in \cite{EspiritoSantoMatthesPintoInhabitation}.)
\end{itemize}

\smallskip
\noindent
\textbf{Comparison with our previous presentations of $\PIL$.} The notion of formula for \pipl\ is the same as the one used in $\lbpol$. The syntax of proof terms deviates from $\lbpol$ \cite[Figure 4]{JESENTCS17} in the following ways: the letters to denote values and covalues are now in lower case, the two expressions to type the cut rules are absent, and the last form of values (the injections) and $\abort$ come with type information, as well as the displayed binding occurrence of variable $x$ in the first form $x^L.e$ of co-terms---all the other syntax elements do not introduce variable bindings, in particular, there is no binding in $\lb p$ or $\coretR xsR$.

Comparing with the typing rules of the presentation of $\PIL$ in \cite[Fig.~2]{JESRMLP-TYPES20}, we  see now one fewer rule: the typing rules for $z^{a^+}.e$ and $x^N.e$ are now given uniformly for a left formula $L$, thanks to the auxiliary operation $\Down L$ (the first rule of the fifth line).  But, as already anticipated, as proof systems, these two presentations of $\PIL$ are isomorphic.  Comparing with \cite[Figures 1--3]{JESENTCS17} (omitting the cut rules) there are more differences. On the one hand, the typing relation of $\PIL$ is slightly reduced: focus negative left sequents of $\PIL$ are restricted to $R$-formulas on the right-hand side,  which is enabled by the fact that in $\lbpol$ the $\mathit{Focus}_L$-rule  (the one typing the $\coretsymb$ construction for proof terms) can be restricted to $R$-formulas on the right-hand side. On the other hand, there are some obvious differences resulting from the fact proof terms of $\PIL$ come with some extra type information as compared to $\lbpol$.

\smallskip
\noindent
\textbf{Uniqueness of types.} The extra type information in proof terms of $\PIL$ ensures \emph{uniqueness of types} in the following sense: given the
shown context $\Gamma$, type $N$ (for sort $s$) and proof term, there is at most one formula that can replace any of the placeholders in
$\Gamma[s:N]\vdash \cdot$, $\Gamma\vdash[v:\cdot]$,
$\Gamma\mid p:\cdot\Longrightarrow \cdot$, $\Gamma\Longrightarrow t:\cdot$ and
$\Gamma\vdash e:\cdot$ so that the obtained sequent is valid. The annotation $R$ in $\coretR xsR$ is not needed for uniqueness to hold.
Thus, the logical sequents are divided into contextual information ($\Gamma$ and $N$ for sort $s$, $\Gamma$ for the other sorts) and the formulas that are uniquely determined by this contextual information and the proof term to be typed: two formulas for sort $p$ and one formula for the other sorts.

\smallskip
\noindent
\textbf{Notational device.} We use the set $S$ of sorts to give a more uniform view of the
different productions of the grammar of $\PIL$ proof terms. E.\,g., we
consider $\thunk\cdot$ as a unary function symbol, which is
sorted\footnote{Since we will \emph{type} proof terms by logical sequents, we prefer to speak of \emph{sorting} of the function symbols.} as $t\to v$, to be written as $\thunk\cdot:t\to v$. As
another example, we see co-pairing as binary function symbol
$[\cdot,\cdot]:p,p\to p$. 
This notational device does not take into
account variable binding, and we simply consider $x^L.\,\cdot$ as
a unary function symbol for every $x$ and every left formula $L$. The
variables $x$ have no special role either in this view, so they are
all nullary function symbols (i.\,e., constants) with sort
$v$. Likewise, for every variable $x$ and every right formula $R$, $\coretR x\cdot R$ is a
unary function symbol sorted as $s\to e$. 
We can thus see the
definition of proof terms of $\PIL$ as based on an infinite signature,
with function symbols $f$ of arities $k\leq2$. The inductive
definition of proof terms of $\PIL$ can then be depicted in the form
of one rule scheme:\\[-1ex]
\[\infer[]{f(T_1,\ldots,T_k):\tau}{f:\tau_1,\ldots,\tau_k\to \tau\quad T_i:\tau_i, 1\leq i\leq k}\]
Later we will write $f(T_i)_i$ in place of $f(T_1,\ldots,T_k)$ and assume that $k$ is somehow known. Instead of writing the $k$ hypotheses $T_i:\tau_i$, we will then just write $\bigwedge_i (T_i:\tau_i)$.

\smallskip
\noindent
\textbf{Examples of inhabitants.} An \emph{inhabitant} of a logical sequent $\sigma$ is any proof term $T$ such that the sequent $\sigma(T)$ is valid. A logical sequent $\sigma$ is \emph{inhabited} if there is an inhabitant of $\sigma$.
Moreover:
\begin{itemize}
	\item An \emph{inhabitant of sort $t$ of $N$} is an inhabitant of the sequent $\Longrightarrow N$. 
	\item An \emph{inhabitant of sort $v$ of $P$} is an inhabitant of the sequent $\vdash[P]$.
	\item An \emph{inhabitant of sort $e$ of $A$} is an inhabitant of the sequent $\vdash A$.
\end{itemize}
Similar definitions will be used throughout for the systems studied in this paper. Instead of inhabitants of sort $t$, we will often just speak of inhabitants.

\begin{example}\label{ex:types} 
Let us consider now specific formulas of $\PIL$ that will be used as running examples in the paper, and can already give an illustration of the richness of ``behaviours'' allowed by the polarity devices of $\PIL$. The starting points are three implicational intuitionistic formulas on a single atom $a$, namely $\Idt := a\impl a$, $\Inft:=(a\impl a) \impl a$  and $\Churcht:=(a\impl a) \impl a \impl a$ (as usual, we assume \emph{right associativity} in presence of a chain of implications).  The last two cases serve as running examples in \cite{JESRMLP-APAL2021}, in the context of a presentation of a proof system for intuitionistic implication,  corresponding to STLC restricted to normal forms.

Let us say that a formula $A$ of \pipl\ is a \emph{polarization of an intuitionistic formula} $B$ when $\fgt B=A$.
  \begin{enumerate}
  \item  In the case of $\Idt$, fixing a uniform \emph{negative} polarization of $a$, we find, e.\,g., the polarization $\Idt^-:=\downshift a^- \impl a^- $, which is a negative type.
    Fixing a uniform \emph{positive} polarization of $a$, we find, again as examples, the negative type $\Idt^{+n}:=a^+ \impl \upshift a^+$ and the positive type $\Idt^{+p}:=\downshift {\Idt^{+n}}$. For the first two examples, there is exactly one inhabitant of sort $t$ in $\PIL$, namely $\lb (x^{a^-}.\coretR x\nil{a^-})$, respectively $\lb (x^{a^+}.\dlv{\ep{\ret x}})$, and in the third example there is exactly one inhabitant of sort $e$ in $\PIL$: $\ret{\thunk{\lb (x^{a^+}.\dlv{\ep{\ret x}})}}$ (which applies $\thunksymb$ then $\retsymb$ to the inhabitant of $\Idt^{+n}$)
    agreeing in all the examples with the number of (normal) inhabitants of $\Idt$ in STLC. The shortest polarization for $\Idt$  is $\Idt^{+-}:=a^+ \impl a^-$. However, this type is not inhabited in $\PIL$ (recall that  axioms on a given atom name require the same positive polarity sign).
\item Three possible polarizations of $\Inft$ are: $\Inft^-:= \downshift (\downshift a^-\impl a^-) \impl a^- $, $\Inft^+:=\downshift (a^+\impl \upshift a^+) \impl \upshift a^+$ and $\Inft^{+--}:= \downshift (a^+\impl a^-) \impl a^-$.  None of these types is inhabited in $\PIL$, in line with the non-existence of inhabitants of $\Inft$ in STLC. However, $\Inft$ has one \emph{infinite solution} in coinductive STLC (\cite[Example 2]{JESRMLP-APAL2021}), and this property is only shared by the polarization $\Inft^-$ (as we will see ahead).
\item We also consider several polarizations of $\Churcht$ (all negative types except where the polarity is indicated): $\Churcht^-:= \downshift (\downshift a^-\impl a^-) \impl \downshift a^- \impl a^-$, $\Churcht^{+n}:=\downshift (a^+\impl \upshift a^+) \impl a^+ \impl \upshift a^+$, $\Churcht^{+p}:=\downshift{\bigl(\downshift (a^+\impl \upshift a^+) \impl \upshift\downshift{(a^+ \impl \upshift a^+)}\bigr)}$ (a positive type) and $\Churcht^{+-}:= \downshift (a^+\impl a^-) \impl a^+ \impl a^-$.
\begin{enumerate}
\item For $\Churcht^-$, we find an infinity of inhabitants of sort $t$ of the form
  $$\lambda (x^{\downshift a^-\impl a^-}.\dlv{\lambda (y^{a^-} .e_k}))$$ ($k\geq 0$), where $e_0:=\coretR y \nil{a^-}$ and
$e_{k+1}:=\coretR{x}{\thunk{\ea{e_k}}::\nil}{a^-}$, which can be thought of as a representation of the \emph{Church numerals} in $\PIL$. Notice that $e_k$ is an inhabitant of $x:\downshift a^-\impl a^-,y:a^-\vdash a^-$.
\item For $\Churcht^{+n}$, we also find an infinity of inhabitants of sort $t$, but here distinct infinite families of inhabitants can be built, such as $\lambda (x^{a^+\impl \upshift a^+}.\dlv{\lambda (y^{a^+} .\dlv{\ep{\tilde e_k}}}))$ ($k\geq 0$), taking $\tilde e_0:=\ret y$ and, either
$\tilde e_{k+1}:=\coretR{x}{y::\cothunk{z^{a^+}.{(\tilde e_k)[y:=z]}}}{a^+}$ or
$\tilde e_{k+1}:=\coretR{x}{y::\cothunk{\_.{\tilde e_k}}}{a^+}$. The family in the first case is closer to the \emph{Church numerals} in the sense that the $\lb$-abstracted variable $y$ (of type $a^+$) is used only once in any family member, whereas the $k$-th element of the family for the second choice requires $k$ uses of $y$. In any case, $\tilde e_k$ is an inhabitant of $x:a^+\impl \upshift a^+,y:a^+\vdash a^+$.
\item For $\Churcht^{+p}$, we can construct analogous sequences of inhabitants of sort $e$, which are all given as
  \[\ret{\thunk{\lambda (x^{a^+\impl \upshift a^+}.\dlv{\ep{\ret{\thunk{\lambda (y^{a^+} .\dlv{\ep{\tilde e_k}})}}}})}}\enspace,\]
  with the two options for defining $\tilde e_k$ just as before for $\Churcht^{+n}$.
\item Very differently, for $\Churcht^{+-}$ only one inhabitant can be constructed, namely $$\lambda (x^{a^+\impl  a^-}.\dlv{\lambda (y^{a^+} .\coretR x{y::\nil}{a^-})})\enspace.$$
\end{enumerate}
  \end{enumerate}
\end{example}


%% file: proof-search-PIPL.tex
\section{Coinductive approach to proof search in $\PIL$ - part I}\label{sec:proof-search-PIPL}

In this and the next section, we adapt our coinductive approach to proof search
from implicational intuitionistic logic to the focused sequent calculus $\PIL$ and the full language of polarized intuitionistic propositional logic. The approach has two parts. The first, covered in the current section, develops a coinductive characterization of the solution space corresponding to the proof search problem determined by a given sequent. In the second part, to be covered in the next section, we develop an alternative, inductive characterization of the same spaces, on which we can base algorithms for decision problems.

In both cases, the focus is on the concept of \emph{solution} -- a generalization of the concept of proof. A proof is a finite tree belonging to the set inductively generated by the inference rules; but it also is the finite tree output by the process of bottom-up application of the inference rules, when all the branches lead to instances of axioms of the proof system, starting from the given logical sequent whose proof is sought. A solution is the possibly infinite tree generated in the same way, when all the branches are only required to never lead to failure -- and failure here is a sequent from where no inference rule is applicable (bottom-up). Hence the proofs are just the finite solutions, but proof search produces in general infinite structures. Surely, an infinite solution is no evidence of the validity of the sequent at its root, and we do not regard it as a cyclic proof, but we take a positive view on solutions, grant them a status of ``first-class citizens'', and also study decision problems about them: as we will see, our methods do not apply better to proofs than to solutions.

This section has 5 subsections. In the first two, we extend $\PIL$ to $\copil$ and then to $\copils$. In $\copil$ we have means to represent individual solutions (co-proof terms) and in $\copils$ we have means to represent solution spaces (forests). The representation of the solution space generated by a sequent is defined and studied in the third subsection. In the fourth subsection we develop a generic and uniform treatment of predicates on forests, to be matched with a finitary counterpart, in part II of our approach. In the last subsection, we pause to illustrate what the tools of this section already achieve.

When moving from implicational logic to $\PIL$, the prolix syntax of the latter, with its high number of syntactic categories and different constructors for proof 
terms, becomes a problem, due to the mostly similar rules that appear in definitions, and mostly similar cases that appear in proofs of meta-theorems. We use the extra notational devices from the end of Section~\ref{sec:background} to ensure a uniform
presentation of similar rules and cases.

\subsection{System $\copil$}\label{sec:copil}
In the footsteps of our coinductive approach to proof search for implicational logic, we start by considering a coinductive extension of $\PIL$ that we call $\copil$. In this extended system, proof terms of $\PIL$ are generalized by allowing a coinductive reading of the grammar of expressions of $\PIL$, and, concomitantly, the typing relation is generalized, by taking a  coinductive interpretation of the typing rules of $\PIL$. The typable expressions of $\copil$ offer a  representation of (individual) solutions of proof search problems (i.\,e., logical sequents ) of $\PIL$.
This refines the development  in \cite{JESRMLP-TYPES20}, where we immediately jumped to the system $\copils$ (recalled ahead), which ``only'' offers the possibility of representing (full) solution spaces.

\begin{definition}[Expressions]
	The values, terms, co-values/spines, co-terms and stable expressions of $\copil$, expressions in the wide sense, are defined by the simultaneous coinductive definition obtained by taking the simultaneous inductive definition of expressions of $\PIL$ and reading it coinductively -- we refrain from repeating the same grammar, written with $::=_\cosign$ just to signal the coinductive reading. 
\end{definition}
The expressions of $\copil$ are still organized into five categories,
introduced by the simultaneous coinductive definition of the sets
$\valuesco$, $\termsco$, $\covaluesco$, $\cotermsco$, and $\expsco$.
However, we will continue to use the sorts $\tau$ taken from the set $S$ that was introduced for $\PIL$. This allows
us to maintain the function-symbol view of $\PIL$ with the same symbols $f$ that keep their sorting.
The definition of the set of
the expressions (in a wide sense) of $\copil$ can thus be expressed very concisely as being obtained by a single rule scheme:
\begin{equation}\label{eq:rule-scheme}
\begin{array}{c}\infer=
	{f(T_1,\ldots,T_k):\tau}{f:\tau_1,\ldots,\tau_k\to \tau\quad \bigwedge_i (T_i:\tau_i)}
\end{array}
\end{equation}
The doubly horizontal line indicates a coinductive reading. 

Although the general idea to obtain expressions is to give a coinductive reading to the grammar of expressions of $\PIL$, actually, for the expressions of interest to us, the coinductive reading will be attached only to certain stable expressions.
\begin{definition}[Co-proof terms]
  The \emph{co-proof terms} of $\copil$ are defined exactly as the expressions, by the rule scheme (\ref{eq:rule-scheme}), except that the coinductive reading is restricted to the case $f=\coretR x\cdot R$, being the rule read inductively for the other cases of $f$.
Co-proof terms are ranged over by the letter $T$. 
\end{definition}
To be clear, we restrict the 
infinitary expressions to obey the following property: infinite branches must go
infinitely often through the 
rule for the unary function symbols of form $\coretR x\cdot R$, of signature $s\to e$.

\smallskip
\noindent
\textbf{Parity condition.} 	This restriction can be	expressed as the \emph{parity condition} (known from parity automata where this is the acceptance condition) based on priority~$2$ for
the rules for all $\coretR x\cdot R$
and priority~$1$ for all the others.
The parity condition requires that the maximum of the priorities seen infinitely often on a path in the
(co-proof term)	construction is even, hence infinite cycling through the other function symbols
is subordinate to infinite cycling through the rule (scheme) for $\coretR x\cdot R$.
Put less technically, we allow infinite branches in the
construction of co-proof terms, but infinity is not allowed to come from infinite use solely of the ``auxiliary'' productions,
thus, in particular ruling out infinite pairing with angle brackets, infinite copairing with brackets or infinite spine composition by way of one of the $::$ constructors -- all of which would never correspond to typable co-proof terms (we are typing co-proof terms with the same finite types of $\PIL$, see below).

Notice that, since our co-proof terms are finitely branching, König's lemma implies that a co-proof term is infinite (i.\,e., not a proof term) iff it has infinitely many occurrences of constructors of priority~$2$.

\smallskip
\noindent
\textbf{Equality.} Since co-proof terms are potentially built from infinitely stacking finitary syntax elements (they are infinitary not in the sense of infinite branching but potentially infinite height),
the notion of equality is not just syntactic equality, but rather \emph{bisimilarity} modulo $\alpha$-equivalence, obtained from the first step above that gives full coinductive reading to the grammar of expressions of $\PIL$.  
Following common mathematical practice, we still use plain equality to denote bisimilarity.

\begin{definition}[Typing system of $\copil$]
In $\copil$, types stay inductive, contexts stay finite, the notion of logical sequent stays unchanged w.\,r.\,t.~$\PIL$ -- but sequent  $\sigma(T)$ is a possibly infinite object, as $T$ can be infinite. The typing rules of \cref{fig:typingpipl} have to be interpreted coinductively. Given a logical sequent $\sigma$ and a co-proof term $T$, the typing relation thus defined determines when the sequent $\sigma(T)$ is \emph{valid}. This will be the case when there is a derivation of this sequent -- a (possibly infinite) tree of sequents -- generated by applying the typing rules bottom-up. When $\sigma(T)$ is valid, the (possibly infinite) co-proof term $T$ is called a \emph{solution} of the logical sequent $\sigma$. Accordingly, the logical sequent $\sigma$ is called \emph{solvable} if there is a solution of $\sigma$.
\end{definition}
Following common practice, we symbolize the coinductive reading of a (typing) rule by the double horizontal line, but, of course,  we
refrain from displaying \cref{fig:typingpipl}  again with double lines. 
The imposed parity condition on co-proof terms potentially limits the bottom-up construction of derivations.\footnote{We mentioned before that infinite pairing, infinite copairing, and infinite spine composition would anyway not be typable since the types remain being constructed from the inductively defined formulas. We will, however, not develop a formal result that the parity condition does not forbid any infinitary derivation. The parity condition serves us to have the right notion of raw syntax.}

Notice that, in $\copil$, when $\sigma(T)$ is valid, we say $T$ is a solution of  $\sigma$ (not an inhabitant). Also, we will say that a co-proof term $T$ is a \emph{solution of a formula} $A$ of sort $\tau\in\{t,e,v\}$ when $T$ is a solution of the logical sequent of the appropriate kind formed with the empty context and $A$ on the right-hand side.

Derivations in $\copil$ subsume derivations in $\PIL$, and we can show by induction on expressions of $\PIL$:
\begin{lemma}\label{lem:conservativeness-typing}
For any $T\in\PIL$, $\sigma(T)$ is valid in $\PIL$ iff   $\sigma(T)$ is valid in $\copil$.
\end{lemma}
Hence, we may say that solutions in $\copil$ subsume typable expressions in $\PIL$, and we may refer to the latter as \emph{finite solutions}, and as \emph{infinite solutions} to those corresponding to  infinitary co-proof terms (thus expressions of $\copil$, but not of  $\PIL$). 

\begin{example}
\label{ex:infinite-solutions}
Let us pick up on \cref{ex:types}.  Here we give examples of infinite solutions of types $\Inft^-$,  $\Churcht^-$, $\Churcht^{+n}$ and $\Churcht^{+p}$, taking inspiration from the infinite families of inhabitants given for the polarizations of $\Churcht$.
\begin{enumerate}
\item Let $T_1:= \lambda (x^{\downshift a^-\impl a^-}. e_\omega )$, where $e_\omega$ stands for the (unique) solution in $T$ of the fixed-point equation $T=\coretR{x}{\thunk{\ea{T}}::\nil}{a^-}$.
The validity of  $x:{\downshift a^-\impl a^-}\vdash e_\omega: a^-$ is established coinductively. Hence, $T_1$ is an infinite solution of $\Inft^-$, which can be associated to the unique infinite solution of $\Inft$ in coinductive STLC (\cite[Example 2]{JESRMLP-APAL2021}). Recall that $\Inft^-$ has no inhabitants in $\PIL$.

\item Similarly, $T_2:= \lambda (x^{\downshift a^-\impl a^-}. \dlv{\lambda (y^{a^-} .
e_\omega)})$ is an infinite solution of $\Churcht^-$  (in the required valid intermediary sequent $x:{\downshift a^-\impl a^-},y:{a^-} \vdash e_\omega: a^-$, the declaration $y:{a^-}$ never gets used).
\item Let $T_3:=\lambda (x^{a^+\impl \upshift a^+}. \dlv{\lambda (y^{a^+} . \dlv{\ep{\omegatwo}})})$,
where $\omegatwo$ stands for the  solution of the fixed-point equation $T=\coretR{x}{y::\cothunk{\_.{T}}}{a^+}$
. We can show coinductively the validity of $x:{a^+\impl \upshift a^+}, y:{a^+},\Gamma\vdash \omegatwo:a^+$ , for any $\Gamma$  composed of $n\geq 0$ declarations $z_i:a^+$ for $i\leq n$ (which are never used). From this it is easy to see that $T_3$ is an infinite solution of $\Churcht^{+n}$.
\item Let $T_4:= \retbig{\thunkbig{\lambda \bigl(x^{a^+\impl \upshift a^+}.\dlvbig{\ep{\ret{\thunk{\lambda (y^{a^+} .\dlv{\ep{\omegatwo}})}}}}\bigr)}}$, with $\omegatwo$ as before. In particular, $x:{a^+\impl \upshift a^+}, y:{a^+}\vdash \omegatwo:a^+$ is valid, and, from this follows easily the validity of $\vdash T_4:\Churcht^{+p}$, in other words, $T_4$ is a solution of $\Churcht^{+p}$ (infinite, of course).
\end{enumerate}
\end{example}

\begin{example}
	For the coinductive reading of syntax and typing, the tagging of $\coretsymb$ with formulas really makes a difference, in what concerns uniqueness of types. Let $E$ be the unique co-proof term of sort $e$ satisfying $E=\coretR x{\cothunk{[y_1^{a_1^+}.E,y_2^{a_2^+}.E]}}R$. Then, $x:\upshift{(a_1^+\lor a_2^+)}\vdash E:R$ is valid, but $R$ is only determined by the tag on $\coretsymb$. Had it not been there, every right formula would be a type of $E$ in this context.
\end{example}

\subsection{System $\copils$}\label{sec:copils}

Now, we extend $\copil$ in order to  capture \emph{choice points} in the search process, and have the means to represent the full collection of solutions of a logical sequent.  This is realised in system $\copils$ (a system introduced in \cite{JESRMLP-TYPES20}), which extends expressions of $\copil$ with \emph{formal sums}.  These are not added to the categories of (co)terms (sorts $t$ and $p$), since in the focused system $\PIL$ (co)terms serve to represent the inversion phase in proof search, where choice is not called for.
\begin{definition}[Expressions]
	The values, terms, co-values/spines, co-terms and stable expressions of $\copils$, expressions in the wide sense, are defined by the simultaneous coinductive definition obtained by taking the simultaneous coinductive definition of expressions of $\copil$ and adding the following three clauses:
	\[
	v::=\!\_{\cosign}\,\, \cdots\mid v_1+\cdots+v_k \qquad s::=\!\_{\cosign}\,\, \cdots\mid s_1+\cdots+s_k \qquad e::=\!\_{\cosign}\,\, \cdots\mid e_1+\cdots+e_k 
	\]
\end{definition}
\noindent where, in all three cases, $k\geq 0$.

Again,
expressions comprise five categories, introduced by the simultaneous coinductive definition of the sets
$\Valuesco$, $\Termsco$, $\Covaluesco$, $\Cotermsco$, and $\Expsco$.
Using still the sorts $\tau$ taken from the set $S$ introduced for $\PIL$ and the sorted function-symbol view, we add to the expressions of $\copil$ finite sums for the classes of values, spines and expressions, denoted with the multiary function symbols $\Sigma^\tau$ for $\tau\in\{v,s,e\}$. We write $\oo$ (possibly with the upper index $\tau\in\{v,s,e\}$ that obviously cannot be
inferred from the summands) for empty sums.
\begin{definition}[Forests]
	The \emph{forests} of $\copils$ are the expressions defined by the rule schemes
\[
\infer[f\neq \coretR x\cdot R]{f(T_1,\ldots,T_k):\tau}{f:\tau_1,\ldots,\tau_k\to \tau\quad \bigwedge_i (T_i:\tau_i)} \qquad 
\infer={\coretR xT R:e}{T:s} \qquad
\infer[\tau\in\{v,s,e\}]{\sum^\tau_i T_i:\tau}{\bigwedge_i(T_i:\tau)}
\]
Forests are ranged over by the letter $T$.
\end{definition}
Therefore, as in co-proof terms of $\copil$, infinite branches in forests must go
infinitely often through the (inherited) $e$-formation rule for $\coretR x\cdot R$. Again, this can be
expressed as the \emph{parity condition} by assigning priority 2 only to (the rule for) $\coretR x\cdot R$, determining that infinity cannot come from
infinite use solely of the other constructors from $\copil$ or the sum operator, hence all the other rules are assigned priority 1.\footnote{The notion of forests thus obtained is more restrictive than in \cite[Sect.~3.1]{JESRMLP-TYPES20}:
  there, infinite branches in forests have to go infinitely often through any of the (inherited) $e$-formation rules coming from $\PIL$,
  i.\,e., use in sort-correct manner any of the unary function symbols $\dlv\cdot$, $\ret\cdot$ and $\coretR x\cdot R$.
  In terms of the \emph{parity condition}, this means that also (the sorting rules for) $\dlv\cdot$ and $\ret\cdot$ have priority 2 in that paper.
  Naturally, the present paper has to align the definitions for $\copil$ and $\copils$, and the notion of co-proof for $\copil$ needs to be more restrictive in order to make possible the coinductive extension and the analysis through forgetful maps of the translations of $\ipl$ into $\pil$, cf.~\cref{sec:applications-IPL}.
  It does not seem to the authors that the slightly richer raw syntax of \cite{JESRMLP-TYPES20} can be exploited in meaningful ways, thus this new restriction is considered as just the right notion of raw syntax.}

The notion of equality for forests is again bisimilarity modulo $\alpha$-equivalence, but treating finite sums specially, as if they were sets, that is, sums are identified up to associativity, commutativity and idempotency.

We now define coinductively the notion of membership on forests (in line to our previous papers on implicational logic, and extending the inductive notion of membership on forests in \cite[Def.~1]{JESRMLP-TYPES20}).
\begin{definition}[Membership]\label{eq:mem-rules}
An expression $T\in\copil$ (not necessarily a co-proof term) is a \emph{member} of a forest $T'\in\copils$ when the
predicate $\mem {T}{T'}$ holds, which is defined coinductively as
follows.\\[-1ex]
\[\infer={\mem{f(T_{i})_i}{f(T'_i)_i}}{\bigwedge_i\mem{T_{i}}{T'_i}}\qquad
  \infer=[\textrm{for some }j]{\mem {T}{T'_1+\ldots+T'_k}}{\mem {T}{T'_j}}
\]
\end{definition}
The intuition of this definition is obviously that the sums expressed
by $\sum^\tau_i$ represent alternatives out of which one is chosen for
a concrete member.

The minimum requirement for this definition to be meaningful is that
the five syntactic categories are respected: if $\mem {T}{T'}$ then
$T\in\tau^\cosign$ iff $T'\in\tau^\cosign_\Sigma$. This property holds since we
tacitly assume that the sum operators are tagged with the respective
syntactic category.

And from our settings of parity in $\copil$ and $\copils$, it is obvious that whenever $\mem {T}{T'}$ holds, then $T$ inherits the fulfillment of the parity condition from the respective one on $T'$. In other words, $T$ is automatically even a co-proof term since we only consider forests $T'$.

\begin{definition} For a forest $T$:
	\begin{enumerate}
		\item The set of members of $T$ is denoted $\ext T$, i.\,e.,
		$\ext T=\{T_0\mid\mem {T_0}T\}$. We call this set the \emph{extension} of $T$.
		\item We call \emph{finite extension} of $T$, denoted by $\finext T$, the set of $\PIL$ proof terms in $\ext T$.
	\end{enumerate}
\end{definition}
In the following definition we collect predicates on forests related to finite extension or extension that are of special interest in this paper:
\begin{definition}\label{def:predicates-on-forests} We define 8 predicates on forest:
\begin{enumerate}
	\item $\exfinmem T$, defined as: $\finext T$ is nonempty; and 
	$\nofinmemsymb$, the complement of $\exfinmemsymb$.
	\item $\finfinmem T$, defined as: $\finext T$ is finite; and 
	$\inffinmemsymb$, the complement of $\finfinmemsymb$.
	\item $\nosolmem T$, defined as: $\ext T$ is empty; and
	$\exsolmemsymb$, the complement of $\nosolmemsymb$.
	\item $\allfinmem T$, defined as: $\ext T\subseteq\finext T$; and
	$\exinfmemsymb$, the complement of $\allfinmemsymb$.
\end{enumerate}
\end{definition}

These predicates play an important role in
Section~\ref{sec:decision-problems-PIPL}. The predicates on the left-hand side will be characterized by inductive definitions, and consequently their complements by coinductive definitions.

\subsection{Representation of solution spaces as forests of $\copils$}\label{sec:solfunction}
Now, we are heading for the infinitary representation of all solutions
of any logical sequent $\jay$ of $\PIL$ as a forest whose members are precisely those solutions (to be confirmed in Prop.~\ref{prop:properties-of-S}).
For all the five categories of logical sequents $\jay^\tau$, we define the associated \emph{solution space}
$\solfunction{\jay^\tau}$ as a forest, more precisely, an element of
$\tau^\cosign_\Sigma$, that is supposed to represent the space of
solutions generated by an exhaustive and possibly
non-terminating search process applied to that given logical sequent
$\jay^\tau$.
This is
by way of the following simultaneous coinductive definition. It is
simultaneous for the five categories of logical sequents.
For each category, there is an exhaustive case analysis on the formula argument.

\begin{definition}[Solution spaces]\label{def:sol}
	We define a forest $\solfunction{\jay^\tau}\in\tau^\cosign_\Sigma$ for every logical sequent $\jay^\tau$, by simultaneous coinduction for all the $\tau\in S$. The definition is found in Fig.~\ref{fig:solfunction-PIPL}, where in the clause for $\solfunction{\Gamma\mid {\Down L}\Longrightarrow A}$, the variable $x$ is supposed to be ``fresh''.
\end{definition} 
In the mentioned clause, since the names of bound variables are considered as immaterial, there is no choice involved in this inversion phase of proof search, as is equally the case for $\solfunction{\Gamma\Longrightarrow \cdot}$ -- as should be expected from the deterministic way inversion rules are dealt with in a focused system like $\PIL$.

\begin{figure}[tb]\caption{Solution spaces for $\PIL$}\label{fig:solfunction-PIPL}
	\[
	\begin{array}{r@{\,\,:=\,\,}lr@{\,\,:=\,\,}l}
		\solfunction{\Gamma\vdash[a^+]}&\sum_{(x:a^+)\in\Gamma}x&\solfunction{\Gamma\vdash[\downshift N]}&\thunk{\solfunction{\Gamma\Longrightarrow N}}\\
		\solfunction{\Gamma\vdash[\falsity]}&\oo^v
		&\solfunction{\Gamma\vdash[P_1\vee P_2]}&\sum_{i\in\{1,2\}}\inj i{P_{3-i}}{\solfunction{\Gamma\vdash[P_i]}}
	\end{array}
	\]
	\[
	\begin{array}{r@{\,\,:=\,\,}lr@{\,\,:=\,\,}l}
		\solfunction{\Gamma\Longrightarrow a^-}&\ea{\solfunction{\Gamma\vdash a^-}}&\solfunction{\Gamma\Longrightarrow P\impl N}&\lb\solfunction{\Gamma\mid P\Longrightarrow N}\\
		\solfunction{\Gamma\Longrightarrow \upshift P}&\ep{\solfunction{\Gamma\vdash P}}&\solfunction{\Gamma\Longrightarrow N_1\wedge N_2}&\langle\solfunction{\Gamma\Longrightarrow N_i}\rangle_i
	\end{array}
	\]
	\[
	\begin{array}{r@{\,\,:=\,\,}lr@{\,\,:=\,\,}l}
		\solfunction{\Gamma[a^-]\vdash R}& \textrm{if $R=a^-$ then $\nil$ else $\oo^s$}\\
		\solfunction{\Gamma[P\impl N]\vdash R}&\solfunction{\Gamma\vdash[P]}::\solfunction{\Gamma[N]\vdash R}\\
		\solfunction{\Gamma[\upshift P]\vdash R}&\cothunk{\solfunction{\Gamma\mid P\Longrightarrow R}}\\
		\solfunction{\Gamma[N_1\wedge N_2]\vdash R}&\sum_{i\in\{1,2\}}(i::\solfunction{\Gamma[N_i]\vdash R})
	\end{array}
	\]
	\[
	\begin{array}{r@{\,\,:=\,\,}lr@{\,\,:=\,\,}l}
		\solfunction{\Gamma\mid {\Down L}\Longrightarrow A}&x^L.\,\solfunction{\Gamma,x:L\vdash A}\\
		\solfunction{\Gamma\mid {\falsity}\Longrightarrow A}&\abortt A&
		\solfunction{\Gamma\mid P_1\vee P_2\Longrightarrow A}&[\solfunction{\Gamma\mid P_i\Longrightarrow A}]_i
	\end{array}
	\]
	\[
	\begin{array}{r@{\,\,:=\,\,}lr@{\,\,:=\,\,}l}
		\solfunction{\Gamma\vdash C}&\dlv{\solfunction{\Gamma\Longrightarrow C}}\\
		\solfunction{\Gamma\vdash a^-}&\sum_{(x:N)\in\Gamma}\coretR x {\solfunction{\Gamma[N]\vdash a^-}}{a^-}\\
		\solfunction{\Gamma\vdash P} & \ret{\solfunction{\Gamma\vdash[P]}}+\sum_{(x:N)\in\Gamma}\coretR x {\solfunction{\Gamma[N]\vdash P}}P
	\end{array}
	\]
\end{figure}

\begin{lemma}[Well-definedness of $\solfunction\jay$]\label{lem:welldefinedness-S}
	For all logical sequents $\jay$, the definition of $\solfunction\jay$ indeed produces a forest.
\end{lemma}
\begin{proof} Well-definedness is
	not at stake concerning productivity of the definition since
	every corecursive call is under a constructor. As is directly
	seen in the definition, the syntactic categories are respected.
	Only the parity condition requires further thought. 
	In Appendix~\ref{sec:appendix-definednessofS},
	we prove it by showing that all the
	``intermediary'' corecursive calls to $\solfunction{\jay'}$ in the
	calculation of $\solfunction{\jay}$ lower the
	``weight'' of the logical sequent, so that infinite branches have to
        go through infinitely many ``principal'' corecursive calls, which are those when a $\coretsymb$ is traversed.

\end{proof}
\begin{example}\label{ex:solution-spaces} 
	Let us illustrate the solution space function $\solletter$ with some of the types considered in \cref{ex:types}.  Below, when a sum has a unique summand we do not make such sum visible, displaying only  its summand (except in the first two illustrations, where the forest obtained without applying this simplification is shown before).
	\begin{enumerate}
		\item  Recall the four polarizations of type $\Idt$ there considered. The solution spaces of the corresponding logical sequents of sort $t$ (in the cases of $\Idt^-, \Idt^{+n},\Idt^{+-}$) and of sort $e$ (in the case of $\Idt^{+p}$) are all  finitary forests, given by:
		\begin{enumerate}
			\item $\solfunction{\Rightarrow\Idt^-}= \lb (x^{a^-}.\sum_{(y:N)\in\Gamma_1}\coretR y{\solfunction{\Gamma_1[N]\!\vdash\! a^-}}{a^-})=\lb (x^{a^-}.\coretR x\nil{a^-})$, with $\Gamma_1:= x:a^-$;
			\item $\solfunction{\Rightarrow\Idt^{+n}}=\lb (x^{a^+}.\dlv{\ep{\ret{\sum_{(y:a^+)\in\Gamma_2}y}}})=\lb (x^{a^+}.\dlv{\ep{\ret x}})$, with $\Gamma_2:= x:a^+$;
			\item $\solfunction{\Rightarrow\Idt^{+-}}=\lb (x^{a^+}.{{\oo^e}})$;
			\item $\solfunction{\vdash\Idt^{+p}}=\ret{\thunk{\lb (x^{a^+}.\dlv{\ep{\ret x}})}}$.
		\end{enumerate}
		Note that the obtained expressions (in their simplified form) can be viewed as the unique inhabitants of these types identified in \cref{ex:types}.
		\item  Recall now the three polarizations of $\Inft$ given in \cref{ex:types}.  The corresponding solution spaces (for sort $t$) are as follows:
		\begin{enumerate}
			\item  $\solfunction{\Rightarrow\Inft^-}= \lambda (x^{\downshift a^-\impl a^-}. T)$, where $T$ stands for the (unique) solution of the fixed-point equation in forests $T=\coretR{x}{\thunk{\ea{T}}::\nil}{a^-}$;  seen as a co-proof term, this expression corresponds to the infinite solution $T_1$ of $\Inft^-$ given in \cref{ex:infinite-solutions};
			\item  $\solfunction{\Rightarrow\Inft^+}=\lambda (x^{a^+\impl \upshift a^+}.\dlv{\ep{\ret{\oo^v} + \coretR{x}{\oo^v::\cothunk{y^{a^+}.{T_1}}}{a^+}}})$, where $T_1=\solfunction{\Gamma\vdash a^+}=\ret{y} + \coretR{x}{y::\cothunk{z^{a^+}.{T_2}}}{a^+}$, and $T_2= \solfunction{\Gamma,z:a^+\vdash a^+}$, with $\Gamma=x:a^+\impl \upshift a^+,y: a^+$; we could continue the unfolding of function $\solletter$ up to any arbitrary depth; however, note that in going from $T_1$ to $T_2$ the only new element is a duplicate declaration of type $a^+$, a phenomenon that we call \emph{decontraction}; ahead we will encounter the \emph{decontraction operation on forests}, which will provide means for a compact way to communicate solution spaces where this phenomenon is observed;
			\item $\solfunction{\Rightarrow\Inft^{+--}}=\lambda (x^{a^+\impl a^-}.\coretR x{\oo^v::\nil}{a^-})$.
		\end{enumerate}
		
		\item The four polarizations of $\Churcht$ of \cref{ex:types}  have solution spaces as follows:
		\begin{enumerate}
			\item $\solfunction{\Rightarrow\Churcht^-}=\lambda (x^{\downshift a^-\impl a^-}.\dlv{\lambda (y^{a^-} .T}))$, where $T$ stands for the solution of the of the fixed-point equation in forests
			$T=\coretR y\nil{a^-}+\coretR{x}{\thunk{\ea{T}}::\nil}{a^-}$.
			\item $\solfunction{\Rightarrow\Churcht^{+n}}=\lambda (x^{a^+\impl \upshift a^+}.\dlv{\lambda (y^{a^+} .\dlv{\ep{T_1}}}))$, 
			where $T_1$ is as above in the calculation of $\solfunction{\Rightarrow\Inft^+}$, and is thus an infinitary solution embodying the decontraction phenomenon;
			\item $\solfunction{\vdash\Churcht^{+p}}=\ret{\thunk{\lambda (x^{a^+\impl \upshift a^+}.\dlv{\ep{\ret{\thunk{\lambda (y^{a^+} .\dlv{\ep{T_1}})}}}})}}$ where again $T_1$ is as in the calculation of $\solfunction{\Rightarrow\Inft^+}$;
			\item $\solfunction{\Rightarrow\Churcht^{+-}}=\lambda (x^{a^+\impl  a^-}.\dlv{\lambda (y^{a^+} .\coretR x{y::\nil}{a^-})})$, where the latter can also be viewed as the unique inhabitant of $\Churcht^{+-}$ given in \cref{ex:types}.
		\end{enumerate}
	\end{enumerate}
\end{example}

Next, we establish that the members of a solution space are exactly the solutions of the sequent at hand. This result generalizes \cite[Prop.~4]{JESRMLP-TYPES20}, which  covers only the case of  finite solutions/inhabitants  of a sequent and thus corresponds to part 2 of the result.
\begin{proposition}[Adequacy of the coinductive representation]\label{prop:properties-of-S}
	\begin{enumerate}
		\item
		For each 
		$\tau\!\in\! S$, $T\in\copil$ of category $\tau$ and  logical sequent $\jay^\tau$,
		$\mem T{\solfunction{\jay}}\!$ iff $\jay(T)$ is valid in $\copil$. 
		\item
		For each 
		$\tau\!\in\! S$, $T\in\PIL$ of category $\tau$ and logical sequent $\jay^\tau$,
		$\mem T{\solfunction{\jay}}\!$ iff $\jay(T)$ is valid in $\PIL$. 
	\end{enumerate}
\end{proposition}
\begin{proof} Item 2 is an immediate consequence of 1 and \cref{lem:conservativeness-typing}. 
	Both implications comprised in item 1 are proved by coinduction,
simultaneously for all the five syntactic categories of co-proof terms. 
The left to right direction goes by coinduction on the (coinductively defined) typing relation  for $\copil$, whereas the other direction profits from the coinductive nature of the membership relation.
	We illustrate the direction from left to right for the case $T=\coretR x{T_1}R$ (for some $x$, $T_1$, $R$). By the assumption and the definitions of the solution space function and the membership predicate, it must be $\jay=\Gamma\vdash R$, with $x:N\in\Gamma$ (for some $\Gamma$, $R$, $N$), and $\mem{T_1}{\solfunction{\Gamma[N]\vdash R}}$. Hence, the coIH 
allows to conclude  that $\Gamma[T_1:N]\vdash R$ is valid in $\copil$, from which follows validity of $\Gamma\vdash \coretR x{T_1}R:R$, as wanted.
	We now  illustrate the direction from right to left again for the case $T=\coretR x{T_1}R$. By the assumption and the typing relation of $\copil$, it must be $\jay=\Gamma\vdash R$, with $x:N\in\Gamma$ (for some $\Gamma$, $N$), and $\Gamma[T_1:N]\vdash R$ is valid in $\copil$. Then, through the coIH 
we get $\mem{T_1}{\solfunction{\Gamma[N]\vdash R}}$, from which follows $\mem{\coretR x{T_1}R}{\coretR x{\solfunction{\Gamma[N]\vdash R}}R}$, hence $\mem{\coretR x{T_1}R}{\solfunction{\Gamma\vdash R}}$, given that $\solfunction{\Gamma\vdash R}=\sum_{(y:M)\in\Gamma}\coretR y{\solfunction{\Gamma[M]\vdash R}}R$.
\end{proof}

\subsection{A class of predicates on forests}\label{sec:predsforests}
Two predicates on forests are \emph{complementary} when a forest belongs to one if and only if it does not belong to the other.
So, viewing the predicates as sets, complementary predicates are complementary sets. 
In Def.~\ref{def:predicates-on-forests} we have seen four pairs of complementary predicates on forests. In this subsection we identify a class of pairs of predicates on forests, encompassing those of Def.~\ref{def:predicates-on-forests}, for which we can spell out a recursive definition of one of the components of the pair, and a corecursive definition of the other component, where the definitions are uniform on the pair of predicates. Such scheme will allow, later on, a uniform treatment of such pairs in the development of their meta-theory. De Morgan duality between least and greatest fixed points will ensure that, in each case, the inductively defined predicate and the coinductively defined predicate are complementary.

We are not only interested in predicates on forests that are expressed through membership, like those in Def.~\ref{def:predicates-on-forests} (one relevant example will be described in \cref{ex:fin} below). Moreover, even for the example predicates listed in Def.~\ref{def:predicates-on-forests}, it will be convenient to see them under our scheme, and thus characterize them as pairs of an inductive and a coinductive subset of all forests.
While the truthfulness of elementhood in these two predicates obviously needs to refer to only one of the two predicates (and, from this point of view, given a pair of complementary predicates is redundant), we will keep both predicates 
as forming a pair under the scheme we will propose (this will be further discussed after \cref{def:dualpairs} and after \cref{def:dualpairsfin}).

In the sequel, when writing (co)recursive definitions, it will be convenient to indicate conjunctive or disjunctive reading of premises by choosing one of the symbols $\bigwedge$ and $\bigvee$ and assigning it to a variable $\lopvar$ or $\lopvaralt$.
When using this variable, this means taking advantage of that logical meaning.
For example, if $\lopvaralt=\bigwedge$, we could write the premise of the first rule of membership as $\lopvaralt_i\,\mem{T_{i}}{T'_i}$.
We freely use dualization $\overline\bigwedge=\bigvee$, $\overline\bigvee=\bigwedge$ also on these variables as $\overline\lopvar$ and $\overline\lopvaralt$ and also allow ourselves to use them as binary connectives.

\begin{definition}[Dual pairs of predicates on $\copils$]\label{def:dualpairs}
  We define the dual pair data and then, given this data, we define the dual pair of predicates.
\begin{enumerate}
\item We consider tuples of the form $(\lopvar,Q,\overline Q,\lopvaralt)$ with $\lopvar$ and $\lopvaralt$ connectives as above and $Q$ and $\overline Q$ two complementary predicates on $\copils$.\label{item:datafordualpair}
  We communicate such tuples as \emph{dual pair data}, with generic letter $D$.
\item Given dual pair data $D=(\lopvar,Q,\overline Q,\lopvaralt)$, inductively define the predicate $\ipredd D$ and coinductively define the predicate $\cpredd D$ on forests as given in \cref{fig:ipred-cpred-rules}.
\end{enumerate}
\end{definition}
We give an example how to read this definition:
We take as $f$ the spine concatenation operation~$::$, seen as function symbol sorted as $v,s\to s$. We assume forests $T_1:v$ and $T_2:s$, i.\,e., $T_1\in\Valuesco$ and $T_2\in\Covaluesco$.
The first inference rule is read inductively and has the conclusion $\ipred D{T_1::T_2}$, where we keep writing $::$ as infix symbol.
If $\lopvar$ happens to be $\bigwedge$, the premise reads $(\ipred D{T_1}\bigwedge\ipred D{T_2})\bigvee(Q(T_1)\bigvee Q(T_2))$.
As is usual practice, the rule would normally be presented as three rules, all having the conclusion $\ipred D{T_1::T_2}$, but the first with two premises $\ipred D{T_1}$ and $\ipred D{T_2}$, the second with the premise $Q(T_1)$ and the third with the premise $Q(T_2)$.
And if $\lopvar$ is $\bigvee$, the premise reads $(\ipred D{T_1}\bigvee\ipred D{T_2})\bigwedge(Q(T_1)\bigwedge Q(T_2))$.
Using distributivity, this rule would normally be transcribed into two rules, both having the conclusion $\ipred D{T_1::T_2}$,
and the $i$-th rule having the three premises $\ipred D{T_i}$, $Q(T_1)$ and $Q(T_2)$.

It is also instructive to logically simplify the premisses for the $f$ rules in case of nullary and unary $f$. In fact, $\lopvar_i\ipred D{T_i}\ovllopvarbinop{\overline\lopvar}_iQ(T_i)$ then shrinks down to $\lopvar=\bigwedge$ and $\ipred D{T_1}\ovllopvarbinop Q(T_1)$, respectively. Likewise, $\overline\lopvar_i\cpred D{T_i}\lopvarbinop\lopvar_i\overline Q(T_i)$ shrinks down to $\lopvar=\bigvee$ and $\cpred D{T_1}\lopvarbinop\overline Q(T_1)$, respectively.
Put more sharply, a nullary function symbol $f$ belongs to $\ipredd D$ iff $\lopvar=\bigwedge$, \emph{a fortiori}, it belongs to $\cpredd D$ iff $\lopvar=\bigvee$.
And empty sums $\oo$ are classified analogously depending on the value of $\lopvaralt$.

Notice that the definition rules are uniform in all function symbols $f$ of $\PIL$, only the summation operation is dealt with separately. The two dual predicates are not interwoven.
By de Morgan duality\footnote{This principle is also recalled in the proof of \cite[Lemma 20]{EspiritoSantoMatthesPintoInhabitation} in the format needed for our present purposes, while being applied there for simpler situations only.} between least and greatest fixed points, $\ipredd D$ and $\cpredd D$ are complementary predicates on $\copils$. We would hope that the re-use of letter $Q$ in this context never leads to confusion with positive formulas.
Concerning item \ref{item:datafordualpair} of \cref{def:dualpairs}, despite being complementary, we still consider $Q$ and $\overline Q$ as potentially being defined independently of each other, particularly they could be other instances of a dual pair of predicates, whose definitions are not interwoven. This will be seen in items \ref{item:finfindef} and \ref{item:allfindef} of \cref{ex:dualpairsmembership} below.
\begin{figure}[tb]\caption{Dual pair of predicates $\ipredd D$ and $\cpredd D$ for dual pair data $D=(\lopvar,Q,\overline Q,\lopvaralt)$}\label{fig:ipred-cpred-rules}
  \[
\infer[]{\ipred D{f(T_i)_i}}{\lopvar_i\ipred D{T_i}\ovllopvarbinop{\overline\lopvar}_iQ(T_i)}\quad\infer
{\ipred D{\sum_iT_i}}{\lopvaralt_i\ipred D{T_i}}\qquad
 \infer=
 {\cpred D{f(T_i)_i}}{\overline\lopvar_i\cpred D{T_i}\lopvarbinop\lopvar_i\overline Q(T_i)}\quad\infer=[]{\cpred D{\sum_iT_i}}{\overline\lopvaralt_i\cpred D{T_i}}
  \]
\end{figure}
Write $\truepred$ resp.~$\falsepred$ for the predicates on $\copils$ that contain all forests resp.~no forest.
Notice that when $(\lopvar,Q,\overline Q)=(\bigwedge,\falsepred,\truepred)$, then the premises of the $f$ rules in \cref{fig:ipred-cpred-rules} logically simplify to $\lopvar_i\ipred D{T_i}$ resp.~$\overline\lopvar_i\cpred D{T_i}$.
This simplification will be understood tacitly, as well as the dual situation $(\lopvar,Q,\overline Q)=(\bigvee, \truepred,\falsepred)$.
In these situations, we even abbreviate the dual pair data to $(\lopvar,\lopvaralt)$, thus omitting $Q$ and $\overline Q$.

In general, $\ipredd D$ is monotone in parameter $Q$, and $\cpredd D$ is monotone in parameter $\overline Q$, more precisely, if $D=(\lopvar,Q,\overline Q,\lopvaralt)$ and $D'=(\lopvar,Q',\overline Q',\lopvaralt)$ are both dual pair data and $Q\subseteq Q'$ (equivalently: $\overline Q'\subseteq \overline Q$), then $\ipredd D\subseteq\ipredd {D'}$ and $\cpredd {D'}\subseteq\cpredd D$.
This is evident from the positive position of $Q$ resp.~$\overline Q$ in the premisses of the defining rules (just under conjunctions and disjunctions).
\begin{example}\label{ex:fin}
	An instance of \cref{def:dualpairs} that yields a predicate on forests unrelated to the (finite) extension is $\ipredd{\bigwedge,\bigwedge}$.
	It designates the forests that are obtained by reading the grammar of forests inductively or, equivalently, as extending the grammar of $\PIL$ proof terms with the operation of finite summation.
	So, the predicate singles out forests that come in a genuine finite description.
	In \cite[Section 2]{JESRMLP-FI2019}, this predicate was studied under the name $\mathsf{fin}$ for the implicational fragment of $\LJT$ (see \cref{subsec:LJT-LJTco} for $\LJT$).
\end{example}

\begin{example}[Dual pairs of predicates for the analysis of membership]\label{ex:dualpairsmembership}
  We give the parameters (i.\,e., the dual pair data) needed for \cref{def:dualpairs} to capture the predicates introduced in Def.~\ref{def:predicates-on-forests}.
  \begin{enumerate}
  \item Let $D=(\bigwedge,\bigvee)$. Then set $\exfinsymb:=\ipredd D$, $\nofinsymb:=\cpredd D$.
  \item Let $D=(\bigwedge,\nofinsymb,\exfinsymb,\bigwedge)$. Then set $\finfinsymb:=\ipredd D$, $\inffinsymb:=\cpredd D$.\label{item:finfindef}
  \item Let $D=(\bigvee,\bigwedge)$. Then set $\nosolsymb:=\ipredd D$, $\exsolsymb:=\cpredd D$.
  \item Let $D=(\bigwedge,\nosolsymb,\exsolsymb,\bigwedge)$. Then set $\allfinsymb:=\ipredd D$, $\exinfsymb:=\cpredd D$.\label{item:allfindef}
  \end{enumerate}
\end{example}

In \cref{fig:exfin-finfin-rules} (which is practically identical to Fig.~2 in the paper \cite{JESRMLP-TYPES20} we are expanding on in the present paper), we show what the first two definitions explicitly amount to (modulo the rearrangements indicated in our reading example right after \cref{def:dualpairs}),
analogously to our previous work \cite{EspiritoSantoMatthesPintoInhabitation}, with which we inductively characterized $\exfinmemsymb$ and $\finfinmemsymb$, and coinductively characterized $\nofinmemsymb$ and $\inffinmemsymb$ (for the implicational fragment of intuitionistic logic considered there).
\begin{figure}[tb]\caption{Predicates $\exfinsymb$, $\nofinsymb$, $\finfinsymb$ and $\inffinsymb$}\label{fig:exfin-finfin-rules}
\[
\begin{array}{c}
\infer[]{\exfin{f(T_i)_i}}{\bigwedge_i\exfin{T_i}}\quad\infer
{\exfin{\sum_iT_i}}{\exfin{T_j}}\qquad
 \infer=
 {\nofin{f(T_i)_i}}{\nofin{T_j}}\quad\infer=[]{\nofin{\sum_iT_i}}{\bigwedge_i\nofin{T_i}}\\[1.5ex]
   \infer[]{\finfin{f(T_i)_i}}{\bigwedge_i\finfin{T_i}}\quad \infer
   {\finfin{f(T_i)_i}}{\nofin{T_j}}\quad\infer[]{\finfin{\sum_iT_i}}{\bigwedge_i\finfin{T_i}}\qquad
    \infer=
    {\inffin{f(T_i)_i}}{\inffin{T_j}&\bigwedge_i\exfin{T_i}}\quad\infer=
    {\inffin{\sum_iT_i}}{\inffin{T_j}}
    \end{array}
\]
\end{figure}
Also for convenience, the second two definitions in \cref{ex:dualpairsmembership} are spelt out concretely in \cref{fig:nosol-allfin-rules}; in this explicit form they are the $\PIL$ adaptations of similar definitions in \cite{JESRMLP-FI2019}.
\begin{figure}[tb]\caption{Predicates $\nosolsymb$, $\exsolsymb$, $\allfinsymb$ and $\exinfsymb$}\label{fig:nosol-allfin-rules}
  \[
      \begin{array}{c}
\infer[]{\nosol{f(T_i)_i}}{\nosol{T_j}}\quad\infer
{\nosol{\sum_iT_i}}{\bigwedge_i\nosol{T_i}}\qquad
 \infer=
 {\exsol{f(T_i)_i}}{\bigwedge_i\exsol{T_i}}\quad\infer=[]{\exsol{\sum_iT_i}}{\exsol{T_j}}\\[1.5ex]
   \infer[]{\allfin{f(T_i)_i}}{\bigwedge_i\allfin{T_i}}\quad \infer
   {\allfin{f(T_i)_i}}{\nosol{T_j}}\quad\infer[]{\allfin{\sum_iT_i}}{\bigwedge_i\allfin{T_i}}\qquad
    \infer=
    {\exinf{f(T_i)_i}}{\exinf{T_j}&\bigwedge_i\exsol{T_i}}\quad\infer=
    {\exinf{\sum_iT_i}}{\exinf{T_j}}
    \end{array}
\]
\end{figure}
In Appendix~\ref{sec:appendix-exfinnofin} it is shown that the characterizations represented in \cref{fig:exfin-finfin-rules} and \cref{fig:nosol-allfin-rules} are indeed adequate, namely that for each inductive predicate $P\in\{\exfinsymb,\finfinsymb,\nosolsymb,\allfinsymb\}$, the predicates $P$ and $P\mathsf{ext}$ hold of the same forests. Consequently, for each coinductive predicate $P\in\{
\nofinsymb,\inffinsymb,\exsolsymb,\exinfsymb\}$, also $P= P\mathsf{ext}$ as sets of forests.
This provides non-obvious inclusions between our predicates: the trivial inclusion $\nofinmemsymb\subseteq\finfinmemsymb$ entails $\nofinsymb\subseteq\finfinsymb$ which does not have an easy direct proof.
The trivial inclusion $\exfinmemsymb\subseteq\exsolmemsymb$ entails $\exfinsymb\subseteq\exsolsymb$, which could also be proven directly by induction on $\exfinsymb$.
Anyway, it is a stepping stone for an interesting inclusion between our predicates
since the remark before \cref{ex:fin} lifts $\exfinsymb\subseteq\exsolsymb$ to $\inffinsymb\subseteq\exinfsymb$ (similarly to what is observed in \cite[Section 3.3]{JESRMLP-FI2019} for the system studied there),
hence we get $\inffinmemsymb\subseteq\exinfmemsymb$, which is an important property of general interest: from infinitely many finite members one can infer the existence of an infinite member.

We have to stress that the dual pairs of predicates classify forests, hence expressions of $\copils$ that satisfy the parity condition. The rules of \cref{fig:exfin-finfin-rules} and \cref{fig:nosol-allfin-rules} -- in particular the coinductive ones -- thus cannot be seen as generative in the sense that forests are being defined by them. Instead, they are rules to infer that an already given forest (satisfying the parity condition) indeed belongs to one of those predicates. We will illustrate this by expressions of sort $s$ that do not satisfy the parity condition: Let $T_1$ and $T_2$ be the unique fixed points of $T_1=1::T_1$ and $T_2=\nil+x::T_2$ for some variable $x$, respectively. The naive reading of the rules for $\nofinsymb$ and $\exsolsymb$ seem to suggest that $T_1$ is coinductively generated by both of them. Likewise, the naive reading of the rules for $\inffinsymb$ and $\exinfsymb$ suggests coinductive generation of $T_2$ again by both of them.
There are unique infinite branches in $T_1$ and in $T_2$.
The only constructors passed are $::$ and $+$ (for $T_2$), but there is no $\coretR y\cdot R$, in particular not an infinity of them.

From the adequacy of the representation of the solution space (\cref{prop:properties-of-S}) and the 
adequacy of the characterizations of the dual pairs of predicates introduced in \cref{ex:dualpairsmembership} 
(\cref{lem:charmemi} and \cref{lem:charmemsol} in \cref{sec:appendix-exfinnofin}), it readily follows:

\begin{corollary}\label{cor:S-props} The following equivalences hold:
	\begin{enumerate}
		\item $\jay$ is inhabited in $\PIL$ iff $\exfin{\solfunction{\jay}}$.
		\item $\jay$ has (only) finitely many  inhabitants in $\PIL$ iff $\finfin{\solfunction{\jay}}$.
		\item $\jay$ is solvable in $\copil$ iff $\exsol{\solfunction{\jay}}$.
		\item $\jay$ has an infinite solution in $\copil$ iff $\exinf{\solfunction{\jay}}$.
	\end{enumerate}
Consequently, for each item, the negation of the left-hand side is equivalent to validity of the respective dual predicate for $\solfunction{\jay}$.
\end{corollary}

\begin{example}\label{ex:nosol-inft-pos}We illustrate how
	\cref{cor:S-props} can be used to argue succinctly about the observation made in \cref{ex:types} that  both  $\Inft^{+}$ and
	$\Inft^{+--}$ have no solutions in $\copil$ (contrary to $\Inft^{-}$). We argue about $\Inft^{+}$. So, it suffices to show $\nosol{\solfunction{\Longrightarrow\Inft^{+}}}$. Recall $\solfunction{\Longrightarrow\Inft^{+}}$ from \cref{ex:solution-spaces}. Inspecting the defining rules of $\nosolsymb$, we need to show both $\nosol{\ret{\oo^v}}$ and $\nosol{\coretR{x}{\oo^v::\cothunk{y^{a^+}.{T_1}}}{a^+}}$.  Both are ultimately consequence of the fact $\nosol{\oo^v}$, the former in one step, the latter via the intermediate observation $\nosol{\oo^v::\cothunk{y^{a^+}.{T_1}}}$.
        Notice that we have no solution despite $\solfunction{\Longrightarrow\Inft^{+}}$ being infinite, which refutes in a strong sense an intuition in the spirit of König's lemma that would expect an infinite solution in an infinite solution space.
        For the implicational fragment of $LJT$, a ``pruned'' solution space can be defined that makes such a result possible \cite[Theorem 25]{JESRMLP-FI2019}.
\end{example}

\subsection{What we obtained so far}\label{sec:firstmetathms}
Derivations of $\PIL$ can be represented by proof terms, runs of bottom-up proof search in the same system can be represented by the co-proof terms of $\copil$, the solution space determined by a logical sequent $\sigma$ can be represented by the forest $\solfunction{\sigma}$ of $\copils$. In this sense, the definition of $\solletter$ embodies the proof search process, and we can base the study of the latter on $\solletter$ and forget about the original logical system. Given Corollary \ref{cor:S-props}, we can take the little formal systems in Figs.~\ref{fig:exfin-finfin-rules} and \ref{fig:nosol-allfin-rules} as the definitions of the concerned predicates. For some purposes, such (co)recursive definitions of the predicates embody all we need to know about them. So, for these purposes, we can conduct a formal way of proceeding, in which our reasoning is solely guided by the dynamics of the definitions of $\solletter$ and of the predicates, without any consideration of -- indeed forgetting -- their original meaning. 

In this subsection, we illustrate this way of proceeding in proving two meta-theorems of $\PIL$, the disjunction and the infinity-or-nothing properties. In both cases, some relevant class of polarized formulas has to be identified.

By inspection of the inference rules of $\pil$, $\Gamma\vdash[P_1\vee P_2]$ is inhabited only if $\Gamma\vdash[P_i]$, for some $i$. With our formal tools, we argue as follows. Let $T=\solfunction{\Gamma\vdash[P_1\vee P_2]}$. By definition of $\solletter$, $T=\sum_{i\in\{1,2\}}\inj i{P_{3-i}}{\solfunction{\Gamma\vdash[P_i]}}$. Using the rules in Fig.~\ref{fig:exfin-finfin-rules}, $\exfin{T}$ only if, for some $i\in\{1,2\}$, $\exfin{\solfunction{\Gamma\vdash[P_i]}}$. Now, how about the stable sequent $\sigma=\Gamma\vdash P_1\vee P_2$? When searching for a proof of $\sigma$, we can focus on a formula in $\Gamma$, instead of the succedent formula. The answer has to consider simultaneously the inhabitation of sequents $\Gamma[N]\vdash P_1\vee P_2$. For such a disjunction property under hypotheses to hold, we have to restrict the formulas in $\Gamma$.

We start by recalling the class of \emph{intuitionistic Rasiowa-Harrop formulas} (see e.\,g.\  \cite{TroelstraS00}), obtained from \ipl\ formulas by forbidding strictly positive occurrences of  disjunction, otherwise said, given inductively by:
\[
{\mathcal R}::= a \mid \falsity \mid A\impl {\mathcal R} \mid {\mathcal R}_1 \wedge {\mathcal R}_2 
\]
with $A$ an arbitrary formula of \ipl.
Let us now
define a polarized counterpart of this class of formulas as the following subclass ${\mathcal L}$ of left formulas, which we call \emph{polarized Rasiowa-Harrop formulas}:
\[
\begin{array}{lcrcl}
\textrm{(polarized Rasiowa-Harrop formulas)} &  & {\mathcal L}& ::= & a^+\mid {\mathcal N}\\
\textrm{(negative Rasiowa-Harrop formulas)} &  & {\mathcal N}& ::= & a^- \mid  \upshift\falsity \mid P\impl {\mathcal N} \mid {\mathcal N}_1 \wedge {\mathcal N}_2 \mid \upshift \Down {\mathcal L}
\end{array}
\]
with $P$ an arbitrary positive formula. Note that the forgetful map is a surjective map both from $\mathcal L$ and from $\mathcal N$ to the class of {intuitionistic Rasiowa-Harrop formulas}. We call \emph{strict} the formulas in the subclasses of ${\mathcal L}$ and ${\mathcal N}$ obtained by forbidding in the grammar of ${\mathcal N}$ the productions $\upshift\falsity$ and $\upshift \Down {\mathcal L}$.

Consider the disjunction property  for \ipl\, (see e.\,g.\  \cite[Theorem 4.2.3]{TroelstraS00} -- under the name \emph{Disjunction property under hypotheses} -- or \cite[Theorem 4.1]{Ferreira17}). The following theorem establishes a polarized version of this property. The polarized version subsumes the version for \ipl\, in the sense that the former delivers the latter via an embedding of \ipl\ into \pipl, as will be seen in \cref{sec:applications-IPL}

\begin{theorem}[Polarized disjunction property under hypotheses]\label{thm:polarized-disjunction-property}
Let  $\Gamma$ be a context built only from  ${\mathcal L}$-formulas, let $P_1$ and $P_2$ be positive formulas of $\PIL$, and let ${\mathcal N}$ be a negative Rasiowa-Harrop formula. 
\begin{enumerate}
\item  If\/ $\Gamma\vdash P_1\vee P_2$ is inhabited,
then, for some $i\in\{1,2\}$: $\Gamma\vdash P_i$ is inhabited; in addition, if all the formulas of\/ $\Gamma$ are strict, then $\Gamma\vdash [P_i]$ is inhabited.
\item If\/ $\Gamma[{\mathcal N}]\vdash P_1\vee P_2$ is inhabited,
  then, for some $i\in\{1,2\}$, $\Gamma[{\mathcal N}]\vdash P_i$ is inhabited, and $\mathcal N$ is not strict.
\end{enumerate}
\end{theorem}
\begin{proof}
Statements (1) and (2) are proved simultaneously. By \cref{cor:S-props}, they can be expressed equivalently by replacing the occurrences of
``$\sigma$ is inhabited'' by $\exfin{\solfunction\sigma}$. Moreover, they can be both brought to the generic form: if $\exfin{T}$ and $T=\solfunction\sigma$ then, for some $i\in\{1,2\}$: first,  $\exfin{\solfunction{\sigma_i}}$, where $\sigma_i$ is obtained from $\sigma$ by replacing the succedent formula $P_1\vee P_2$ by $P_i$; second, some refinement. The proof is by induction on $\exfin T$. We will silently use Fig.~\ref{fig:exfin-finfin-rules} all the time.

In statement (1), the definition of $\solletter$ gives:
\[
T=\ret{\solfunction{\Gamma\vdash[P_1\vee P_2]}}+ \sum_{(x:N)\in\Gamma}\coretR x {\solfunction{\Gamma[N]\vdash P_1\vee P_2}}{P_1\vee P_2} \enspace.
\]
The hypothesis $\exfin{T}$ gives two cases. (i) $\exfin{\Gamma\vdash[P_1\vee P_2]}$. By the same reasoning seen before, we get $\exfin{\solfunction{\Gamma\vdash [P_i]}}$, for some $i\in\{1,2\}$, hence $\exfin{\ret{\solfunction{\Gamma\vdash [P_i]}}}$ and the result $\exfin{\solfunction{\Gamma\vdash P_i}}$; or (ii) there is $(x:N)\in\Gamma$ such that  $\exfin{\solfunction{\Gamma[N]\vdash P_1\vee P_2}}$, to which we can apply the induction hypothesis (by the assumption on $\Gamma$, $N$ is negative Rasiowa-Harrop) to get $\exfin{\solfunction{\Gamma[N]\vdash P_i}}$, for some $i\in\{1,2\}$, hence  the result. If all the formulas of $\Gamma$ are strict, then, in case (ii), $N$ is strict. The induction hypothesis says that $\exfin{\solfunction{\Gamma[N]\vdash P_1\vee P_2}}$ is impossible, as this would imply that $N$ is not strict. So, no $N$ works in case (ii) and case (i) is forced. As we have seen, in case (i) we have $\exfin{\solfunction{\Gamma\vdash [P_i]}}$.

Regarding statement (2), we sketch the different cases.

Case ${\mathcal N}=a^-$. Then $T=\solfunction{\Gamma[a^-]\vdash P_1\vee P_2}=\oo^s$, and $\exfin{\oo^s}$ does not hold.

Case ${\mathcal N}=\upshift\falsity$. Then $T=\solfunction{\Gamma[\upshift\falsity]\vdash P_1\vee P_2}=\cothunk{\abortt{P_1\vee P_2}}$ and $\exfin{T}$. Predicate $\exfinsymb$ holds of $\cothunk{\abortt{P_i}}$ as well. 
And by definition of strictness, $\upshift\falsity$ is not strict.

Case ${\mathcal N}=\upshift \Down {\mathcal L}_0$. The hypothesis $\exfin{T}$ gives $\exfin{\cothunk{x^{\mathcal L_0}.\solfunction{\Gamma, x:{\mathcal L_0}\vdash P_1\vee P_2}}}$,  hence $\exfin{\solfunction{\Gamma, x:{\mathcal L_0}\vdash P_1\vee P_2}}$, from which the induction hypothesis
gives, for some $i\in\{1,2\}$,
$\exfin{\solfunction{\Gamma, x:{\mathcal L_0}\vdash P_i}}$
(note that the enlarged context still only contains ${\mathcal L}$-formulas),
which readily allows to conclude $\exfin{\solfunction{\Gamma[\upshift \Down{\mathcal L_0}]\vdash P_i}}$. By definition of strictness, $\upshift \Down {\mathcal L}_0$ is not strict.

Case ${\mathcal N}=P\impl{\mathcal N'}$. Then $\solfunction{\Gamma[P\impl{\mathcal N'}]\vdash P_1\vee P_2}=\solfunction{\Gamma\vdash[P]}::\solfunction{\Gamma[{\mathcal N'}]\vdash P_1\vee P_2}$, hence the hypothesis $\exfin{T}$ gives $\exfin{\solfunction{\Gamma\vdash[P]}}$ and $\exfin{\Gamma[{\mathcal N'}]\vdash P_1\vee P_2}$. By induction hypothesis, $\exfin{\Gamma[{\mathcal N'}]\vdash P_i}$ and ${\mathcal N'}$ is not strict. Hence ${\mathcal N}$ is not strict either; in addition, we easily reconstruct $\exfin{\solfunction{\Gamma[P\impl{\mathcal N'}]\vdash P_i}}$.

Case ${\mathcal N}={\mathcal N_1}\wedge{\mathcal N_2}$. Similar to the last one.
\end{proof}

Let us now illustrate an ``infinity or nothing'' property of $\PIL$, that is a family of sequents for which the existence of one inhabitant implies the existence of infinitely many inhabitants. Roughly speaking, our illustration explores  the presence of a declaration $x:\upshift P$ in contexts, where $P$ can be a rather general positive formula, namely a \emph{not fully absurd positive formula}. This subclass of positive formulas is given by the grammar:
\[
\nfasymb ::= \Down L \mid \nfasymb \vee P \mid P \vee \nfasymb
\]
with $L$ (resp. $P$) an arbitrary left (resp.\ positive) formula. In other words, such a formula $\nfasymb$ is a non-void disjunction of positive formulas (parenthesized at will), out of which one is of the form $\Down L$ (i.\,e., $a^+$ or $\downshift N$).

The following lemma is useful:

\begin{lemma}\label{lem:weak-pos}
For any positive formula $P$, if\/ $\Gamma\vdash R$ is inhabited, then $\Gamma\mid P\Rightarrow R$ is inhabited.
\end{lemma}
\begin{proof}
We want to prove that $\exfin{\solfunction{\Gamma\vdash R}}$ implies $\exfin{\solfunction{\Gamma\mid P\Rightarrow R}}$. The proof is by induction on $P$. The cases where $P$ is a positive atom or a shift from a negative formula use the fact that weakening of left formulas is admissible in $\PIL$. (Recall contexts of $\PIL$ allow only left formulas.)
\end{proof}

\begin{theorem}[Infinity or nothing property]\label{thm:infinity-or-nothing}
Let $\nfasymb$ be a not fully absurd positive formula. Let $\Gamma$ be a context s.\,t.~$(x:\upshift \nfasymb)\in\Gamma$. For any right formula  $R$, the stable sequent $\Gamma\vdash R$ has the infinity-or-nothing property.
\end{theorem}
\begin{proof} We will silently use Fig.~\ref{fig:exfin-finfin-rules} all the time. First, we prove an auxiliary result: the sequent $\Gamma\mid \nfasymb' \Rightarrow R$ has the infinity-or-nothing property, where $\nfasymb'$ is a not fully absurd positive formula. Let $\sigma = (\Gamma\mid \nfasymb' \Rightarrow R)$ and $T=\solfunction{\sigma}$. We want to prove that $\exfin T$ implies $\inffin T$. The proof is by coinduction on $T$.  We illustrate the case $\nfasymb'=\Down L$ (the other cases bring no new difficulties). By definition of $\solletter$, $T=y^L.T_1$, for $T_1=\solfunction{\Gamma, y:L\vdash R}$. Given that $x:\upshift \nfasymb\in\Gamma$, one of the summands of $T_1$ is $\coretR x {\cothunk{T_2}}R$, for $T_2={\solfunction{\Gamma, y:L\mid \nfasymb \Rightarrow R}}$. Now, the hypothesis $\exfin T$ gives $\exfin{T_1}$, hence $\exfin{T_2}$, due to \cref{lem:weak-pos}. From $\exfin{T_2}$,  the  coIH relative to $T_2$ gives $\inffin{T_2}$. From this and $\exfin{T_2}$, we can get $\inffin{\coretR x {\cothunk{T_2}}R}$, hence $\inffin{T_1}$. This and $\exfin{T_1}$, finally, give $\inffin T$.

Now the proof of the theorem. Let $\sigma=(\Gamma\vdash R)$ and $T=\solfunction{\sigma}$. Suppose $\exfin T$. By definition of $\solletter$, $T_1=\coretR x {\cothunk{T_2}}R$ is one of the summands of $T$, where $T_2={{\solfunction{\Gamma\mid \nfasymb\Rightarrow R}}}$. The hypothesis and \cref{cor:S-props} give $\exfin T$, hence $\exfin{T_2}$ (by \cref{lem:weak-pos}). Therefore,
the auxiliary result gives $\inffin{T_2}$. From this and $\exfin{T_2}$ we can justify $\inffin{T_1}$, which gives $\inffin T$.
\end{proof}

This ends the illustration of the development of meta-theory of $\PIL$. In the next sections, we find algorithmic counterparts to function $\solletter$ and the predicates in Figs.~\ref{fig:exfin-finfin-rules} and \ref{fig:nosol-allfin-rules}, with which we solve decision problems.

\section{Coinductive approach to proof search in $\PIL$ - part II}\label{sec:proof-search-PIPL-partII}

This section is a kind of finitary mirror of the previous one, and accordingly is organized in a similar fashion. First we define a system for the finitary representation of solution spaces. Next we develop this representation. Next we extend the uniform treatment of dual pairs of predicates on forests, started in the previous section, to a similar treatment for predicates on finitary forests. Finally, we present specific applications to the meta-theory of $\pil$, namely decidability of several predicates concerning proof search.

\subsection{System $\pils$}\label{sec:pils}
We are going to present a finitary version of $\copils$ in the form of a system $\pils$ of \emph{finitary forests} that are again generically denoted by letter $T$.
Even though they are finitary forests, we will define an \emph{interpretation} of them as forests of $\copils$.
Due to this interpretation, finitary forests have both finite and infinite members and can represent the search for both inhabitants and solutions in $\pil$.

\begin{definition} The finitary forests of $\pils$ are inductively defined by the following schemes:
	\[\infer[]{f(T_1,\ldots,T_k):\tau}{f:\tau_1,\ldots,\tau_k\to \tau\quad \bigwedge_i(T_i:\tau_i)}\qquad
	\infer[\tau\in\{v,s,e\}]{\sum^\tau_i T_i:\tau}{\bigwedge_i(T_i:\tau)}\qquad
	\infer[]{X^\rho:e}{}\quad \infer[]{\gfp X^\rho.T:e}{T:e}
	\]
\end{definition}
We are again making extensive use of our notational device introduced in Section~\ref{sec:background}. The letter $f$ ranges over the function symbols in this specific view on $\PIL$. Summation is added analogously as for $\copils$, and there are two more constructions for the category of expressions.
$X$ is assumed to range over a countably infinite set of
\emph{fixpoint variables} and $\rho$ ranges over R-stable
sequents, as said before. The conventions regarding sums $\sum_i$
in the context of forests are also assumed for finitary
forests.  

We stress that this is an all-inductive definition, and that
w.\,r.\,t.~$\PIL$, the same finite summation mechanism is added as for
$\copils$, but that the coinductive generation of stable expressions
is replaced by formal fixed points whose binding and bound/free
variables are associated with R-stable sequents $\rho$ whose proof
theory is our main aim.

We will write $f^*$ to stand for a function symbol $f$ or
the prefix ``$\gfp X^\rho.$'' of a finitary forest, the latter being
seen as special unary function symbol (of sort $e\to e$).

Below are some immediate adaptations of definitions in our previous
paper \cite{EspiritoSantoMatthesPintoInhabitation}. However, they are
presented in the new uniform notation. Moreover, the notion of
guardedness only arises here, due to the wider formulation of finitary
forests we are now employing, that allows fixed-point formation for any finitary forest of
the category of stable expression.
\begin{definition}
For a finitary forest $T$, let $\FPV(T)$ denote the set of freely occurring typed fixed-point variables in
$T$, which can be described by structural recursion:
\[\begin{array}{rcl}
	\multicolumn{3}{c}{
		\FPV(f(T_i)_i)=\FPV(\sum_iT_i)=\bigcup_i\FPV(T_i)\qquad\FPV(X^\rho)=\{X^\rho\}}\\
	\FPV(\gfp\,{X^\rho}.T)&=&\FPV(T)\setminus
	\{X^{\rho'}\mid\mbox{$\rho'$ R-stable sequent and $\rho\leq\rho'$}\}
\end{array}\]
$T$ is \emph{closed} if $\FPV(T)=\emptyset$. Notice the non-standard definition that considers $X^{\rho'}$ also
bound by $\gfp X^\rho$, as long as $\rho\leq\rho'$. This special
view on binding necessitates to study the following restriction on
finitary forests: A finitary forest is called \emph{well-bound} if, for
any of its subterms $\gfp\,{X^\rho}.T$ and any free occurrence of
$X^{\rho'}$ in $T$, $\rho\leq\rho'$.
\end{definition}
\begin{definition}
      To any free occurrence of an $X^{\rho}$ in $T$ is associated a
\emph{depth}: for this, we count the function symbols on the
path from the occurrence to the root and notably do not count
the binding operation of fixed-point variables and the sum operations. So, $X^\rho$
only has one occurrence of depth $0$ in $X^\rho$,  likewise in $\gfp Y^{\rho'}.X^\rho$.
	
	We say a
	finitary forest $T$ is \emph{guarded} if for any of its
	subterms $T'$ of the form $\gfp\,{X^{\rho}}.T''$, it holds
	that every free occurrence in $T''$ of a fixed-point variable
	$X^{\rho'}$ that is bound by this fixed-point constructor has
	a depth of at least $1$ in $T''$.
\end{definition}

\begin{definition}[Interpretation of finitary forests as forests]
	For a finitary forest $T$, the interpretation $\interps T$ is a forest given
	by structural recursion on $T$:\label{def:interpretation}
\[
  \begin{array}{rclcrcl}
    \interps{f(T_1,\ldots,T_k)}&=&f(\interps{T_1},\ldots,\interps{T_k})&&\interps{X^{\rho}}& = &\solfunction{\rho}\\
    \interps{T_1+\ldots+T_k}&=&\interps{T_1}+\ldots+\interps{T_k}&&\interps{\gfp\,{X^{\rho}}.T}&= & \interps T
	\end{array}
      \]
\end{definition}
This definition may look too simple to handle the
interpretation of bound fixed-point variables adequately, and in our previous paper
\cite{EspiritoSantoMatthesPintoInhabitation} we called an analogous definition ``simplified semantics''
to stress that point.
However, as in that previous paper, we
can study those finitary forests for which the definition is ``good
enough'' for our purposes of capturing solution spaces:
we say a 
finitary forest $T$ is \emph{proper} if for any of its subterms
	$T'$ of the form $\gfp\,{X^{\rho}}.T''$, it holds that
	$\interps{T'}=\solfunction{\rho}$.

\subsection{Finitary representation of solution spaces}\label{sec:finrep}

\begin{definition}[Finitary solution spaces for $\PIL$]\label{def:finrep}
	Let $\Xi:=\vect{X:\rho}$ be a vector
	of $m\geq 0$ declarations $(X_i:\rho_i)$ where no
	fixed-point variable name occurs twice. The definition
        of the finitary forest $\finrep{\jay}{\Xi}$ is as follows.
	If for some $1\leq i\leq m$, $\rho_i=:({\Gamma_i}\vdash{R_i})\leq\jay$  
	(i.\,e., $\sigma=\Gamma\vdash R_i$ and $\Gamma_i\leq\Gamma$), then
	$\finrep{\jay}{\Xi}=X_i^{\jay}$,
	where $i$ is taken to be the biggest such index (notice that the produced  $X_i$ will not necessarily appear with the $\rho_i$ associated to it in $\Xi$).
	Otherwise,
	$\finrep{\jay}{\Xi}$ is as displayed in Fig.~\ref{fig:finrepPIPL}. 
	Then, $\finrepempty{\jay}$ denotes $\finrep{\jay}{\Xi}$ with empty
	$\Xi$. 
\end{definition}
\begin{figure}[tb]\caption{All other cases of the finitary representation of solution spaces for $\PIL$.}\label{fig:finrepPIPL}
\[
	  \begin{array}{r@{\,\,:=\,\,}lr@{\,\,:=\,\,}l}
	 \finrep{\Gamma\vdash[a^+]}{\Xi}&\sum_{(x:a^+)\in\Gamma}x&
	 	\finrep{\Gamma\vdash[\downshift N]}{\Xi}&\thunk{\finrep{\Gamma\Longrightarrow N}{\Xi}}\\
\finrep{\Gamma\vdash[\falsity]}{\Xi}&\oo^v&
	\finrep{\Gamma\vdash[P_1\vee P_2]}{\Xi}&\sum_{i\in\{1,2\}}\inj i{P_{3-i}}{\finrep{\Gamma\vdash[P_i]}{\Xi}}
	  \end{array}
\] 
\[
	  \begin{array}{r@{\,\,:=\,\,}lr@{\,\,:=\,\,}l}
	 \finrep{\Gamma\Longrightarrow a^-}{\Xi}&\ea{\finrep{\Gamma\vdash a^-}{\Xi}}&
	 \finrep{\Gamma\Longrightarrow P\impl N}{\Xi}&\lb\finrep{\Gamma\mid P\Longrightarrow N}{\Xi}\\
	\finrep{\Gamma\Longrightarrow \upshift P}{\Xi}&\ep{\finrep{\Gamma\vdash P}{\Xi}}&
	\finrep{\Gamma\Longrightarrow N_1\wedge N_2}{\Xi}&\langle\finrep{\Gamma\Longrightarrow N_i}{\Xi}\rangle_i
		  \end{array}
\]
\[ \begin{array}{rcl}
   \finrep{\Gamma[a^-]\vdash R}{\Xi}&:=& \textrm{if $R=a^-$ then $\nil$ else $\oo^s$}\\
    \finrep{\Gamma[\upshift P]\vdash R}{\Xi}&:=&\cothunk{\finrep{\Gamma\mid P\Longrightarrow R}{\Xi}}\\
    \finrep{\Gamma[P\impl N]\vdash R}{\Xi}&:=&\finrep{\Gamma\vdash[P]}{\Xi}::\finrep{\Gamma[N]\vdash R}{\Xi}\\
    \finrep{\Gamma[N_1\wedge N_2]\vdash R}{\Xi}&:=&\sum_{i\in\{1,2\}}(i::\finrep{\Gamma[N_i]\vdash R}{\Xi})
   
   \end{array}
\]
\[
	  \begin{array}{rcl}
	 \finrep{\Gamma\mid \Down L\Longrightarrow A}{\Xi}&:=&x^{L}.\,\finrep{\Gamma,x:L\vdash A}{\Xi}\qquad\qquad\qquad\;\, (\textrm{$x$ fresh})\\
	\finrep{\Gamma\mid P_1\vee P_2\Longrightarrow A}{\Xi}&:=&[\finrep{\Gamma\mid P_i\Longrightarrow A}{\Xi}]_i\\
	  \finrep{\Gamma\mid {\falsity}\Longrightarrow A}{\Xi}&:=&\abortt A

			  \end{array}
\]	
\[
	\begin{array}{rcll}
			\finrep{\Gamma\vdash C}{\Xi}&:=&\dlv{\finrep{\Gamma\Longrightarrow C}{\Xi}}\\
			\finrep{\Gamma\vdash a^-}\Xi&:=&\gfp\,{Y^\rho}.\sum_{(x:N)\in\Gamma}\coretR{x}{\finrep{\Gamma[N]\vdash a^-}{\,\Xi,\,Y\!:\!\rho}}{a^-}
	
		\quad (\rho\!=\!\Gamma\!\vdash\!a^-,\textrm{$Y$ fresh})\\
	\finrep{\Gamma\vdash P}\Xi&:=&
	\gfp\,{Y^\rho}.
	\begin{array}[t]{cl}
	& \ret{\finrep{\Gamma\vdash[P]}{\,\Xi,\,Y\!:\!\rho}}\qquad\qquad\qquad\;\; (\rho\!=\!\Gamma\!\vdash\!P,\textrm{$Y$ fresh})\\
		+&\sum_{(x:N)\in\Gamma}\coretR{x} {\finrep{\Gamma[N]\vdash P}{\,\Xi,\,Y\!:\!\rho}}P
	\end{array}
	\end{array}
\]
\end{figure}

\begin{example}\label{ex:finitary-solution-spaces} 
Here we provide the finitary represention of the solution space of each of the logical sequents  considered in \cref{ex:solution-spaces} obtained by function $\finrepsymb$. Recall that in \cref{ex:solution-spaces} the solutions spaces are presented in the form of potentially
infinite forests, calculated through the corecursive function $\solletter$. It is instructive to compare the representations delivered by $\finrepsymb$ and $\solletter$ in each case, and observe how the $\gfp$-construction and its relaxed form of binding  allow to keep a lockstep relationship.
\begin{enumerate}
\item $\finrepempty{\Rightarrow\Idt^-}=\lb (x^{a^-}.\gfp\,{X^{\rho_1}}.\coretR x\nil{a^-})$, with $\rho_1:= x:a^-\vdash a^-$;
\item $\finrepempty{\Rightarrow\Idt^{+n}}=\lb (x^{a^+}.\dlv{\ep{\gfp\,{X^{\rho_2}}.\ret x}})$, with $\rho_2:= x:a^+\vdash a^+$;
\item $\finrepempty{\Rightarrow\Idt^{+-}}=\lb (x^{a^+}.{\gfp\,{X^{\rho_3}}.{\oo^e}})$, with $\rho_3:= x:a^+\vdash a^-$;
\item $\finrepempty{\vdash\Idt^{+p}}=\gfp\,{X^{\rho_4}}.\ret{\thunk{\lb (x^{a^+}.\dlv{\ep{\gfp\,{Y^{\rho_2}}.\ret x}})}}$, with $\rho_4:=\, \vdash\Idt^{+p}$ (and $\rho_2$ as above). 
\item  $\finrepempty{\Rightarrow\Inft^-}= \lambda (x^{\downshift a^-\impl a^-}. \gfp\,{X^{\rho_5}}.\coretR{x}{\thunk{\ea{X^{\rho_5}}}::\nil}{a^-})$ ($\rho_5:=x:{\downshift a^-\impl a^-}\vdash a^-$);  
\item  $\finrepempty{\Rightarrow\Inft^+}=\lambda (x^{a^+\impl \upshift a^+}.\dlv{\ep{\gfp\,{X^{\rho_6}}.(\ret{\oo^v} + \coretR{x}{\oo^v::\cothunk{y^{a^+}.{T}}}{a^+})}})$, where $T:=\finrepempty{\rho_7}=\gfp\,{Y^{\rho_7}}.(\ret{y} + \coretR{x}{y::\cothunk{z^{a^+}.{Y^{\rho_8}}}}{a^+})$,
with  $\Gamma:=x:a^+\impl \upshift a^+$, $\rho_6:=\Gamma\vdash a^+$, $\rho_7:=\Gamma, y:a^+\vdash a^+$, $\rho_8:=\Gamma,y:a^+,z:a^+\vdash a^+$; note that $\rho_8$ is an inessential extension of $\rho_7$, and therefore the relaxed form of binding provided by the $\gfp$-construction makes the only occurrence of $Y^{\rho_8}$ bound by $\gfp\,{Y^{\rho_7}}$;
\item $\finrepempty{\Rightarrow\Inft^{+--}}=\lambda (x^{a^+\impl a^-}.\gfp\,{X^{\rho_9}}.\coretR x{\oo^v::\nil}{a^-})$, with $\rho_9:=x:{a^+\impl a^-}\vdash a^-$.
\item $\finrepempty{\Rightarrow\Churcht^-}=\lambda (x^{\downshift a^-\impl a^-}.\dlv{\lambda (y^{a^-} .T}))$, where 
$$T=\gfp\,{X^{\rho_{10}}}.(\coretR y\nil{a^-}+\coretR{x}{\thunk{\ea{X^{\rho_{10}}}}::\nil}{a^-})\enspace,$$ with $\rho_{10}:=x:{\downshift a^-\impl a^-, y:{a^-}}\vdash a^-$;  
\item $\finrepempty{\Rightarrow\Churcht^{+n}}=\lambda (x^{a^+\impl \upshift a^+}.\dlv{\lambda (y^{a^+} .\dlv{\ep{T}}}))$, 
where $T=\finrepempty{\rho_7}$ (as above, in the calculation of $\finrepempty{\Rightarrow\Inft^+}$), and thus also profits from the specific  form of binding provided by the $\gfp$-construction;
\item $\finrepempty{\vdash\Churcht^{+p}}=$
$$\gfp\,{X^{\rho_{11}}}.\ret{\thunk{\lambda (x^{a^+\impl \upshift a^+}.\dlv{\ep{\gfp\,{Z^{\rho_{12}}}.\ret{\thunk{\lambda (y^{a^+} .\dlv{\ep{T}})}}}})}}\enspace,$$
where again $T=\finrepempty{\rho_7}$, and $\rho_{11}:=\, \vdash\Churcht^{+p}$, and  $\rho_{12}:= x:a^+\impl \upshift a^+\vdash \downshift(a^+\impl \upshift a^+)$;
\item $\finrepempty{\Rightarrow\Churcht^{+-}}=\lambda (x^{a^+\impl  a^-}.\dlv{\lambda (y^{a^+} .\gfp\,{X^{\rho_{13}}}.\coretR x{y::\nil}{a^-})})$, setting $\rho_{13}$ to ${x:a^+\impl a^-}, {y:a^+}\vdash a^-$.
  \end{enumerate}
\end{example}


Analogously to the similar result for
implicational logic \cite[Lemma 24]{JESRMLP-APAL2021}, one can show
that $\finrep{\jay}{\Xi}$ is well-defined (the above recursive
definition terminates)---some details are given in Appendix~\ref{sec:appendix-terminationofF}.
Notice that the ``if-guard'' in the above definition presupposes that $\sigma$ is an R-stable sequent,
hence for other forms of sequents, one necessarily has to apply the (mostly recursive) rules of
Fig.~\ref{fig:finrepPIPL}.
\begin{theorem}[Equivalence of representations for $\PIL$]\label{thm:equiv-PIPL-simpl}
  Let $\jay$ be a logical sequent and $\Xi$ as in Def.~\ref{def:finrep}. We have:
  \begin{enumerate}
  \item $\finrep\jay\Xi$ is guarded.\label{thm:equiv-PIPL-simpl.0}
  \item 
  $\finrep\jay\Xi$ is well-bound and $\finrepempty{\jay}$ is closed.
  \label{thm:equiv-PIPL-simpl.0-1}
  \item $\finrep\jay\Xi$ is proper.\label{thm:equiv-PIPL-simpl.1}
  \item $\interps{\finrep\jay\Xi}=\solfunction\jay$; hence the coinductive and the finitary representations are equivalent.\label{thm:equiv-PIPL-simpl.2}
  \end{enumerate}
\end{theorem}
\begin{proof}
	The proof is by structural induction on $\finrep\jay\Xi$. Items \ref{thm:equiv-PIPL-simpl.0} and \ref{thm:equiv-PIPL-simpl.0-1} are proved independently (the former is an easy induction, the latter on well-boundness uses in the two cases which generate  $\gfp$-constructions
the lemma ``if $X^{\rho'}$ occurs free in $\finrep\jay\Xi$, then, for some $\rho\leq\rho'$, $X:\rho\in\Xi$'', 
also proved by structural induction on $\finrep\jay\Xi$, and from that lemma follows immediately that $\finrepempty{\jay}$ is closed). As in the proof of \cite[Thm.~19]{EspiritoSantoMatthesPintoInhabitation},
	item  \ref{thm:equiv-PIPL-simpl.1} uses item \ref{thm:equiv-PIPL-simpl.2}, which can be proved independently, but some effort is saved if the two items are proved simultaneously. 
 \end{proof}
      

%% file: analysis-PIPL.tex
\subsection{Analysis of predicates on forests with the finitary system $\pils$}\label{sec:analysis-PIPL}

We are going to provide a counterpart, on the side of finitary forests, to the dual pairs of predicates on forests of \cref{def:dualpairs}.
The predicates there are naturally decidable in a number of cases of interest to us, and we have a general result (\cref{prop:genfinchar} below) to relate such predicates to those on forests through the interpretation according to \cref{def:interpretation}.
By virtue of part \ref{thm:equiv-PIPL-simpl.2} of \cref{thm:equiv-PIPL-simpl}, this opens the way to decide predicates on the solution spaces.

\begin{definition}[Dual pairs of predicates on $\pils$]\label{def:dualpairsfin}
  We define the finitary dual pair data and then,  given this data, we define the finitary dual pair of predicates.
  \begin{enumerate}
  \item Given dual pair data $D=(\lopvar,Q,\overline Q,\lopvaralt)$ according to \cref{def:dualpairs}.\ref{item:datafordualpair}, we consider tuples of the form $(D,\Pi,\overline\Pi,P)$, where $\Pi$ and $\overline\Pi$ are two (complementary) predicates on $\pils$ such that for all finitary forests $T$, $\Pi(T)$ iff $Q(\interps T)$, and $\overline\Pi(T)$ iff $\overline Q(\interps T)$, and where $P$ is a predicate on $R$-stable $\PIL$ sequents.\footnote{We would hope that the re-use of letter $P$ in this context never leads to confusion with positive formulas.}\label{item:dataforfinitarydualpair}
    We communicate such tuples as \emph{finitary dual pair data}, with generic letter $D^+$, and when using  $D^+$, we consider symbol $D$ to represent its contained dual pair data.\footnote{Beware that notation $D^+$ is not meant to indicate a positive polarity, but we hope that no confusion will arise.}
  \item Given finitary dual pair data $D^+=(D,\Pi,\overline\Pi,P)$, with dual pair data $D=(\lopvar,Q,\overline Q,\lopvaralt)$, we inductively define the predicate $\ipredfd {D^+}$ and coinductively define the predicate $\cpredfd {D^+}$ on finitary forests as given in \cref{fig:ipredf-cpredf-rules}.
  \end{enumerate}
\end{definition}
Notice the simplicity of the rules governing finitary forests of the form $\gfp X^\rho.T$. It fits well with \cref{def:interpretation} that gives their also very simple interpretation, as can be seen in the proofs in \cref{sec:appendix-mainpropgencoindanalysis}.

By de Morgan duality between least and greatest fixed points, also $\ipredfd {D^+}$ and $\cpredfd {D^+}$ are complementary predicates.
Moreover, an induction on the inductive construction of finitary proof terms shows immediately that there can only be one solution $\cpredfd {D^+}$ of the fixed-point equations expressed by the inference rules of \cref{fig:ipredf-cpredf-rules}, hence the suggested greatest fixed point is in fact the unique one, which makes superfluous the doubly horizontal lines indicating a coinductive reading.
Still worded differently, both $\ipredfd {D^+}$ and $\cpredfd {D^+}$ can be seen as defined recursively, by structural recursion on its finitary forest argument. Therefore:
\begin{proposition}[Generic decidability I]\label{prop:decidability-complementary-predicates}
	Let $D$ and $D^+$ be as in \cref{def:dualpairsfin}. If $P$ and $\Pi$ are decidable, then $\ipredfd {D^+}$ and $\cpredfd {D^+}$ are decidable predicates on finitary forests.
\end{proposition}

\smallskip
\noindent
\textbf{Remark on the usefulness of having a dual pair and on \cref{prop:decidability-complementary-predicates}.} The presence of the complementary predicates $\Pi$ and $\overline\Pi$ on $\pils$ instead of $Q\circ\interps{\_}$ and $\overline Q\circ\interps{\_}$ (we defended earlier that we assume the presence of both $Q$ and $\overline Q$) makes very good sense if we want to read the rules of \cref{fig:ipredf-cpredf-rules} as giving decision procedures for $\ipredfd {D^+}$ and $\cpredfd {D^+}$.
For this to work, we would assume $P$, $\Pi$ and $\overline\Pi$ to be given as decision procedures as well.
And not that, e.\,g., $Q\circ\interps{\_}$ ``happens'' to be decidable.
And through our constructions, we get naturally both $\Pi$ and $\overline\Pi$, and thus may indicate a decision procedure for $\overline\Pi$ distinct from the one given by negation of $\Pi$. In the light of this remark, we will interpret \cref{prop:decidability-complementary-predicates} not only as a result about preservation of decidability but consider, e.\,g., $\ipredfd {D^+}$ as a decision procedure given decision procedures $P$ and $\Pi$.

\begin{figure}[tb]\caption{Dual pair of predicates $\ipredfd {D^+}$ and $\cpredfd {D^+}$ for finitary dual pair data $D^+=((\lopvar,Q,\overline Q,\lopvaralt),\Pi,\overline\Pi,P)$}\label{fig:ipredf-cpredf-rules}
  \[
    \begin{array}{c}
\infer[]{\ipredf {D^+}{X^\rho}}{P(\rho)}\quad
\infer[]{\ipredf {D^+}{\gfp X^\rho.T}}{\ipredf {D^+}{T}}\quad
\infer[]{\ipredf {D^+}{f(T_i)_i}}{\lopvar_i\ipredf {D^+}{T_i}\ovllopvarbinop{\overline\lopvar}_i\Pi(T_i)}\quad
\infer{\ipredf {D^+}{\sum_iT_i}}{\lopvaralt_i\ipredf {D^+}{T_i}}\\[1.5ex]
\infer={\cpredf {D^+}{X^\rho}}{\neg P(\rho)}\quad
\infer={\cpredf {D^+}{\gfp X^\rho.T}}{\cpredf {D^+}{T}}\quad
\infer= {\cpredf {D^+}{f(T_i)_i}}{\overline\lopvar_i\cpredf {D^+}{T_i}\lopvarbinop\lopvar_i\overline\Pi(T_i)}\quad
\infer=[]{\cpredf {D^+}{\sum_iT_i}}{\overline\lopvaralt_i\cpredf {D^+}{T_i}}
    \end{array}
  \]
\end{figure}
We will re-use $\truepred$ resp.~$\falsepred$ to indicate also the trivial (constant) predicates on $\pils$, do the tacit logical simplifications of the premises of the inference rules and also abbreviate the finitary dual pair data $((\bigwedge,\lopvaralt),\falsepred,\truepred,P)$ and $((\bigvee,\lopvaralt),\truepred,\falsepred,P)$ to $((\bigwedge,\lopvaralt),P)$ and $((\bigvee,\lopvaralt),P)$, respectively.

\begin{example}\label{ex:nbinf-binf+stuck-nstuck}
  Consider the finitary analogues of the first and third item in \cref{ex:dualpairsmembership}:
  \begin{enumerate}
  \item Let $D=(\bigwedge,\bigvee)$. Then set $\nbinfpsymb P:=\ipredfd {(D,P)}$, $\binfpsymb P:=\cpredfd {(D,P)}$.
  \item Let $D=(\bigvee,\bigwedge)$. Then set $\stuckpsymb P:=\ipredfd {(D,P)}$, $\nstuckpsymb P:=\cpredfd {(D,P)}$.
  \end{enumerate}
\end{example}
The inference rules are detailed in \cref{fig:nbinf-binf+stuck-nstuck}, already using inductive rules throughout. As mentioned before, these predicates could have been defined by recursion over the finitary forests as well. Thus, the predicates $\nbinfpsymb P$, $\binfpsymb P$, $\stuckpsymb P$ and $\nstuckpsymb P$ are decidable if $P$ is. We remark that the rules for $\stuckpsymb P$ and $\nstuckpsymb P$ are rather close to \cite[Figure 6]{JESRMLP-FI2019}.
\begin{figure}[tb]\caption{$\nbinfpsymb P$ and $\binfpsymb P$ predicates, $\stuckpsymb P$ and $\nstuckpsymb P$ predicates}\label{fig:nbinf-binf+stuck-nstuck}
\[
\begin{array}{c}
\infer[]{\nbinfp P{X^\rho}}{P(\rho)}\quad
\infer[]{\nbinfp P{f^*(T_i)_i}}{\bigwedge_i\nbinfp P{T_i}}\quad\infer[]{\nbinfp P{\sum_iT_i}}{\nbinfp P{T_j}}\quad
\infer[]{\binfp P{X^\rho}}{\neg P(\rho)}\quad
\infer[]{\binfp P{f^*(T_i)_i}}{\binfp P{T_j}}\quad
  \infer[]{\binfp P{\sum_iT_i}}{\bigwedge_i\binfp P{T_i}}\\[1.5ex]
  \infer[]{\stuckp P{X^\rho}}{P(\rho)}\quad
\infer[]{\stuckp P{f^*(T_i)_i}}{\stuckp P{T_j}}\quad\infer[]{\stuckp P{\sum_iT_i}}{\bigwedge_i\stuckp P{T_i}}\quad
\infer[]{\nstuckp P{X^\rho}}{\neg P(\rho)}\quad
\infer[]{\nstuckp P{f^*(T_i)_i}}{\bigwedge_i\nstuckp P{T_i}}\quad
  \infer[]{\nstuckp P{\sum_iT_i}}{\nstuckp P{T_j}}
\end{array}
\]
\end{figure}
The missing analogues of the second and fourth item in \cref{ex:dualpairsmembership} need deeper preparations so as to provide meaningful parameters $\Pi$ and $\overline\Pi$: as required in \cref{def:dualpairsfin} we need to characterize the composition of the semantic interpretation function with the parameters $Q$ in use, and in \cref{ex:dualpairsmembership}, those are $\nofinsymb$ and $\nosolsymb$.

The next proposition is key to obtaining the decidability results in section \ref{sec:decision-problems-PIPL}. It gives, for a predicate $\ipredd{D}$ on forests, a characterization of the predicate $\ipredd{D}\circ\solletter$ on logical sequents, and of the predicate $\ipredd{D}\circ\interps{\_}$ over finitary forests in terms of predicates $\ipredfd {D^+}$.

The next proposition refers to the property of closure under decontraction. Recall the concept of inessential extension, given in \cref{sec:background}, and denoted $\rho\leq\rho'$: the sequent $\rho$ can be seen as obtained from $\rho'$ by a number of applications of the structural inference rule of contraction. The operation of decontraction $[\rho'/\rho]T$ causes in the forest $T$ an effect of ``duplication''. Roughly, each free occurrence in $T$ of $x$ as a value (resp.~as the head-variable of $\coretR{x}{s}R$) becomes a summand $x'$ (resp.~$\coretR{x'}{s}R$) of a sum over the $x'$ which are duplicates $(x':L)\in\rho'$ of $(x:L)\in\rho$. A predicate $P$ over forests is closed under decontraction if $P(T)$ and $\rho\leq\rho'$ imply $P([\rho'/\rho]T)$. In this paper, we treat closure under decontraction as a technical condition and relegate the details to \cref{sec:appendix-decontraction}.\footnote{For an in-depth treatment of decontraction in the context of intuitionistic implicational logic, see \cite{JESRMLP-APAL2021}.}

\begin{proposition}
  [Generic finitary characterization]\label{prop:genfinchar}
  Given finitary dual pair data $D^+=(D,\Pi,\overline\Pi,P)$ with dual pair data $D=(\lopvar,Q,\overline Q,\lopvaralt)$ for the dual pair of predicates $\ipredd D$ and $\cpredd D$, we consider the dual pair of predicates $\ipredfd {D^+}$ and $\cpredfd {D^+}$ according to \cref{def:dualpairsfin}.
	\begin{enumerate}
		\item If $P\subseteq\ipredd D\circ\solletter$ and $\ipredf{D^+}T$ then $\ipred D{\interps T}$.\label{prop:genfinchar.1}
		\item Assume that $\overline Q$ is closed under decontraction.
                  Let $T\in\pils$ be well-bound, guarded and proper.\label{prop:genfinchar.2} If\/
		$\cpredf {D^+}T$ and for all $X^\rho\in\FPV(T)$, $\ipred D{\solfunction\rho}$ implies $P(\rho)$,
		then $\cpred D{\interps T}$.
              \item \label{prop:genfinchar.3}
                Assume that $\overline Q$ is closed under decontraction.
                Let $D' := (D,\Pi,\overline \Pi,\falsepred)$.
		For any $T\in\pils$  well-bound, guarded, proper and closed,  $\ipredf {D'} T$ iff\/  $\ipred D{\interps T}$.
                In particular, for any logical sequent $\jay$,  $\ipredf {D'}{\finrepempty{\jay}}$ iff\/ $\ipred D{\solfunction{\jay}}$.
              \item  \label{prop:genfinchar.4}
                Assume that $\overline Q$ is closed under decontraction.
                Let $\Pcansymb:=\ipredfd{D'}\circ\finrepsymb$ and let $D'' := (D,\Pi,\overline \Pi,\Pcansymb)$.
                Then, for any $T\in\pils$, $\ipredf {D''} T$ iff\/  $\ipred D{\interps T}$.
	\end{enumerate}
      \end{proposition}
\begin{proof}
  \ref{prop:genfinchar.1}. is proved by induction on the predicate
  $\ipredfd{D^+}$ (or, equivalently, on $T$). The base case for
  fixpoint variables needs the proviso on $P$, and all other cases
  are easy by the induction hypothesis (the case $\gfp X^\rho.T$ is even simpler).

  \ref{prop:genfinchar.2}. This needs a special notion of depth of
  observation for the truthfulness of $\cpredd{D}$ for forests. A more
  refined statement has to keep track of this observation depth in
  premise and conclusion, even taking into account the depth of
  occurrences of the bound fixed-point variables of $T$. This is
  presented with details in \cref{sec:appendix-mainpropgencoindanalysis}.
  Closure under decontraction of $\overline Q$ enters closure under decontraction of an auxiliary predicate used in the proof.

  \ref{prop:genfinchar.3}. For $P=\falsepred$ resp.~for closed $T$, the
  extra condition on $P$ in part~\ref{prop:genfinchar.1}
  resp.~part~\ref{prop:genfinchar.2} is trivially satisfied.
  We now use
  that $\ipredd D$ and $\cpredd D$ are complements, as are
  $\ipredfd {D'}$ and $\cpredfd {D'}$.
  The final statement comes from the case $T=\finrepempty{\jay}$, using all four parts of \cref{thm:equiv-PIPL-simpl} (with empty vector of declarations $\Xi$).

  \ref{prop:genfinchar.4}. By (\ref{prop:genfinchar.3}), $\Pcansymb$ satisfies the condition on $P$ in (\ref{prop:genfinchar.1}), hence we get the direction from left to right.
  For the direction from right to left, we prove the contraposition that $\cpredf {D''} T$ implies $\cpred D{\interps T}$, by induction on $T$. Case $T=X^\rho$: By inversion, $\cpredf {D''} {X^\rho}$ implies $\neg\Pcan\rho$, i.\,e., $\neg\ipred D{\solfunction{\rho}}$ by the final statement in (\ref{prop:genfinchar.3}), in other words, $\cpred D{\interps{X^\rho}}$.
  The other cases are mostly captured by the induction hypothesis.
  We only show the case $T=f(T_i)_i$: For every $i$, the induction hypothesis gives the implication $\cpredf {D''} {T_i}\Rightarrow\cpred D{\interps {T_i}}$.
  Also for every $i$, the definition of finitary dual pairs gives the implication $\overline\Pi(T_i)\Rightarrow\overline Q(\interps{T_i})$.
  Together, they justify the implication $\overline\lopvar_i\cpredf {D''}{T_i}\lopvarbinop\lopvar_i\overline\Pi(T_i)\Rightarrow\overline\lopvar_i\cpred D{\interps{T_i}}\lopvarbinop\lopvar_i\overline Q(\interps{T_i})$, which in turns justifies the implication to be shown in this case.
\end{proof}
\begin{corollary}[Generic decidability II]\label{cor:generic-decidability}
Let $D$, $D'$, and $D''$ be as in \cref{prop:genfinchar}, and assume
$\Pi$ given by a decision procedure, and let $\overline Q$ be closed under decontraction. Then:
\begin{enumerate}
	\item Predicate $\ipredd{D}\circ\solletter$ on sequents of $\LJP$ is decided by the algorithm $\ipredfd{D'}\circ\finrepsymb$.
	\item Predicate $\ipredd{D}\circ\interps{\_}$ on finitary forests is decided by the algorithm $\ipredfd{D''}$.
\end{enumerate}
\end{corollary}
\begin{proof}
	Part (1) follows from part (\ref{prop:genfinchar.3}) of \cref{prop:genfinchar}, and the facts $\finrepsymb$ is computable and $\ipredfd{D'}$ is a decision procedure (the latter fact is a consequence of \cref{prop:decidability-complementary-predicates} -- recall $D' := (D,\Pi,\overline \Pi,\falsepred)$ and $\falsepred$ is a decision procedure).
Part (2)  follows from part (\ref{prop:genfinchar.4}) of \cref{prop:genfinchar} and $\ipredfd{D''}$ being a decision procedure, again a consequence of \cref{prop:decidability-complementary-predicates}, with the help of the two facts used in part (1)  (recall $D'' := (D,\Pi,\overline \Pi,\ipredfd{D'}\circ\finrepsymb)$).
\end{proof}
Remember the predicates in \cref{cor:S-props} are all of the form $\ipredd{D}\circ\solletter$. That the respective predicates $\Pi$ are given by decision procedures will be shown below, and \cref{sec:appendix-exfinnofin} proves that our four running examples satisfy the extra condition on $\overline Q$.


%% file: decision-problems-PIPL.tex
\subsection{Deciding problems on inhabitants and on solutions in the systems $\PIL$ and $\copil$}\label{sec:decision-problems-PIPL}

We are now ready to adapt to $\PIL$ our method \cite{EspiritoSantoMatthesPintoInhabitation}, expanded in \cite{JESRMLP-FI2019}, able to address decidability of existence problems and of finiteness problems, but that has been 
only available for intuitionistic implication (while the extension to $\PIL$ in the paper \cite{JESRMLP-TYPES20} which the present paper expands upon does not consider solutions). 

\begin{theorem}[Deciding the existence of inhabitants in $\PIL$ and of solutions in $\copil$]
  \label{thm:decide-existence}
Given logical sequent $\jay$ in $\PIL$ (equivalently, $\copil$):
\begin{enumerate}
\item \textbf{Existence of inhabitants:} $\jay$ has an inhabitant in $\PIL$ iff\/ 
  $\nbinfp \falsepred{\finrepempty\sigma}$.
Hence ``$\sigma$ is
  inhabited'' is decided by deciding
  $\nbinfp \falsepred{\finrepempty\sigma}$. In other words, the
  inhabitation problem for $\PIL$ is decided by the computable
  predicate $\nbinfpsymb\falsepred\circ\finrepsymb$.
\item \textbf{Existence of solutions:} $\jay$ has a solution in $\copil$ iff\/
  $\nstuckp \falsepred{\finrepempty\sigma}$.
Hence ``$\sigma$ is
  solvable'' is decided by deciding
  $\nstuckp \falsepred{\finrepempty\sigma}$. In other words, the
  solvability problem for $\copil$ is decided by the computable
  predicate $\nstuckpsymb\falsepred\circ\finrepsymb$.
\end{enumerate}
\end{theorem}
\begin{proof}
Part 1. By \cref{cor:S-props} (part 1), the left-hand side is equivalent to 
$\exfin{\solfunction\sigma}$, which in turn is equivalent to  
$\nbinfp \falsepred{\finrepempty\sigma}$ by part 3 of \cref{prop:genfinchar}. 
To see the latter, recall from \cref{ex:dualpairsmembership} and \cref{ex:nbinf-binf+stuck-nstuck}, 
$\exfinsymb=\ipredd D$ for $D=(\bigwedge,\falsepred,\truepred,\bigvee)$, and
$\nbinfpsymb \falsepred:=\ipredfd{D'}$ for $D'=(D,\falsepred,\truepred,\falsepred)$.
Computability follows from \cref{cor:generic-decidability} (part 1), with the help of the trivial facts that $\Pi=\falsepred$ is a decision procedure, and that $\overline Q=\truepred$ is closed under decontraction.
Part 2 follows analogously, but resorting instead to part 3 of \cref{cor:S-props}, and the facts: 
$\exsolsymb=\cpredd D$ for $D=(\bigvee,\truepred,\falsepred,\bigwedge)$; 
$\nstuckpsymb \falsepred:=\cpredfd{D'}$ for $D'=(D,\truepred,\falsepred,\falsepred)$; 
$\Pi=\truepred$ is a decision procedure and $\overline Q=\falsepred$ is closed under decontraction;
and, of course, the fact the two pairs of predicates $(\cpredd D,\ipredd D)$, and $(\cpredfd{D'},\ipredfd{D'})$ are both complementary.
\end{proof}

As in the generic \cref{prop:genfinchar}, when passing from part \ref{prop:genfinchar.3} to part \ref{prop:genfinchar.4}, building on the previous theorem, we will consider other meaningful settings for parameter $P$ in the predicates 
of \cref{ex:nbinf-binf+stuck-nstuck}, and this will be essential in the proof of \cref{thm:decide-finiteness} below. 
Specifically, let us consider: (i)  the predicates $\nbinfcansymb$ and $\binfcansymb$ on
  $\pils$  defined by $\nbinfcansymb:=\nbinfpsymb \Pcansymb$ and
  $\binfcansymb:=\binfpsymb \Pcansymb$ for $\Pcansymb=\nbinfpsymb\falsepred\circ\finrepsymb$;
 and  (ii) the predicates $\nstuckcansymb$ and $\stuckcansymb$ on
  $\pils$  defined by $\nstuckcansymb:=\nstuckpsymb \Pcansymb$ and
  $\stuckcansymb:=\binfpsymb \Pcansymb$ for $\Pcansymb=\nstuckpsymb\falsepred\circ\finrepsymb$.
\begin{lemma}[Sharp finitary characterizations and their decidability]\label{lem:exfin-exsol-sharp}
  For all $T\in\pils$, 
\begin{enumerate}
\item $\nbinfcan T$ iff  $\exfin{\interps T}$. 
Morever,  $\nbinfcansymb$ is a decision procedure.
\item $\nstuckcan T$ iff  $\exsol{\interps T}$. 
Morever, $\nstuckcansymb$ is a decision procedure. 
\end{enumerate}
\end{lemma}
\begin{proof}
These are plainly instances of part \ref{prop:genfinchar.4} of \cref{prop:genfinchar} and \cref{cor:generic-decidability} (part 2).
Let us see how to do the instantiation to prove the first claim of the first part. 
Recall $\nbinfpsymb \falsepred:=\ipredfd{D'}$ for $D'=(D,\falsepred)$, with $D=(\bigwedge,\bigvee)$ (beware of the use of the logical simplifications in expressing (finitary) dual pair data, which in particular give $\Pi=\falsepred$ and $\overline Q=\truepred$).  
Let $P_*$ be $\nbinfpsymb\falsepred\circ\finrepsymb$.
So, $D''=(D,P_*)$, hence $\ipredfd{D''}=\nbinfcansymb$.
Additionally, observe that in this case $\ipredsymb_D$ amounts to $\exfinsymb$ (recall the definition of the latter from  \cref{ex:dualpairsmembership}). 
Then, the second claim of part 1 follows immediately from part 2 of \cref{cor:generic-decidability}, with the help of the trivial facts $\Pi=\falsepred$ is a decision procedure and $\overline Q=\truepred$ is closed under decontraction.
The second part follows analogously, with the required instantiation taking, of course, $\nstuckpsymb\falsepred\circ\finrepsymb$ for $\Pcansymb$.
\end{proof}

Now we turn our attention to the two remaining decision problems on logical sequents emerging from our collection of dual pairs of predicates of \cref{ex:dualpairsmembership}, namely: (i) the problem of the finiteness of the number of inhabitants in $\PIL$ ; (ii) the problem of the existence of an infinite solution in $\copil$. The pursued decidability results will be achieved by mimicking the development for deciding the existence of inhabitants in $\PIL$ and of solutions in $\copil$, but will additionally require concepts and results offered by the latter.

Recall the parameter tuple $D=(\bigwedge,\nofinsymb,\exfinsymb,\bigwedge)$ used for the definition of $\finfinsymb$ and $\inffinsymb$.
We extend it by $\Pi:=\binfcansymb$ and $\overline\Pi:=\nbinfcansymb$ to provide the data for the definition of $\FFpsymb P:=\ipredsymbf_P$ and $\NFFpsymb P:=\cpredsymbf_P$ as instance of \cref{def:dualpairsfin}: $\Pi$ and $\overline\Pi$ qualify thanks to \cref{lem:exfin-exsol-sharp} (part 1). For convenience, the obtained inference rules are spelt out in \cref{fig:FF-NFF}.\footnote{Notice that the treatment of $\gfp X^\rho.T$ differs w.\,r.\,t.~\cite[Figure 6]{JESRMLP-TYPES20}, so as to become an instance of the generic scheme. For the instance $\FFpsymb P$, this change is rather irrelevant, but not so for the proof of \cref{prop:genfinchar}.}
\begin{figure}[tb]\caption{$\FFpsymb P$ and $\NFFpsymb P$ predicates}\label{fig:FF-NFF}
\[
\begin{array}{c}
\infer[]{\FFp P{X^\rho}}{P(\rho)}\qquad
\infer[]{\FFp P{\gfp X^\rho.T}}{\FFp P{T}}\qquad
\infer[]{\FFp P{f(T_i)_i}}{\bigwedge_i\FFp P{T_i}}\qquad
\infer[]{\FFp P{f(T_i)_i}}{\binfcan{T_j}}\qquad
\infer[]{\FFp P{\sum_iT_i}}{\bigwedge_i\FFp P{T_i}}\\[1.5ex]
\infer[]{\NFFp P{X^\rho}}{\neg P(\rho)}\qquad
\infer[]{\NFFp P{\gfp X^\rho.T}}{\NFFp P{T}}\qquad
\infer[]{\NFFp P{f(T_i)_i}}{\NFFp P{T_j}&\bigwedge_i\nbinfcan{T_i}}\qquad
\infer[]{\NFFp P{\sum_iT_i}}{\NFFp P{T_j}}
\end{array}
\]
\end{figure}

Recall  now the parameter tuple $D=(\bigwedge,\nosolsymb,\exsolsymb,\bigwedge)$ used for the definition of $\allfinsymb$ and $\exinfsymb$.
  We extend it by $\Pi:=\stuckcansymb$ and $\overline\Pi:=\nstuckcansymb$ to provide the data for the definition of $\AFpsymb P:=\ipredsymbf_P$ and $\NAFpsymb P:=\cpredsymbf_P$ as instance of \cref{def:dualpairsfin}: $\Pi$ and $\overline\Pi$ qualify thanks to \cref{lem:exfin-exsol-sharp} (part 2). The ready-to-use formulations of the inference rules for $\AFpsymb P$ and $\NAFpsymb P$ are analogous to those presented in \cref{fig:FF-NFF} for $\FFpsymb P$ and $\NFFpsymb P$. The only difference is that $\stuckcansymb$ and $\nstuckcansymb$ replace $\binfcansymb$ and $\nbinfcansymb$ as side conditions, respectively.

  \begin{theorem}[Deciding finiteness in $\PIL$ and existence of an infinite solution in $\copil$]
  \label{thm:decide-finiteness}
Given logical sequent $\jay$ in $\PIL$ (equivalently, $\copil$):
\begin{enumerate}
\item \textbf{Finiteness of the number of inhabitants:} $\jay$ has finitely many inhabitants in $\PIL$ iff\/ 
  $\FFp \falsepred{\finrepempty\sigma}$.
Hence the problem ``$\sigma$ has finitely many inhabitants'' is decided by deciding
 the computable predicate $\FFpsymb\falsepred\circ\finrepsymb$.
\item \textbf{Existence of infinite solutions:} $\jay$ has an infinite solution in $\copil$ iff\/
  $\NAFp \falsepred{\finrepempty\sigma}$.
Hence the problem ``$\sigma$ has an infinite solution'' is decided by deciding
 the computable predicate $\NAFpsymb\falsepred\circ\finrepsymb$.
\end{enumerate}
\end{theorem}
\begin{proof}
We argue about part 1 (part 2 follows analogously).
By \cref{cor:S-props} (part 2), the left-hand side is equivalent to 
$\finfin{\solfunction\sigma}$, which in turn is equivalent to  
$\FFp \falsepred{\finrepempty\sigma}$ by part 3 of \cref{prop:genfinchar} (recall 
$\finfinsymb:=\ipredd D$ and $\FFpsymb \falsepred:=\ipredfd{D'}$ 
for $D=(\bigwedge,\nofinsymb,\exfinsymb,\bigvee)$, and $D'=(D,\binfcansymb,\nbinfcansymb,\falsepred)$). 
Computability is once more a consequence of computability of $\finrepsymb$ and part 1 of \cref{cor:generic-decidability}, with the help of the facts that $\binfcansymb$ is given as a decision algorithm (a consequence of part 1 of \cref{lem:exfin-exsol-sharp}) and that $\exfinsymb$ is closed under decontraction (argued for in \cref{sec:appendix-decontraction}).
In arguing in part 2 about computability of the predicate $\NAFpsymb\falsepred\circ\finrepsymb$, the two helpful facts are that $\stuckcansymb$ is given as a decision algorithm (a consequence of part 2 of \cref{lem:exfin-exsol-sharp}) and that $\exsolsymb$ is closed under decontraction (argued for in \cref{sec:appendix-decontraction}).
\end{proof}
In essence, the preceding theorem is about decidability of questions of finiteness: this is obvious concerning the set of inhabitants, but existence of an infinite solution for $\jay$ is the negation of ``all solutions of $\jay$ are finite'', that is reflected in the name $\allfinsymb$ for the respective predicate on forests and $\AFpsymb P$ for the respective parametrized predicate on finitary forests. In \cite{JESRMLP-FI2019}, the present authors argued about both the predicates $\finfinsymb$ and $\allfinsymb$ as instances of a more abstract concept of finiteness (but only for implicational intuitionistic logic).


%% file: applications-IPL.tex
\section{Applications to intuitionistic propositional logic with all connectives}\label{sec:applications-IPL}
One of the interests of polarized logic is that it can be used to analyze other logics \cite{LiangMillerTCS09}. This is also true of $\PIL$ and we illustrate it now, deriving meta-theory and algorithms for
decision problems for two proof systems for intuitionistic propositional logic (\ipl)
with all connectives, which correspond to variants of Herbelin's systems $\LJT$ and $\LJQ$ \cite{HerbelinPhD}.  
Again, methodologically, we work primarily at the level of co-proof terms and solutions.
We consider coinductive versions of $\LJT$ and $\LJQ$, and corecursive translations of such coinductive systems into $\copil$.
We call these the \emph{negative}, resp.\ \emph{positive}, translation, as \ipl\, formulas are mapped to negative, resp.\ positive, formulas of \pipl. They encompass recursive translations of our variants of $\LJT$ and $\LJQ$ into $\LJP$. We will prove that the translations are
\emph{full embeddings}, in a sense to be made precise. 
This will readily enable to transfer results from $\copil$ (resp.~$\LJP$) to our coinductive (resp.~inductive) versions of $\LJT$  and $\LJQ$. 

The section contains two subsections, one dedicated to $\LJT$ and the negative translation, the other to $\LJQ$ and the positive translation. 
To go through
the details of the section, it is useful to go back to \cref{sec:background} and recapitulate 
the propositional formulas we consider for \ipl.

\input{app-LJT}
\input{app-LJQ-unoptimised}


%% file: app-LJT.tex
\subsection{Systems $\LJT$, $\coLJT$ and their negative embedding into polarized logic}\label{subsec:LJT-LJTco}

\smallskip
\noindent
\textbf{System $\LJT$.}
The best known variant of the focused sequent calculus $LJT$ for \ipl~is the one for implication only \cite{HerbelinCSL94}.
Variants including conjunction and disjunction as well can be found in \cite{HerbelinPhD,DP96}.
We present our own variant, denoted $\LJTour$ when comparing it with other formulations. Normally, we will omit the upper index $e$ that indicates the presence of the syntactic class of expressions whose purpose is to constrain the term-forming operations in a way that allows us to give a faithful and full embedding into $\PIL$.

\emph{Proof terms} of $\LJT$ are organized in three syntactic categories as follows:\\[-1ex]
\[
\begin{array}{lcrcl}
\textrm{(terms)} &  & t& ::= & \lb x^A.t\mid\pairljt{t_1}{t_2}\mid e\\
\textrm{(expressions)} &  & e& ::= & \focl xsR\mid \inj 1 {A}{t}\mid \inj 2 {A}{t}\\
\textrm{(spines)} &  & s & ::= & \nil\mid t::s\mid 1::s\mid 2::s\mid\abortt R\mid [x_1^{A_1}.e_1,x_2^{A_2}.e_2]
\end{array}
\]
where $x$ ranges over a countable set of variables. We will refer to $e_1$ and $e_2$ in the latter form of spines as \emph{arms}. Proof terms in any category are ranged over by $T$. Notice that we restrict the upper index in $\abortt R$ to right formulas $R$ already in the syntax, not only through the typing rules. Additionally, note that, in a slight departure from the version of $\LJT$ introduced in
\cite[Section 5.1]{JESRMLP-TYPES20}, there is now type information in the construction $\focl xs R$ (an $R$-formula), which will become important in the coinductive extension of $\LJT$ where it enables the definition of the negative translation for the raw syntax of (co-)proof terms.

There are three forms of \emph{sequents}, $\Gamma\Longrightarrow t:A$ and $\Gamma\vdash e:R$ and $\Gamma[s:A]\vdash R$, where, as usual, $\Gamma$ is a context made of associations of variables with formulas. Therefore, a logical sequent $\jay$ in $\LJT$ may have three forms: $\Gamma\Longrightarrow A$ and $\Gamma\vdash R$ and $\Gamma[A]\vdash R$. The latter two forms
require a right formula to the right of the turnstile. Typing a proof term $T$ means finding a logical sequent $\sigma$ so that the sequent $\sigma(T)$  (defined analogously to $\PIL$) can be derived from the typing rules. 
The full definition of the typing rules of $\LJT$ is given in Fig.~\ref{fig:typingljt}. A sequent $\sigma(T)$ that is derived from the typing rules is called a \emph{valid} sequent.
As for $\PIL$, uniqueness of types is guaranteed: the type annotations provide means to ensure that there is at most one
formula that can replace the placeholders in
$\Gamma\Longrightarrow t:\cdot$, $\Gamma\vdash e:\cdot$ and $\Gamma[s:A]\vdash \cdot$ and give a valid sequent. The annotation with $R$ in $\focl xsR$ is not needed for uniqueness to hold.
We use this annotation for qualifying an expression as \emph{atomic} iff it is of the form $\focl xsa$, i.\,e., it must be in the first of the three cases in the grammar, and even with $R=a$.

\begin{figure}[tb]\caption{Typing rules of $\LJT$}\label{fig:typingljt}
	\[
	\begin{array}{c}
\infer[]{\Gamma\Longrightarrow\lb x^A.t:A\impl B}{\Gamma, x:A\Longrightarrow t:B} \quad\infer[]{\Gamma\Longrightarrow\pairljt{t_1}{t_2}: A_1\wedge A_2}{\Gamma\Longrightarrow t_i: A_i\quad\textrm{for $i=1,2$}}\quad\infer[]{\Gamma\Longrightarrow e:R }{\Gamma\vdash e:R}\quad\infer[]{\Gamma,x:A\vdash
	 \focl xsR:R}{\Gamma,x:A[s:A]\vdash R}\\[2.5ex]\infer[i\in\{1,2\}]{\Gamma\vdash
	\inj i{A_{3-i}}t:A_1\vee A_2}{\Gamma\Longrightarrow t:A_i}\quad\infer[]{\Gamma[t::s:A\impl B]\vdash R}{\Gamma\Longrightarrow t:A\quad\Gamma[s:B]\vdash R}\quad\infer[]{\Gamma[\nil:a]\vdash a}{}\\[2.5ex]
	\infer[]{\Gamma [\abortt R:\falsity]\vdash R}{}\quad \infer[i\in\{1,2\}]{\Gamma[{i}::s:A_1\wedge A_2]\vdash R}{\Gamma[s:A_{i}]\vdash R}\quad\infer[]{\Gamma[x_1^{A_1}.e_1,x_2^{A_2}.e_2:A_1\vee A_2]\vdash R}{\Gamma, x_i:A_i\Longrightarrow e_i:R\quad\textrm{for $i=1,2$}}\\[2.5ex]
	\end{array}   
	\]
\end{figure}

The characteristic feature of the design of $\LJTour$ is the restriction of the type of spines to right formulas.
Since the type of $\nil$ is atomic, spines have to be ``long''; and the arms of spines cannot be lambda-abstractions nor pairs, which is enforced by restricting the arms of spines to be expressions, rather than general terms: this is the usefulness of separating the class of expressions from the class of terms.  In the typing rules, the restriction to right formulas is generated at the \emph{select rule}
(the typing rule for $xs$); and the long form is forced by the identity axiom (the typing rule for $\nil$) because it applies to atoms only.

We could not find in the literature the restriction of cut-free LJT we consider here, but Ferrari and Fiorentini \cite{FF19} consider a presentation of \ipl\ that enforces a similar use of right formulas, in spite of being given in natural deduction format and without proof terms. It is easy to equip this natural deduction system with proof terms and map it into $\LJT$: the technique is fully developed in \cite{JESENTCS17} for polarized logic, but goes back to \cite{DP96}. Since the just mentioned system \cite{FF19} is complete for provability, so is $\LJT$.

\smallskip
\noindent
\textbf{System $\coLJT$.}
The coinductive extension of $\LJT$, denoted $\coLJT$, is obtained (in analogy to the coinductive extension of $\PIL$ in \cref{sec:copil})
by taking:  (i)  a coinductive reading of the grammar of expressions of $\LJT$;  and (ii) a  coinductive interpretation of the typing rules of $\LJT$. So, as in $\LJT$, expressions of $\coLJT$ are still organized into three categories, introduced by the simultaneous coinductive definition of the sets $t^{co}$, $e^{co}$ and $s^{co}$. However (as  in $\copil$), the coinductive reading of the grammar of \emph{co-proof terms} of $\coLJT$ is only attached to the sub-class of expressions of sort $e$, and of the co-proof term formers of sort $e$ only those forming  expressions $\focl xsR$  get priority 2. Hence,  
infinite branches must arise by going infinitely often through expressions of the form $\focl xsR$.
Similarly to $\copil$, the notion of equality on co-proof terms is  \emph{bisimilarity} modulo $\alpha$-equivalence.  The  \emph{logical sequents} of $\coLJT$ are the same as those of $\LJT$ (in parallel to the sharing of logical sequents between $\copil$ and $\LJP$).
In analogy to $\copil$,  derivations in $\coLJT$ subsume those of $\LJT$, and an easy induction on expressions of $\LJT$ shows:
\begin{lemma}\label{lem:conservativeness-typing-LJT}
For any $T\in\LJT$, $\sigma(T)$ is valid in $\LJT$ iff   $\sigma(T)$ is valid in $\coLJT$.
\end{lemma}
As for $\copil$, a co-proof term $T\in\coLJT$ is a \emph{solution} of a logical sequent $\sigma$ when $\sigma(T)$ is derivable in $\coLJT$, and $T$ is called a \emph{finite solution} (resp. \emph{infinite solution}) when $T\in\LJT$ (resp. $T\not\in\LJT$.)
Contrary to $\LJT$, if in construction $\focl xsR$ the $R$-formula is dropped, we immediately lose uniqueness of typing: for example, at $x:A_1\vee A_2\vdash E:\cdot$, with $E$ the unique infinite expression satisfying $E=x[x_1^{A_1}. E,x_2^{A_2}. E]$,  the placeholder can be replaced by any $R$-formula and still give a valid sequent.

\begin{figure}\caption{Negative translation $\unt{(\_)}$ of $\coLJT$ (resp. $\LJT$) into $\copil$ (resp. $\PIL$)
and auxiliary translations $\urt{(\_)}$ and $\unte{(\_)}$}\label{fig:negative-translation}
\[
\begin{array}{c}
\begin{array}[4]{rclcrcl}
\unt{(A\supset B)}&=&\downshift \unt A\supset \unt B&\quad&\urt{(A\vee B)} &=&\downshift \unt A\vee \downshift \unt B\\
\unt{(A\wedge B)}&=&\unt A\wedge \unt B&&\urt\falsity &=&\falsity\\
\unt R&=&\Up \urt R&&\urt a&=&{a^-}
\end{array}
\\ \\[-2ex]
\begin{array}{rclcrcl}
\unt{(\lb x^{A}. t)}&=&
\left\{ 
\begin{array}[2]{ll}
\lb (x^{\unt A}. \unte e) &\textrm{if $t=e$ atomic}\\
\lb (x^{\unt A}. \dlv{\unt t}) &\textrm{else}\\
\end{array}
\right .
&\quad&\unte{(\focl xsR)}&=&\coretR x {\unt s}{\urt R}\\
\unt{\pairljt{t_1}{t_2}}&=&\pairljt{\unt{t_1}}{\unt{t_2}}&&\unte{\inj i A{t}}&=&\ret{\inj i {\downshift {\unt A}}{\thunk{\unt t}}}\\[0.5ex]
\unt{e}&=&
\left\{ 
\begin{array}[2]{ll}
\ea {\unte e} \quad\textrm{if $e$ is atomic}\\
\ep {\unte e} \quad\textrm{else}\\
\end{array}
\right .
\end{array}
\\ \\[-2ex]
\begin{array}{rclcrcl}
\unt\nil&=&\nil&&\unt{(\abortt{R})}&=&\cothunk{\abortt{\urt R}}\\
\unt{(t::s)}&=&\thunk {\unt t}::\unt s&&\unt{[x_1^{A_1}.e_1,x_2^{A_2}.e_2]}&=&\cothunk{[x_1^{\unt{A_1}}.{\unte{e_1}},x_2^{\unt{A_2}}.{\unte{e_2}}]}\\
\unt{(i::s)}&=&i::\unt s
\end{array}
\end{array}
\]
\end{figure}
\smallskip
\noindent
\textbf{Translation of formulas and of logical sequents.}
At the level of
formulas, the main translation $\unt A$ of any intuitionistic formula $A$ is a negative formula of \pipl\ (hence the naming of the translation),
and all atoms of \pipl\ have negative polarization.
This translation is given in \cref{fig:negative-translation}, and 
uses an auxiliary translation of right intuitionistic formulas $R$:  $\urt R$ is a right formula (and specifically, $\urt P$ is a positive formula). 
Here, we use the  abbreviation $\Up R$ (for $R$ a right formula of $\PIL$) introduced in \cref{sec:background}. Recall that $\Up R$ is always a negative formula of \pipl. 
Notice that translation $\urt{(\_)}$ can easily be extended to a translation of all intuitionistic formulas into right formulas by stipulating $\urt N:=\downshift \unt N$.
Notice also that the formula translations $\unt{(\_)}$ and $\urt{(\_)}$ are both sections of the forgetful map from \pipl\ formulas to \ipl\ formulas (see \cref{sec:background}). 

The translations $\unt{(\_)}$ and $\urt{(\_)}$ of \ipl\ formulas readily allow to define a translation of logical sequents of $\coLJT$ into logical sequents of $\copil$ as follows:
$\unt{(\Gamma\Longrightarrow A)}=(\unt\Gamma\Longrightarrow\unt A)$ and $\unt{(\Gamma\vdash R)}=(\unt\Gamma\vdash \urt R)$ and $\unt{(\Gamma[A]\vdash R)}=(\unt\Gamma[\unt A]\vdash \urt R)$. Here $\unt{\Gamma}$ indicates application of $\unt{(\_)}$ to all formulas in $\Gamma$.  Notice that this is also a translation of logical sequents of $\LJT$ into logical sequents of $\PIL$, because the logical sequents of  $\LJT$ (resp.\ $\PIL$) and of $\coLJT$ (resp.\ $\copil$) are the same.

\smallskip
\noindent
\textbf{Translation of (co-)proof terms.}
The translation of co-proof terms
is also presented in \cref{fig:negative-translation}.  It maps 
co-proof terms of sort $t$ (resp.~$s$, $e$)  
of $\coLJT$ to co-proof terms of the same sort 
of $\copil$, and this makes use of an auxiliary map $\unte{(\_)}$  when translating
co-proof terms of sort $e$.
(The notion of atomic expressions of $\LJT$ is extended to co-proof terms of sort $e$ of $\coLJT$ verbatim.)
Note that in the translation $\unt{(\_)}$ of co-proof terms, for every constructor of priority 2 in the source, a constructor with priority~2 appears in the target. A consequence is that an infinite co-proof term of $\coLJT$ is translated into an infinite co-proof term of $\copil$.
Note also that in the clauses defining the translation, all corecursive calls are guarded by $\copil$-constructors,  and are thus legitimate.

An easy induction over $\LJT$-proof terms shows that the negative translation $\unt{(\_)}$ maps 
proof terms of $\LJT$ into proof terms of $\PIL$ of the same sort.  
Therefore, the restriction of $\unt{(\_)}$ and $\urt{(\_)}$ to proof terms of $\LJT$ defines a translation into $\PIL$, which can also be obtained directly, by taking the defining clauses of these translations as simultaneous recursive definitions along the grammar of $\LJT$-proof terms. The  translation $\unt{(\_)}$ of $\LJT$ presented in \cite[Section 5.1]{JESRMLP-TYPES20} is defined only for \emph{legal proof terms}.  The reason why in the translation given here there is no need to resort to such class of proof terms is due to the type information now available in the construction $\focl xsR$. Also, to make evident the legitimacy of both the corecursive and the recursive interpretations of the clauses defining $\unt{(\_)}$ at the level of expressions, here we avoided the auxiliary notation $\mathsf{DLV}$, used in  \cite[Section 5.1]{JESRMLP-TYPES20}  for the translation of lambda abstractions. Still, the negative translations presented here and in \cite[Section 5.1]{JESRMLP-TYPES20} coincide for typable terms of $\LJT$  (up to the typing annotation in constructions $\focl xsR$ and $\coretR xsR$). Recall that any typable expression in the version of  $\LJT$ in \emph{op.\ cit.} is a legal proof term.

\begin{example}
  In continuation of \cref{ex:types}, we observe that $\Idt^-=\unt{\Idt}$ and that its inhabitant of sort $t$ given there is nothing but $\unt{(\lb x^a.\focl {x\,}\nil a)}$.
  We also observe $\Churcht^-=\unt{\Churcht}$ and, setting $x^0y:=\focl{y\,}\nil a$ and $x^{k+1}y:= \focl x{(x^ky::\nil)}a$, it is easy to see that $e_k=\unte{(x^ky)}$.
  Therefore, the $k$-th inhabitant of $\Churcht^-$ given in \cref{ex:types} is $\unt{(\lb x^{a\impl a}\lb y^a.x^ky)}$.
\end{example}
\begin{example}
In continuation of \cref{ex:infinite-solutions}, 
consider the $\coLJT$-co-proof term $x^\omega$ of sort $e$ given as the unique solution of $x^\omega= \focl x{(x^\omega::\nil)}a$. We coinductively type $x:a\impl a\vdash x^\omega:a$. 
Also recall the $\copil$-co-proof term $e_\omega$ from  \cref{ex:infinite-solutions}, given  by $e_\omega=\coretR{x}{\thunk{\ea{e_\omega}}::\nil}{a^-}$.
An easy coinduction shows $\unte{(x^\omega)}=e_\omega$.
 In $\coLJT$, $\Inft$ has the infinite solution $\lambda x^{a\impl a}.x^\omega$, and applying $\unt {(\_)}$ to this co-proof term gives $T_1$ of \cref{ex:infinite-solutions}, described there as infinite solution of $\Inft^-$, and the latter formula is $\unt{\Inft}$.
  The type $\Churcht$ has an $\omega$-th solution in $\coLJT$ beyond the inhabitants considered above: $\lambda x^{a\impl a}\lambda y^a.x^\omega$.
  Its negative translation is $T_2$ of \cref{ex:infinite-solutions}, described as a solution of $\Churcht^-$.
 
\end{example}

\smallskip
\noindent
\textbf{Properties of the negative translation.} We will argue first about the properties of the translation of $\coLJT$ into $\copil$, and obtain as a corollary
the same properties for the translation of $\LJT$ into $\PIL$.
\begin{proposition}[Soundness]
\label{prop:soundness-negative-translation}
The negative embedding is sound in the sense that for any co-proof term $T$ of\/ $\coLJT$ and sequent $\sigma$ of\/ $\coLJT$ such that $\sigma(T)$ is valid in $\coLJT$, $\unt{\sigma}(T^\bullet)$ is valid in $\copil$, for the appropriate $\bullet\in\{\untsymb,\untesymb\}$, that is:
\begin{enumerate}
\item If\/ $\Gamma\Longrightarrow t:A$ is valid in $\coLJT$ then $\unt\Gamma\Longrightarrow\unt t:\unt A$ is valid in $\copil$.
\item If\/ $\Gamma\vdash e:R$ is valid in $\coLJT$ then $\unt\Gamma\vdash\unte e:\urt R$ is valid in $\copil$.
\item If\/ $\Gamma[s:A]\vdash R$ is valid in $\coLJT$ then $\unt\Gamma[\unt s:\unt A]\vdash \urt R$ is valid in $\copil$.
\end{enumerate}
\end{proposition}
\begin{proof} 
The three items are proved simultaneously by coinduction on the typing relation of $\copil$.
We illustrate the case where $t=\lb x^B.t_0$. 
The assumption implies $A= B\horseshoe C$, and (a) $\Gamma,x:B\Longrightarrow t_0:C$ is valid in $\coLJT$.  Also,  $\unt A=\downshift \unt B\horseshoe\unt C$.  
Sub-case $C=a$, it must be  $t_0=e$ with $e=\focl ysa$ (for some $y,s$), and (b) $\Gamma,x:B\vdash e:a$ is valid in $\coLJT$.
So, $\unt t=\lb(x^{\unt B}.\unte e)$, and  with two inferences of $\copil$ we obtain $\unt\Gamma\Longrightarrow \unt t:\unt A$  from  $\unt\Gamma, x:{\unt B} \vdash {\unte e}:\urt a$ (recall $\unt a=\urt a$).  Since $\unt\Gamma, x:{\unt B}=\unt{(\Gamma, x:{ B})}$, the latter follows then from the coinductive hypothesis (guarded by the mentioned inferences of $\copil$) and (b).
Sub-case $C$ is not atomic. On the one hand, $t_0$ cannot be of the form $\focl ysa$, hence $\unt t=\lb(x^{\unt B}.\dlv{\unt{t_0}})$.  On the other hand, $\unt C$ is (negative) composite, hence in three steps we obtain $\unt\Gamma\Longrightarrow \unt t:\unt A$  from  $\unt\Gamma, x:{\unt B} \Longrightarrow {\unt {t_0}}:\unt C$, and so, to conclude, it suffices to use  the coinductive hypothesis together with (a).
\end{proof}

In order to strengthen the previous proposition and argue about other properties, 
we need to understand better its image,
which we will call the \emph{$\untsymb$-fragment} of $\copil$. Consider the following subclasses of \pipl-formulas:\\[-1ex] 
\[
\begin{array}{rrcl}
(\textrm{$\untsymb$-formulas})
  &M,N&::=&a^- \mid \upshift P\mid \downshift N\supset M\mid N\wedge M\\
(\textrm{positive $\urtsymb$-formulas})
&P&::=&\falsity\mid \downshift N\vee \downshift M \\
  (\textrm{$\urtsymb$-formulas})
  &R&::=&a^-\mid P
\end{array}
\]
An $\untsymb$-formula is a negative formula; a positive $\urtsymb$-formula is a positive formula; an $\urtsymb$-formula is a right formula.
The names of these subclasses of formulas make sense in that, for any formula $A$, $\unt A$ is an $\untsymb$-formula, for any positive formula $P$, $\urt P$ is a positive $\urtsymb$-formula and, for any right formula $R$, $\urt R$ is an $\urtsymb$-formula, easily established by induction over \ipl\ formulas.
We already knew that the two translations of \ipl\ formulas are right inverses of the forgetful map. But if we restrict their codomain to these subclasses, an easy induction on formulas also shows that those restrictions are left inverses to the respective forgetful maps, obtained by restricting the domain accordingly, in symbols, this is just:
for every $\untsymb$-formula $N$, $\unt{\fgt N}=N$, for every positive $\urtsymb$-formula $P$, $\urt{\fgt P}=P$ and for every $\urtsymb$-formula $R$, $\urt{\fgt R}=R$. For this, we also establish that, for any positive $\urtsymb$-formula $P$, $\fgt P$ is a positive formula.
Thus, the negative translation, at the level of formulas, is a bijection from intuitionistic formulas to $\untsymb$-formulas, from positive intuitionistic formulas to positive $\urtsymb$-formulas; and from right intuitionistic formulas to $\urtsymb$-formulas.

It will also be necessary to characterize the image of the translation at the level of co-proof terms, that is the co-proof terms in the $\untsymb$-fragment of $\copil$.
This will be done through the unary predicates $\ntermsymb$, $\nexpsymb$ and $\nspinesymb$ on  co-proof terms of $\copil$ (of sorts $t$, $e$ and $s$, respectively), whose  simultaneous coinductive definition is given in \cref{fig:imageLJTco} of \cref{app:appendix-negative-embedding}.
An immediate coinduction on $\ntermsymb$ (together with $\nexpsymb$ and $\nspinesymb$)  shows that:
 for any $t\in\coLJT$,  $\ntermsymb$ holds of $\unt{t}$; and,  for any $e\in\coLJT$,  $\nexpsymb$  holds of $\unte{e}$; and, for any $s\in\coLJT$, $\nspinesymb$ holds of $\unt{s}$. As in \cref{app:appendix-negative-embedding}, we will therefore say that $\unt{t}$ is an $\untsymb$-term, $\unte{e}$ is a $\untesymb$-expressions, and $\unt{s}$ is an $\untsymb$-spine.

There is a \emph{forgetful} map $\fgt{\_}$, from $\untsymb$-terms (resp.~$\untesymb$-expressions, $\untsymb$-spines) to co-proof terms of sort $t$ (resp.~$s$, $e$) of $\coLJT$, 
that essentially erases co-proof term decorations. This map is given corecursively in \cref{fig:forgetful-map}
of \cref{app:appendix-negative-embedding}.
An easy induction on $\PIL$-expressions shows that the restriction of the forgetful map to $\untsymb$-terms, $\untesymb$-expressions and $\untsymb$-spines in $\PIL$  defines a forgetful map into $\LJT$.
As for formulas, the forgetful map from the identified subclasses of $\copil$-expressions gives a way to invert the translation of $\coLJT$ co-proof terms, namely, an easy coinduction on  bisimilarity for $\coLJT$ co-proof terms gives:  $\fgt{\unt t}=t$, $\fgt{\unte e}=e$, and $\fgt{\unt s}=s$. In particular, the translations $\unt{(\_)}$ and $\unte{(\_)}$ of co-proof terms of $\coLJT$  are injective. 
Of course, the restrictions of these translations to proof-terms of $\LJT$ are also injective and a right inverse to the restriction of the forgetful map to $\PIL$-expressions.
Furthermore, this forgetful map into $\coLJT$ preserves
validity (see \cref{lem:soundness-forgetful-map-to-coLJT} in \cref{app:appendix-negative-embedding}). \footnote{
	Notice that preservation of validity by the forgetful map (\cref{lem:soundness-forgetful-map-to-coLJT}), and the fact that the translation is a right inverse of the forgetful map, allow the strengthening of
	\cref{prop:soundness-negative-translation} into an equivalence.} 

Let $\sigma$ be a logical sequent of $\coLJT$. By the \emph{$\sigma$-restriction} of $\unt{(\_)}$ we mean the restriction of the translation to the solutions of $\sigma$. \cref{prop:soundness-negative-translation} implies that $\unt{(\_)}$ sends solutions of $\sigma$ to solutions of $\unt{\sigma}$. As observed above, each $\sigma$-restriction is an injective map and, for this reason, we say that the negative translation is an \emph{embedding}.

\begin{proposition}[Full embedding]\label{prop:prep-faithfulness-negative-translation}
For $\Gamma, A,R$ in $\coLJT$ and for $t,e,s\in\copil$:
\begin{enumerate}
\item
  If\/ $\unt\Gamma\Longrightarrow t:\unt A$ is valid in $\copil$, then (i) $\nterm t$ and (ii) $\unt{\fgt t}=t$.
\item
  If\/ $\unt\Gamma\vdash e:\urt R$ is valid in $\copil$, then (i) $\nexp e$ and (ii) $\unte{\fgt e}=e$.
\item
  If\/ $\unt\Gamma[s:\unt A]\vdash \urt R$ is valid in $\copil$, then (i) $\nspine s$ and (ii) $\unt{\fgt s}=s$.
\end{enumerate}
Hence the negative translation is \emph{full}, in the sense that each $\sigma$-restriction is surjective.
\end{proposition}
\begin{proof}
First one proves simultaneously parts (i) of each item, by coinduction on $\ntermsymb$ (given simultaneously with $\nexpsymb$ and $\nspinesymb$). Then, one proves simultaneously parts (ii) of each item, by coinduction on bisimilarity for co-proof terms of $\copil$.  In each of the items we do a case analysis on types, namely $A$, (resp.\, $R$, $A$), for item 1 (resp.\, 2, 3).
We illustrate item 1 in the case $A= B\horseshoe C$.  So, $\unt A=\downshift \unt B\horseshoe\unt C$, $t=\lb(x^{\unt B}.e)$ (for some $e\in \copil$), and (a) $\unt\Gamma, x:\unt B \vdash e:\unt C$ is valid. Sub-case $C=a$. Hence, $\unt C=a^-$, $e=\coretR ys{a^-}$ (for some $y,s$). Observe $\unt C=\urt C(=a^-)$.  To argue for (i), it suffices to use the coinduction hypothesis and (a) to get $\nexp e$. Regarding (ii), observe $ \unt{\fgt t}=
\unt{ (\lb x^{ \fgt{\unt B} }. {\focl y {\fgt{s}}{a}})  }
=\lb (x^{ \unt B} . \unte{({\focl y {\fgt{s}}{a}})})  
=\lb (x^{\unt B}.\unte {\fgt e })=t$, where the last step uses the coinduction hypothesis and (a). 
Sub-case $C$ is not an atom. Hence, $\unt C$ is (negative) composite, $e=\dlv{t_0}$ (for some $t_0\in\copil$), and (b) $\unt\Gamma,x:\unt B \Longrightarrow t_0:\unt C$ is valid. To prove (i) we can use the coinduction hypothesis and (b), which give $\nterm{t_0}$. Regarding (ii), observe $ \unt{\fgt t}=
\unt{ (\lb x^{ \fgt{\unt B} }. \fgt{t_0})  }
=\lb (x^{\unt B}.\dlv{\unt {\fgt {t_0}}})=t$, where the last step uses the coinduction hypothesis and (b), and in the penultimate step $\fgt {t_0}$ cannot be of the form $\focl ysa$, as this would contradict $\Gamma,x:B \Longrightarrow \fgt{t_0}:C$ (recall $C$ is  not an atom), a fact that follows from (b) and  \cref{lem:soundness-forgetful-map-to-coLJT} (with the help of right inversion by the forgetful map at the level ot types).

Regarding surjectivity, the only question is whether $|t|$ (for instance) is in the domain of the $\sigma$-restriction. But this is true, due to the preservation of validity by the forgetful map and the fact that, at the level of formulas, the negative translation is a right inverse to the forgetful map.
\end{proof}

\begin{corollary} [Reduction of decision problems]\label{cor:reduction-problems} Let $\sigma$ be a logical sequent of $\coLJT$.
	\begin{enumerate}
		\item $\jay$ is solvable in $\coLJT$ iff $\jay^{\untsymb}$ is solvable in $\copil$. \footnote{Due to this, we may say the negative translation is \emph{faithful} -- here the word ``faithful'' is used in its logical sense, as in~\cite{TroelstraS00}.}
                \item  $\jay$ has an infinite solution in $\coLJT$ iff  $\jay^{\untsymb}$ has an infinite solution in $\copil$. 
	\end{enumerate}
\end{corollary}
\begin{proof}
(1) The implication from left to right follows from \cref{prop:soundness-negative-translation}. The implication from right to left follows from \cref{prop:prep-faithfulness-negative-translation}, the fact that the forgetful map into $\coLJT$ preserves validity, and the fact that, at the level of formulas, the negative translation is right inverse to the forgetful map.

(2) Can be argued for as (1) with the help of the facts that both the negative translation and the forgetful map preserve infinity of co-proof terms (remarked before and in  \cref{prop:prep-faithfulness-negative-translation}, respectively).
\end{proof}

Next we extract results for the negative translation between the inductive systems.

\begin{theorem}[Properties of the translation of $\LJT$ into $\LJP$]\label{thm:props-embedding-LJT}
The negative translation of\/ $\LJT$ is a full embedding. As a consequence, for $\sigma$ a logical sequent of\/ $\LJT$: inhabitation of $\sigma$ in $\LJT$ is equivalent to inhabitation of $\unt{\sigma}$ in $\pil$, and the number of inhabitants of $\sigma$ in $\LJT$ is finite iff the number of inhabitants of $\unt{\sigma}$ in $\pil$ is finite.
\end{theorem}
\begin{proof}
Propositions \ref{prop:soundness-negative-translation} and \ref{prop:prep-faithfulness-negative-translation} hold with the removal of the upper ``co'' indices. We argue about \cref{prop:soundness-negative-translation}. The three items follow analogously. We detail for item 1.  
From the assumption, \cref{lem:conservativeness-typing-LJT} gives that $\Gamma\Longrightarrow t:A$ is valid in $\coLJT$.  Hence, by \cref{prop:soundness-negative-translation},  
$\unt\Gamma\Longrightarrow\unt t:\unt A$ is valid in $\copil$.  
We already observed that translation $\unt{(\_)}$ maps $\LJT$-terms into terms of $\PIL$,  hence $\unt t$ is a term of $\PIL$, therefore, by \cref{lem:conservativeness-typing}, $\unt\Gamma\Longrightarrow\unt t:\unt A$ is valid in $\PIL$.
\end{proof}

\smallskip
\noindent
\textbf{Applications to meta-theory.}
Let us illustrate how the negative translation can obtain the disjunction property of \ipl\ for the following subclass of Rasiowa-Harrop formulas, that we call \emph{left Rasiowa-Harrop formulas}:
\[
{\mathbf L}::= a \mid
A\impl {\mathbf L} \mid {\mathbf L}_1 \wedge {\mathbf L}_2 
\]
with $A$ an arbitrary formula of \ipl. Note that this is a subclass of intuitionistic left formulas and  its grammar coincides with the one for Rasiowa-Harrop formulas (given in \cref{sec:firstmetathms}), once the production relative to $\falsity$ is omitted. Additionally, note that the next proposition already entails the \emph{disjunction property of \ipl}. Nonetheless, in the next section we will see that the embedding of $\LJQ$ into $\PIL$ can obtain the disjunction property of \ipl\, for the full class of Rasiowa-Harrop formulas.

\begin{proposition}[Disjunction property under left Rasiowa-Harrop hypotheses]
\label{prop:disj_prop_LJT}
In case $\Gamma$ only contains left Rasiowa-Harrop formulas, if\/ $\Gamma\Rightarrow A_1\vee A_2$ is inhabited in $\LJT$, then one of the logical sequents $\Gamma\Rightarrow A_1$ and $\Gamma\Rightarrow A_2$ is inhabited.
\end{proposition}
\begin{proof}
Soundness of the negative translation
(\cref{thm:props-embedding-LJT})
gives inhabitation of $\unt\Gamma\Rightarrow \unt{(A_1\vee A_2)}$ in $\PIL$, from which follows $\unt\Gamma\vdash \downshift \unt{A_1}\vee \downshift\unt{A_2}$. 
An easy induction shows $\unt{\mathbf L}$ is
a strict, polarized Rasiowa-Harrop formula 
when ${\mathbf L}$ is
an intuitionistic left Rasiowa-Harrop formula. 
Thus, the assumption on $\Gamma$ implies that $\unt\Gamma$ only contains
strict, polarized Rasiowa-Harrop formulas,
 and so \cref{thm:polarized-disjunction-property}
applies,
giving that $\unt\Gamma\vdash[\downshift\unt{A_i}]$ must be inhabited for some $i\in\{1,2\}$.
Therefore, $\unt\Gamma\Rightarrow \unt{A_i}$ is inhabited, and, finally,
faithfulness
(also observed in \cref{thm:props-embedding-LJT})
gives inhabitation of $\Gamma\Rightarrow {A_i}$ in $\LJT$. 
\end{proof}

The negative embedding  can also immediately obtain an infinity-or-nothing property for $\LJT$ from the one established for $\LJP$ (\cref{thm:infinity-or-nothing}):
\begin{proposition}[Infinity-or-nothing property of $\LJT$ under a disjunctive hypothesis]
\label{prop:inf-nothing-LJT}
If\/ $x:A_1\vee A_2\in\Gamma$, then, for any $A$,  $\Gamma\Rightarrow A$ has the infinity-or-nothing property in $\LJT$.
\end{proposition}
\begin{proof}
Follows by induction on $A$. The interesting case is $A=R$. (The other cases follow easily from the induction hypothesis.)
So, it suffices to argue about infinity-or-nothing for $\Gamma\vdash R$, or equivalently (by \cref{thm:props-embedding-LJT}) for $\sigma=(\unt\Gamma\Rightarrow\urt R)$. Given that $\unt{(A_1\vee A_2)}=\upshift(\downshift\unt{A_1}\vee \downshift\unt{A_2})$ and $\downshift\unt{A_1}\vee \downshift\unt{A_2}$ is a not fully absurd positive formula, and that $\urt R$ is a positive formula, \cref{thm:infinity-or-nothing} applies, giving infinity-or-nothing for $\sigma$.
\end{proof}
Let us pause to relate to the observation in  \cite[Section 1.7]{SchererRemy15} that \emph{focused proofs fail to be canonical}. In \emph{op.\,cit.}, $z:a,x:a\impl(b\vee c)\vdash b\vee c$ is given as an example of a logical sequent with a single canonical proof, but infinitely many focused inhabitants. As an immediate consequence of \cref{prop:inf-nothing-LJT}, we can conclude that in $\LJT$ the ``phenomenon of a single canonical inhabitant vs.\ infinitely many focused inhabitants'' appears already with the simpler sequent $x:b\vee c\vdash b\vee c$. However, note that under the focused discipline explained in  \cite[Section 1.7]{SchererRemy15} this simpler  example is not possible because inversion of a disjunction on the left-hand side does not create a duplicate, contrary to what happens in $\LJT$, which already makes a duplicate of a disjunction (in fact of any formula)  at its selection for left focusing.   

\smallskip
\noindent
\textbf{Application to decision problems.}
\Cref{thm:props-embedding-LJT} and \cref{cor:reduction-problems}
readly enable decision algorithms for problems in $\LJT$ and $\coLJT$ via our decision algorithms for problems in $\LJP$ and $\copil$.
Specifically, we will extract algorithms that decide the
\emph{existence of inhabitants} and the \emph{finiteness of the number of inhabitants} problems in $\LJT$  and the \emph{existence of solutions} and the \emph{existence of infinite solutions} problems in $\coLJT$.  
In all the four procedures the first step is the recursive calculation of the negative translation of the logical sequent at hand. This step is then followed by the composition of two recursive functions (as for $\PIL$ and $\copil$): first, $\finrepsymb$ 
calculates the finitary representation of the full solution space; second, recursing on the structure of this representation, the appropriate predicate on finitary forests is decided. 

\begin{theorem}[Decision algorithms for $\LJT$ and $\coLJT$]
\label{thm:decision-algorithms-LJT}
Given $\jay$ in $\LJT$ (equivalently, $\coLJT$):
\begin{enumerate}
\item \textbf{Existence of inhabitants:} $\jay$ has an inhabitant in $\LJT$ iff\/ 
$\nbinfp{\falsepred}{\finrepempty{\unt\jay}}$;
\item \textbf{Finiteness of the number of inhabitants:} $\jay$ has finitely many inhabitants in $\LJT$ iff\/ $\FFp{\falsepred}{\finrepempty{\unt\jay}}$;
\item \textbf{Existence of solutions:} $\jay$ has a solution in $\coLJT$ iff\/
 $\nstuckp \falsepred{\finrepempty{\unt\jay}}$;
\item \textbf{Existence of infinite solutions:} $\jay$ has an infinite solution in $\coLJT$ iff\/ 
 $\NAFp \falsepred{\finrepempty{\unt\jay}}$.
\end{enumerate}
In each of the four items, the right-hand-side of the equivalence is the obtained decision algorithm\footnote{Recall from  \cref{sec:decision-problems-PIPL}  that, although predicates $\nbinfpsymb\emptyset$, etc
are given inductively, they can be equivalently given by recursion over the structure of finitary forests.}.
\end{theorem}
\begin{proof} Let us argue about item 1: $\jay$ is inhabited in $\LJT$ 
iff $\unt\jay$ is inhabited in $\PIL$ (\cref{thm:props-embedding-LJT})
iff $\nbinfp{\falsepred}{\finrepempty{\unt\jay}}$ (part 1 of \cref{thm:decide-existence}). 
The other items follow analogously,  with a first equivalence resorting
to  \cref{thm:props-embedding-LJT}  or to one of the parts of \cref{cor:reduction-problems}, and
with a second equivalence resorting
to part 2 of \cref{thm:decide-existence} or to one of the parts of \cref{thm:decide-finiteness}.
\end{proof}


%% file: app-LJQ-unoptimised.tex
\subsection{Systems $\LJQ$, $\coLJQ$ and their positive embedding into polarized logic}\label{subsec:LJQ-LJQco}

\smallskip
\noindent
\textbf{System $\LJQ$.}
$\LJQ$ is a well-known focused sequent calculus with a long history in proof theory \cite{HerbelinPhD,DyckhoffLengrand2006,DyckhoffLengrand2007}. A fundamental feature of $\LJQ$ is that a  left-implication inference requires the side premise to be axiomatic or to result from a right inference. As a logical system, $\LJQ$ captures reasoning by forward chaining and, as a computational system 
it has a connection with call-by-value computation. 
Here we treat a cut-free variant of $\LJQ$ with the full set of propositional connectives 
that we still denote by $\LJQ$ or, when necessary to disambiguate, by $\LJQouru$.
(the superscript $u$ intends to signal that in left rules there is unnecessary repetition of the main formula in most premises). 
This system essentially follows the original formulation of $\LJQ$ by Herbelin  \cite{HerbelinPhD}[Sec. 5.2.1], but also integrates an additional focusing constraint found in the system $\LJQ'$  \cite{DyckhoffLengrand2006}  (explained below).  

\emph{Proof terms} of $\LJQ$ are organized into two syntactic categories of \emph{terms} $t$ and \emph{values} $v$:
\[ 
\begin{array}[4]{rrcl}
\textrm{(terms)}&t &::= &\crc{v}\mid  \impleftljqt xvyBtA\mid \conjleftljqt xyBtiA\mid \disjleftljqt x{y_1}{B_1}{t_1}{y_2}{B_2}{t_2}A\mid \abortljqt xA\\
\textrm{(values)}&v &::= &  x\mid\lb x^A.t\mid\pairljq{t_1}{t_2}\mid\inj{i}A{v}\\

\end{array}
\]

So, the set of sorts of $\LJQ$ is $S=\{t,v\}$, and there is a form of \emph{sequent} for each of them, namely $\Gamma\Longrightarrow t:A$ (of sort $t$) and $\Gamma\vdash [v:A]$ (of sort $v$),  where, as usual, $\Gamma$ is a context made of associations of variables with formulas. Therefore, a logical sequent $\jay$ in $\LJQ$ may have two forms:  $\Gamma\Longrightarrow A$ or $\Gamma\vdash [A]$, with sorts as for the sequents.
Also for $\LJQ$, typing a proof term $T$ means finding a logical sequent $\sigma$ so that the sequent $\sigma(T)$ (defined analogously to $\PIL$) can be derived from the typing rules.
The full definition of the typing rules of $\LJQ$ is given in Fig.~\ref{fig:typing-idLJQouru}.
As for the other systems in this paper, a sequent $\sigma(T)$ that is derived from the typing rules is called a \emph{valid} sequent.

\begin{figure}[tb]\caption{Typing rules of $\LJQ$ }\label{fig:typing-idLJQouru}
$$
\begin{array}{c}
\infer[]{\Gamma,x:a\vdash[x:a]}{}\qquad\infer[]{\Gamma\vdash [\lb x^{A}.t:A\horseshoe B]}{\Gamma,x:A\Longrightarrow t:B}\\\\
\infer[]{\Gamma\vdash[\pairljq{t_1}{t_2}:A_1\wedge A_2]}{\Gamma\Longrightarrow t_i:A_i&\textrm{for $i\in\{1,2\}$}}\qquad\infer[i\in\{1,2\}]{\Gamma\vdash[\inj i{A_{3-i}}{v}:A_1\vee A_2]}{\Gamma\vdash[v:A_i]}\\ \\
\qquad\infer[]{\Gamma\Longrightarrow \crc{v}:A}{\Gamma\vdash[v:A]}\qquad
\qquad\infer[]{\Gamma,x:A\horseshoe B\Longrightarrow \impleftljqt xvyBtC:C}{\Gamma,x:A\horseshoe B\vdash [v:A]&\Gamma,x:A\horseshoe B,y:B\Longrightarrow t:C}\\ \\
\infer[i\in\{1,2\}]{\Gamma,x:A_1\wedge A_2\Longrightarrow \conjleftljqt x y{A_i}tiA:A}{\Gamma,x:A_1\wedge A_2,y:{A_i}\Longrightarrow t:A}\\ \\
\infer[]{\Gamma,x:A_1\vee A_2\Longrightarrow \disjleftljqt x{y_1}{A_1}{t_1}{y_2}{A_2}{t_2}A:A}{\Gamma,x:A_1\vee A_2,y_i:A_i\Longrightarrow t_i:A & \textrm{for $i\in\{1,2\}$}}\qquad
\infer[]{\Gamma,x:\falsity\Longrightarrow \abortljqt xA:A}{}
\\ \\
\end{array}
$$
\end{figure}
As for $\PIL$ and $\LJT$, type annotations guarantee uniqueness of types in the sense that there is at most one formula that can replace the placeholders in $\Gamma\Longrightarrow t:\cdot$ and $\Gamma\vdash [v:\cdot]$ and yield a valid sequent (proved by an easy simultaneous induction on $t$ and $v$).
The type annotations $A$ in $\impleftljqt xvyBtA$, $\conjleftljqt x y{A_i}tiA$ and $\disjleftljqt x{y_1}{A_1}{t_1}{y_2}{A_2}{t_2}A$ are not needed for uniqueness to hold.

Here is a close comparison of $\LJQouru$ with the original version of $\LJQ$ in
\cite{HerbelinPhD}[Sec. 5.2.1], with the variant formulation $\LJQ'$ in
\cite{DyckhoffLengrand2006}[Sec. 2] (both without proof terms), and with the version informed by proof terms of the implicational fragment of $\LJQ$
in \cite{DyckhoffLengrand2006}[Sec. 5] and in \cite{DyckhoffLengrand2007}[Sec. 4].  
Despite having proof terms, as already mentioned,  $\LJQouru$ essentially follows Herbelin's formulation of $\LJQ$, which agrees with the version of implicational $\LJQ$ in \cite{DyckhoffLengrand2006,DyckhoffLengrand2007}.  In these formulations and in 
$\LJQouru$, left rules repeat the main formula in the premises.
The constraint of a focused premise in right-disjunction inferences that we adopt is found only in $\LJQ'$, as well as the left rule for the $\falsity$ connective.  However, left rules of $\LJQ'$ do not repeat the main formula in premises, and  $\LJQ'$ also differs of $\LJQouru$ in the treatment of conjunction:  the right rule for conjunction of $\LJQ'$ requires focused premises; for introducing a conjunction on the left, $\LJQ'$ has single rule, which requires the premise to have simultaneously the two conjuncts  in context.  Still, it is an easy exercise to obtain completeness of $\LJQouru$ from completeness of $\LJQ'$, as each rule of $\LJQ'$ is either a rule of $\LJQouru$ or is easily derivable in $\LJQouru$ (with the help of weakening).
 
\smallskip
\noindent
\textbf{System $\coLJQ$.}
The coinductive extension of $\LJQ$, denoted $\coLJQ$, is obtained (in analogy to the coinductive extensions of $\PIL$ and of $\LJT$)
by taking:  (i)  a coinductive reading of the grammar of expressions of $\LJQ$;  and (ii) a  coinductive interpretation of the typing rules of $\LJQ$. In the case of $\coLJQ$, the coinductive reading of the grammar of \emph{co-proof terms} is only attached to the sub-class of expressions of sort $t$, with all the co-proof terms formers of sort $t$, except for $\crc\cdot$, getting priority 2. So, in an infinite branch of an $\coLJQ$-co-proof term, \emph{left eliminators} (i.e.\, constructors attached to the left rules) must occur infinitely often.
Again, the notion of equality on co-proof terms is  \emph{bisimilarity} modulo $\alpha$-equivalence.  Also, note that the concept of \emph{logical sequent} stays unchanged from $\LJQ$.
Analogously to the example we showed for $\coLJT$, the unique co-proof term $T$ of sort $t$ satisfying $T=\disjleftljqt x{y_1}{A_1}{T}{y_2}{A_2}{T}A$ gets type $A$ in context $x:A_1\lor A_2$, but if the term had not been tagged with the formula $A$, every type would have been possible.

In analogy to $\copil$,  derivations in $\coLJQ$ subsume those of $\LJQ$, and an easy induction on expressions of $\LJQ$ shows:
\begin{lemma}\label{lem:conservativeness-typing-LJQ}
For any $T\in\LJQ$, $\sigma(T)$ is valid in $\LJQ$ iff   $\sigma(T)$ is valid in $\coLJQ$.
\end{lemma}
As for $\copil$, a co-proof term $T\in\coLJQ$ is a \emph{solution} of a logical sequent $\sigma$ when $\sigma(T)$ is valid in $\coLJQ$, and $T$ is called a \emph{finite solution} (resp. \emph{infinite solution}) when $T\in\LJQ$ (resp. $T\not\in\LJQ$.)

\smallskip
\noindent
\textbf{Translation of formulas and of logical sequents.}
At the level of formulas,  the main translation $\upt A$ of any intuitionistic formula $A$ is a positive formula of \pipl,
and all atoms of \pipl\ have positive polarization.
The translation $\uptsymb$  makes use of an auxiliary definition 
of a left formula $\ult A$, for any \ipl\ formula $A$. 
In the translation of an intuitionistic formula $A$, whereas
$\uptsymb$ is used to translate positive occurrences of subformulas of $A$, $\ultsymb$ is used to translate negative occurrences of subformulas of $A$.   The definitions are given in \cref{fig:coLJQ-positive-translation}. For a lighter notation, we use
the abbreviation $\Down L$
(for $L$ a left formula of \pipl)
 introduced in \cref{sec:background}
(which always stands for a positive formula of \pipl).

Notice that the two translations $\upt{(\_)}$ and $\ult{(\_)}$ of \ipl\ formulas are sections of the forgetful map $\fgt{\_}$ from \pipl\ formulas to \ipl\ formulas  described in \cref{sec:background}. This follows by an immediate induction on formulas of \ipl, proving simultaneously  $\fgt{\upt A}=A$ and $\fgt{\ult A}=A$. In particular, these translations are injective.

The translations $\upt{(\_)}$ and $\ult{(\_)}$ of \ipl\ formulas readily allow to define a translation of logical sequents of $\coLJQ$ into logical sequents of $\copil$ as follows: $\upt{(\Gamma\Longrightarrow A)}:=\ult{\Gamma}\vdash {\upt A}$ and
$\upt{(\Gamma\vdash[A])}:=\ult{\Gamma}\vdash[\upt A]$. Here $\ult{\Gamma}$ indicates application of $\ult{(\_)}$ to all formulas in $\Gamma$.  Notice that this is also a translation of logical sequents of $\LJQ$ into logical sequents of $\PIL$, as $\LJQ$ (resp. $\PIL$) and $\coLJQ$ (resp. $\copil$) have the same logical sequents.

\smallskip
\noindent
\textbf{Translation of (co-)proof terms.}
The translation of co-proof terms of $\coLJQ$ into $\copil$
maps an $\coLJQ$-co-proof term $t$ (of sort $t$)
to $\upt t$, a co-proof term of $\copil$ of sort $e$,
with the help of an auxiliary translation $\urntv v$ (for  $v$ a $\coLJQ$-value), producing a co-proof term in $\copil$ of sort $v$.
The definition of these two maps is simultaneous, by corecursion on $\coLJQ$ 
co-proof terms,  and is  also given in \cref{fig:coLJQ-positive-translation}. 
For a lighter notation, we use macro facilities (denoting co-proof terms of $\copil$) in \cref{fig:coLJQ-macros}. Note that, for every constructor with priority 2 in the source, a constructor with priority 2 appears in the target (which is immediately seen once the macro expansions of \cref{fig:coLJQ-macros} are applied in the target),
 consequently an infinite co-proof term of $\coLJQ$ is translated into an infinite co-proof term of $\copil$.
Note also that in the clauses defining the translation of $\coLJQ$-co-proof terms,
all corecursive calls are guarded by $\copil$-constructors, and are thus legitimate.  Additionally, observe that an obvious induction shows the translation to map $\LJQ$-terms to $\PIL$-stable expressions, and $\LJQ$-values to $\PIL$-values. Therefore, the restriction of $\upt{(\_)}$ and $\urntv{(\_)}$ to proof-terms of $\LJQ$ defines a translation into $\PIL$, which can also be obtained directly, by taking the defining clauses of these translations as simultaneous recursive definitions over $\LJQ$-expressions.

\begin{figure}[tb]\caption{Macros in $\copil$ for translation of $\coLJQ$}\label{fig:coLJQ-macros}
    \[\begin{array}{rcl}
    \lbforqut{x^L}e & := & \lbforqunomacro{x^L}e\enspace,\\
    \pairforqut{e_1}{e_2} & := & \pairforqunomacro{e_1}{e_2}\enspace,\\[0.6ex]
    \elimimpforqt xv{y^L}eR & := & \elimimpforqnomacrot xv{y^L}eR\enspace,\\[0.6ex]
    \elimandforqt x{y^{L}}eiR& := & \elimandforqnomacrot x{y^{L}}eiR \enspace,\\[0.6ex]
   \elimorforqt x{y_1^{L_1}}{e_1}{y_2^{L_2}}{e_2}R & := & \elimorforqnomacrot x{y_1^{L_1}}{e_1}{y_2^{L_2}}{e_2}R\enspace,\\[0.6ex]
   \elimabsforqt xR& := & \elimabsforqnomacrot xR \enspace.

      \end{array}\]
  \end{figure}
\begin{figure}[tb]\caption{Translation $\upt{(\_)}$ of $\coLJQ$ (resp. $\LJQ$) into $\copil$ (resp. $\PIL$)
and auxiliary translations $\ult{(\_)}$ and $\urntv{(\_)}$}\label{fig:coLJQ-positive-translation}
$$
\begin{array}{c}
\begin{array}{rclcrcl}
\upt a& = &a^+&\quad&\ult a& = &a^+\\
\upt{(A\horseshoe B)} & = &\downshift(\Down\ult A \horseshoe \upshift {\upt B})&\quad&\ult{(A\horseshoe B)} & = &\upt A \horseshoe \upshift \Down{\ult B}\\
\upt{(A\wedge B)}& = &\downshift(\upshift{\upt A}  \wedge \upshift{\upt B})&&\ult{(A\wedge B)} & = &(\upshift \Down{\ult A}) \wedge (\upshift \Down{\ult B})\\
\upt{(A\vee B)}& = &\upt A \vee \upt B  &\quad&\ult{(A\vee B)} & = &\upshift(\Down{\ult A} \vee \Down{\ult B})\\ 
\upt{\falsity}& = &\falsity&&\ult\falsity& = &\upshift \falsity\\
\end{array}\\\\
\begin{array}{rclcrcl}
\upt{\crc{v}} & = & \ret{{\urntv{v}}}&\quad&\urntv x & = &x\\
\upt{(\impleftljqt xvyBtA)} &=&\elimimpforqt x{\urntv v}{y^{\ult B}}{\upt t}{\upt{A}}&\quad& \urntv{(\lb x^A.t)} &=& \lbforqut{x^{\ult A}}{\upt t}\\ 
\upt{(\conjleftljqt xyBtiA)} &=&\elimandforqt x{y^{\ult B}}{\upt t}i{\upt{A}}& &\urntv{\langle t_1,t_2\rangle} &=&  \pairforqut{\upt{t_1}}{\upt{t_2}}\\
\upt{(\disjleftljqt x{y_1}{A_1}{t_1}{y_2}{A_2}{t_2}A)} &=&\elimorforqt x{y_1^{\ult A_1}}{\upt{t_1}}{y_2^{\ult{A_2}}}{\upt{t_2}}{\upt{A}}&&\urntv{(\inj i A v)} &=& {\inj i {\upt A}{\urntv v}}\\
\upt{(\abortljqt xA)} &=&\elimabsforqt x{\upt{A}}\\
\end{array}
\end{array}
$$
\end{figure}

\begin{example}
  In continuation of \cref{ex:types}, we observe that $\Idt^{+p}=\upt{\Idt}$ and that its inhabitant of sort $e$ given there is nothing but $\upt{\crc{\lb x^a.\crc x}}$.
  We also observe $\Churcht^{+p}=\upt{\Churcht}$ and, setting $x^0y:=\crc y$ and $x^{k+1}y:= 
\impleftljqt xyza{x^kz}a$ (to be defined simultaneously for all variables $y$), it is easy to see that $\tilde e_k=\upt{(x^ky)}$ -- for the first choice in the definition of $\tilde e_{k+1}$.
Therefore, the $k$-th inhabitant of $\Churcht^{+p}$ given in \cref{ex:types} is $\upt{\crc{\lb x^{a\impl a}.\crc{\lb y^a.x^ky}}}$.
  For the second choice in that definition, one would have to replace
$\impleftljqt xyza{x^kz}a$ by $\impleftljqt xy{\_}{}{x^ky}a$
in the definition of $x^{k+1}y$ and would arrive at the same conclusion.
\end{example}
\begin{example}
 In continuation of \cref{ex:infinite-solutions}, 
consider the $\coLJQ$-co-proof term $\omegafour$ of sort $t$ given as the unique solution of $T=  \impleftljqt xy{\_}{}{T}a$. We can show validity of $x:a\impl a, y:a\Longrightarrow \omegafour:a$ (by showing coinductively that $x:a\impl a, y:a,\Gamma\Longrightarrow \omegafour:a$ is valid for any $\Gamma$ composed of $n\geq 0$ declarations $z_i:a$ for $i\leq n$).
Also recall the $\copil$-co-proof term $\omegatwo$ from  \cref{ex:infinite-solutions},   which  stands for the  solution of the fixed-point equation $T=\coretR{x}{y::\cothunk{\_.{T}}}{a^+}$.
An easy coinduction shows $\upt{\left(\omegafour\right)}=\omegatwo$.
  The type $\Churcht$ also has an $\omega$-th solution in $\coLJQ$ beyond the inhabitants considered above, namely: $\crc{\lambda x^{a\impl a}.\crc{\lambda y^a.\omegafour}}$.
  Its positive translation is the solution $T_4$ of $\Churcht^{+p}$ given in \cref{ex:infinite-solutions}.
\end{example}

If we had defined $\upt{(\cdot)}$ adding double shifts in the clauses for disjunction and absurdity (that is, $\upt{(A\vee B)}=\downshift\upshift(\upt{A}\vee\upt{B})$ and $\upt{\perp}=\downshift\upshift\perp$) then we would have had $\upt{A}=\Down{\ult{A}}$ and then we could have just given the recursive definition of $\ult{A}$, using $\upt{A}$ as a macro. Therefore, we could have done the same if we were just interested in translating negative formulas (which show no occurrence of $\vee$ or $\perp$). The separation of $\upt{(\cdot)}$ and $\ult{(\cdot)}$ thus brings some optimization, in the form of omission of some double shifts in specific places -- but not in others: notice how $\upt{(\cdot)}$ translates the antecedent of an implication with $\Down{\ult{(\cdot)}}$ instead of $\upt{(\cdot)}$, hence with potentially more double shifts. The following example discusses this question.

\begin{example}[On the double shifts in the positive translation]
  Two subsequent shifts have no effect on polarity, but influence proof search. Consider proving $\Gamma\vdash \downshift\upshift P$ versus $\Gamma\vdash P$ (notice $P$ can be $\upt{A}$). Every solution of the former sequent, if it starts by focusing on the succedent formula, has the form $\ret{\thunk{\ep{e}}}$, with $e$ a solution of the latter sequent. Here the influence of double shifts is minimal and their omission is welcome.\footnote{This contrasts with solutions of $\Gamma\Longrightarrow\upshift{\downshift N}$, which can be of the form $\ep{\ret{\thunk t}}$ with $t$ a solution of $\Gamma\Longrightarrow N$, but can also be of the form $\ep{\coretR{x}{s}{\downshift N}}$, truly profiting from the double shifts to open a new alternative.}
  But sometimes the influence is crucial. Consider the \ipl\ formula $A:=(a\vee a)\impl a$. Its translation $\upt A$ is $\downshift{(\downshift{\upshift{(a^+\vee a^+)}}\impl \upshift{a^+})}$, so it features a double shift in the antecedent of the implication: 
If, instead of $\upt A$, we consider $B:=\downshift{((a^+\vee a^+)\impl \upshift{a^+})}$ as alternative positive translation of $A$, we would be faced with the following:
  While there are infinitely many inhabitants of $\Longrightarrow A$ in $\LJQ$, there is only the inhabitant $\ret{\thunk{\lb([x_i^{a^+}.\dlv{\ep{\ret{x_i}}}]_{i})}}$ of $\vdash B$ in $\pil$ (both these inhabitation questions can be checked by elementary analysis\footnote{Here are some details. The solution of $\vdash B$ has the form $\ret{\thunk{\lb p}}$ with $\mid p:a^+\vee a^+\Longrightarrow\upshift a^+$. Hence 
  $
  p=[x_i^{a^+}.\dlv{\ep{e_i}}]_i
  $, with $x_i:a^+\vdash e_i:a^+$. Hence $e_i=\ret{x_i}$ is forced.}).
  This would be in contrast to 
a full embedding,  which we are heading for,  analogously to 
\cref{thm:props-embedding-LJT}
 for the negative translation (where such double shifts are less frequently observed), and confirmed below in \cref{thm:props-embedding-LJQ}.
  In particular, that result guarantees that $\vdash\upt A$ has infinitely many inhabitants in $\pil$ -- which, again, can be confirmed by an elementary analysis, for this particular example.\footnote{Any solution of $\vdash \upt{A}$ has the form $\ret{\thunk{\lb p}}$ with $\mid p:\downshift\upshift(a^+\vee a^+)\Longrightarrow\upshift a^+$.  Let $L=\upshift(a^+\vee a^+)$. Then
  \[
  p=y^L.\dlv{\ep{\coretR{y}{\cothunk{[x_i^{a^+}.e_i]_i}}{a^+}}}\enspace,
  \]
  with $y:L,x_i:a^+\vdash e_i:a^+$. Here $e_i=\ret{x_i}$ would stop the search, but this solution is not forced. We can, again and again, choose $e_i=\coretR{y}{\cothunk{\cdots}}{a^+}$.}
\end{example}

\smallskip
\noindent
\textbf{Properties of the positive translation.} 
Again, we
argue first about the properties of the coinductive translation
and obtain as a corollary the same properties of the inductive translation.
\begin{proposition}[Soundness]\label{prop:soundness-positive-translation}
The positive translation is sound in the sense that for any co-proof term $T$ of\/ $\coLJQ$ and sequent $\sigma$ of\/ $\coLJQ$ such that $\sigma(T)$ is valid in $\coLJQ$, $\upt{\sigma}(T^\bullet)$ is valid in $\copil$, for the appropriate $\bullet\in\{\uptsymb,\urntvsymb\}$,
that is:
\begin{enumerate}
\item If\/ $\Gamma\Longrightarrow t:A$ is valid in $\coLJQ$ then $\ult\Gamma\vdash\upt t:\upt A$ is valid in $\copil$.
\item If\/ $\Gamma\vdash [v:A]$ is valid in $\coLJQ$ then $\ult\Gamma\vdash[\urntv v:\upt A]$ is valid in $\copil$.
\end{enumerate}
\end{proposition}
\begin{proof}
The two items are proved simultaneously by coinduction on the typing relation of $\copil$.
We illustrate the case where $v=\lb x^B.t_0$. So, our assumption implies $A= B\horseshoe C$, and the validity of $\Gamma,x:B\Longrightarrow t_0:C$  in $\coLJQ$.  Also, $\urntv v=\thunk{\lb(x^{\ult B}.\dlv{\ep{\upt t_0}})}$ and $\upt A=\downshift(\Down \ult B\horseshoe\upshift\upt C)$. Therefore, we may conclude the validity of $\ult\Gamma\vdash[\urntv v:\upt A]$, applying four inferences of $\copil$ from  $\ult\Gamma, x:{\ult B} \vdash {\upt t_0}:\upt C$.  But, since $\ult\Gamma, x:{\ult B}=\ult{(\Gamma, x:{ B})}$, the validity of the latter follows from the coinductive hypothesis (guarded by the mentioned inferences of $\copil$), provided $\Gamma,x:B\Longrightarrow t_0:C$ is valid in $\coLJQ$, which we already observed to have.
\end{proof}

As for $\coLJT$, 
to strenghen the previous property and obtain other properties of the positive translation,
 we need to analyze its image. Consider the following subclasses of formulas of \pipl:
\[
\begin{array}{rrcl}
(\text{$\uptsymb$-formulas})&P&::=&{a^+}\mid \downshift(\Down L \horseshoe \upshift P)\mid \downshift(\upshift P_1  \wedge \upshift P_2)\mid P_1\vee P_2\mid  \falsity\\
(\text{$\ultsymb$-formulas})&L&::=&{a^+}\mid P\horseshoe \upshift \Down L\mid (\upshift \Down{L_1})\wedge (\upshift \Down{L_2})\mid\upshift(\Down{L_1}\vee \Down{L_2})\mid\upshift \falsity\\
\end{array}
\]

Note that the subclasses of  $\uptsymb$-formulas and  $\ultsymb$-formulas are, respectively, subclasses of the classes of positive and left formulas of \pipl.
Also, the names of these classes of formulas make good sense in that, for any formula $A$, an immediate induction shows that: $\upt A$ is a $\uptsymb$-formula, and $\ult A$ is an $\ultsymb$-formula. We already knew that the two translations of \ipl\ formulas are right inverses of the forgetful map. But if we restrict their codomain to these subclasses, an easy induction on formulas also shows that those restrictions are left inverses to the respective forgetful maps, obtained by restricting the domain accordingly, in symbols, this is just:
for every $\uptsymb$-formula $P$, $\upt{\fgt P}=P$ and for every $\ultsymb$-formula $L$, $\ult{\fgt L}=L$. Hence
the subclasses  of $\uptsymb$-formulas, and of $\ultsymb$-formulas are all in bijection with all of the \ipl\ formulas.

It will also be useful to characterize the image of the translation at the level of co-proof terms.
This will be done through the unary predicates $\pexpsymb$ and $\vvalsymb$ on  co-proof terms of $\copil$ (of sorts $t$ and $v$, respectively), whose  simultaneous coinductive definition is given in \cref{fig:imageLJQco} of \cref{app:appendix-positive-embedding}.  We call $\uptsymb$-expressions (resp. $\urntvsymb$-values) those co-proof terms $e$ (resp.\ $v$) with $\pexp e$ (resp.\ $\vval v$).
This naming makes sense in that  $\uptsymb$-expressions (respectively, $\urntvsymb$-values) are  stable expressions (respectively, values)  of $\copil$, and an immediate coinduction on $\pexpsymb$ (and $\vvalsymb$)  shows that:
 for any $t\in\coLJQ$,  $\upt{t}$ is a $\uptsymb$-expression and, for any $v\in\coLJQ$, $\urntv{v}$ is a $\urntvsymb$-value.

There is an obvious forgetful map $\fgtQ{\_}$ from the identified subclasses of co-proof terms of $\copil$ to  co-proof terms of $\coLJQ$, given corecursively in \cref{fig:coLJQ-forgetful-map} of \cref{app:appendix-positive-embedding}.  It maps $\uptsymb$-expressions (resp. $\urntvsymb$-values) to expressions of $\coLJQ$ of sort $t$ (resp. $v$). 
As for formulas, the forgetful map from the identified subclasses of $\copil$-expressions gives a way to invert the translation of $\coLJQ$ co-proof terms, namely, an easy coinduction on  bisimilarity for $\coLJQ$ co-proof terms gives:  $\fgtQ{\upt e}=e$, and $\fgtQ{\urntv v}=v$. In particular, the translations $\upt{(\_)}$ and $\urntv{(\_)}$ of co-proof terms of $\coLJQ$  are injective. 
Of course, the restrictions of these translations to proof-terms of $\LJQ$ are also injective and a right inverse to the restriction of the forgetful map to $\PIL$-proof terms.
Additionally, this forgetful map preserves validity
 (see \cref{lem:soundness-forgetful-map-to-coLJQ} in \cref{app:appendix-positive-embedding}). 
\footnote{As in the case of the negative translation, notice that preservation of validity by the forgetful map into $\coLJQ$ (\cref{lem:soundness-forgetful-map-to-coLJQ}), and the fact that the positive translation is a right inverse of this forgetful map, allow the strengthening of \cref{prop:soundness-positive-translation} into an equivalence.} 

Like for the negative translation,  we may say the positive translation is an embedding. In fact, for each  logical sequent $\sigma$ of $\coLJQ$,  the $\sigma$-restriction of $\upt{(\_)}$ (which by \cref{prop:soundness-positive-translation} maps solutions of $\sigma$ to solutions of $\upt{\sigma}$) is injective, since $\upt{(\_)}$  itself is injective (as observed above). 

\begin{proposition}[Full embedding]\label{prop:full-faithfulness-positive-translation}
For $\Gamma, A$ in $\coLJQ$ and for $e,v\in\copil$:
\begin{enumerate}
\item
  If\/ $\ult\Gamma\vdash e:\upt A$ is valid in $\copil$, then (i) $\pexp e$ and (ii) $\upt{(\fgtQ e)}=e$.
\item 
  If\/ $\ult\Gamma\vdash[v:\upt A]$ is valid in $\copil$, then, (i) $\vval v$ and (ii) $\urntv{(\fgtQ v)}=v$.
\end{enumerate}
Hence the positive translation is \emph{full}, in the sense that each $\sigma$-restriction is surjective.
\end{proposition}
\begin{proof} 
First one proves simultaneously parts (i) of each item, by coinduction on $\pexpsymb$ (given simultaneously with $\vvalsymb$). Then, one proves simultaneously parts (ii) of each item, by coinduction on bisimilarity for co-proof terms of $\copil$.
In each of the items we do a case analysis on type $A$.

Regarding surjectivity, the only question is whether $\fgtQ e$ (for instance) is in the domain of the $\sigma$-restriction. This is guaranteed by preservation of validity by the forgetful map into $\coLJQ$ (\cref{lem:soundness-forgetful-map-to-coLJQ}) and the fact that, at the level of formulas, the positive translation is a right inverse to the forgetful map.
\end{proof}

\begin{corollary} [Reduction of decision problems]
\label{cor:reduction-problems-LJQ}\quad
\begin{enumerate}
\item $\jay$ is solvable in $\coLJQ$ iff $\jay^{\uptsymb}$ is solvable in $\copil$. 
\item  $\jay$ has an infinite solution in $\coLJQ$ iff  $\jay^{\uptsymb}$ has an infinite solution in $\copil$. 
\end{enumerate}
\end{corollary}
\begin{proof}
Both items are proved in lock-step with the respective proofs  of \cref{cor:reduction-problems}. 

(1) The implication from left to right follows from \cref{prop:soundness-positive-translation}. The converse follows from \cref{prop:full-faithfulness-positive-translation}, preservation of validity by the forgetful map into $\coLJQ$, and the fact that, at the level of formulas, the positive translation is right inverse to the forgetful map.
 
(2) The proof of (1) only needs to be refined by appealing to preservation of infinity of co-proof terms by $\upt{(\_)}$ and by the forgetful map into $\coLJQ$ (the latter shown in \cref{app:appendix-positive-embedding}).
\end{proof}
Next we extract results for the inductive translation.
\begin{theorem}[Properties of the translation of $\LJQ$ into $\LJP$]\label{thm:props-embedding-LJQ}
The positive translation is a full embedding. 
As a consequence, for $\sigma$ a logical sequent of\/ $\LJQ$: 
(i) inhabitation of $\sigma$ in $\LJQ$  is equivalent to inhabitation of $\upt{\sigma}$ in $\LJP$, and 
(ii) the number of inhabitants of $\sigma$ in $\LJQ$ is finite iff  the number of inhabitants of $\upt{\sigma}$ in $\LJP$ is finite.
\end{theorem}
\begin{proof}
Propositions \ref{prop:soundness-positive-translation} and 
\ref{prop:full-faithfulness-positive-translation} hold with all upper ``co'' indices removed. We argue about  \cref{prop:soundness-positive-translation}, and detail for item 1. 
The assumption and  \cref{lem:conservativeness-typing-LJQ} give validity of  $\Gamma\Longrightarrow t:A$  in $\coLJQ$.  Hence, by \cref{prop:soundness-positive-translation},  
$\ult\Gamma\vdash\upt t:\upt A$ is valid in $\copil$.  
Therefore, given that $\upt t$ is a stable expression of $\PIL$,  by \cref{lem:conservativeness-typing}, follows validity of $\ult\Gamma\vdash \upt t:\upt A$ in $\PIL$.
\end{proof}

\smallskip
\noindent
\textbf{Applications to meta-theory.}
As anticipated,
the positive translation offers a 
simple proof of the  disjunction property of \ipl\ for the full class of
Rasiowa-Harrop formulas:

\begin{proposition}[Disjunction Property under assumptions for \ipl]
\label{prop:disj_prop_ipl_LJQ}
If\/ $\Gamma$ only contains Rasiowa-Harrop formulas, and $\Gamma\Rightarrow A_1\vee A_2$ is inhabited in $\LJQ$, then  one of the logical sequents $\Gamma\Rightarrow A_1$ and $\Gamma\Rightarrow A_2$ is inhabited.
\end{proposition}
\begin{proof}
The argument is analogous to the one proving \cref{prop:disj_prop_LJT}.
From soundness of the positive translation (\cref{thm:props-embedding-LJQ}) follows inhabitation of $\ult\Gamma\vdash \upt{A_1}\vee \upt{A_2}$ in $\PIL$. 
An easy induction shows that
if  ${\mathcal R}$ is  an intuitionistic Rasiowa-Harrop formula,
 $\ult{\mathcal R}$ is  a polarized Rasiowa-Harrop formula. 
Thus, the assumption implies $\ult\Gamma$ only contains
polarized Rasiowa-Harrop formulas,
 and so \cref{thm:polarized-disjunction-property}
 gives inhabitation of $\ult\Gamma\vdash  \upt{A_i}$ in $\PIL$,
for some $i\in\{1,2\}$. 
Therefore, \cref{thm:props-embedding-LJQ} gives inhabitation of $\Gamma\Rightarrow {A_i}$ in $\LJQ$, for that $i$.  
\end{proof}

With the positive translation,
one also immediately obtains an infinity-or-nothing property (even more simply than in the case of the negative translation):
\begin{proposition}[Infinity-or-nothing property of $\LJQ$ under a disjunctive hypothesis]
\label{prop:inf-nothing-LJQ}
If\/ $x:A_1\vee A_2\in\Gamma$, then, for any $A$,  $\Gamma\Rightarrow A$ has the infinity-or-nothing property in $\LJQ$.
\end{proposition}
\begin{proof}
By \cref{thm:props-embedding-LJQ}, it suffices to argue about infinity-or-nothing for $\sigma=(\ult\Gamma\vdash\upt A)$. Given that $\ult{(A_1\vee A_2)}=\upshift(\Down\ult{A_1}\vee \Down\ult{A_2})$ and $\Down\ult{A_1}\vee \Down\ult{A_2}$ is a not fully absurd positive formula, and that $\upt A$ is a positive formula, \cref{thm:infinity-or-nothing} applies, giving infinity-or-nothing for $\sigma$. 
\end{proof}

\smallskip
\noindent
\textbf{Application to decision problems.}
We are now ready to extract from our decision algorithms 
for $\LJP$ and $\copil$ decision algorithms to decide 
\emph{existence of inhabitants} and \emph{finiteness of the number of inhabitants} in $\LJQ$  and \emph{existence of solutions} and \emph{existence of infinite solutions} in $\coLJQ$.  
These algorithms are exactly as the corresponding decision algorithms for $\LJT$ and $\coLJT$, the sole difference being that the first step is now the calculation of the positive translation of the logical sequent $\jay$ at hand. 

\begin{theorem}[Decision algorithms for $\LJQ$ and $\coLJQ$]
\label{thm:decision-algorithms-LJQ}
Given $\jay$ in $\LJQ$ (equivalently, $\coLJQ$):
\begin{enumerate}
\item \textbf{Existence of inhabitants:} $\jay$ has an inhabitant in $\LJQ$ iff\/
 $\nbinfp{\falsepred}{\finrepempty{\upt\jay}}$;
\item \textbf{Finiteness of the number of inhabitants:} $\jay$ has finitely many inhabitants in $\LJQ$ iff\/
 $\FFp{\falsepred}{\finrepempty{\upt\jay}}$;
\item \textbf{Existence of solutions:} $\jay$ has a solution in $\coLJQ$ iff\/
 $\nstuckp \falsepred{\finrepempty{\upt\jay}}$; 
\item \textbf{Existence of infinite solutions:} $\jay$ has an infinite solution in $\coLJQ$ iff\/
 $\NAFp \falsepred{\finrepempty{\upt\jay}}$. 
\end{enumerate}
In each of the four items, the right-hand-side of the equivalence is the obtained decision algorithm.
\end{theorem}
\begin{proof} 
Again, the proof is
simple, and it is analogous to the proof of \cref{thm:decision-algorithms-LJT}. 
Part 1: $\jay$ is inhabited in $\LJQ$ iff $\upt\jay$ is inhabited in $\PIL$ (\cref{thm:props-embedding-LJQ}); 
iff $\nbinfp{\falsepred}{\finrepempty{\upt\jay}}$ (part 1 of \cref{thm:decide-existence}). 
The other three parts follow analogously,  with a first equivalence resorting to   \cref{thm:props-embedding-LJQ} or to one of the parts of \cref{cor:reduction-problems-LJQ}, and
with a second equivalence resorting  to  part 2 of \cref{thm:decide-existence} or to one of the parts of \cref{thm:decide-finiteness}.
\end{proof}


%% file: final.tex
\section{Final remarks}\label{sec:final}

\smallskip
\noindent
\textbf{On the contribution.}
We have shown that proof search in a focused sequent calculus for polarized intuitionistic logic \cite{LiangMillerTCS09,SimmonsTOCL2014,JESENTCS17} can be brought into the scope of
coinductive proof search.
The vehicle for the study conducted here is 
$\PIL$, a slight variation of the system proposed by the first author \cite{JESENTCS17}. Coinductive proof search coped well with the new case study, despite the wealth of connectives and inherent sophistication of the proof system. Together with the basic result about the equivalence of the coinductive and finitary representations of solution spaces, we: (1) showed how to develop meta-theory (disjunction and infinity-or-nothing properties) with the coinductive syntax of $\copils$; (2) obtained decidability of some predicates 
concerning inhabitants or solutions through recursive predicates defined over the finitary syntax.

Next we showed that $\PIL$ is a platform for the study other logics and proof systems.
We illustrated this view with the focused proof systems  $\LJT$ and $\LJQ$,
which allow proofs with very distinct flavors, namely proofs by backward chaining and forward chaining, respectively. This was achieved by means of the faithful interpretations of each of the two proof systems into $\LJP$, with a negative and a positive translation, respectively. The translations allow the inheritance of some meta-theory from $\LJP$, and the reduction of decision problems about $\LJT$ or $\LJQ$ to decision problems about $\LJP$, including the finiteness problem. This is possible because our interpretations are defined as translations of proof terms (see below for the co-proof terms), and are proved to be \emph{full embeddings}, establishing a bijection between the proofs of a given sequent in $\LJT$ or $\LJQ$ and the translated sequent in $\LJP$. Mere faithfulness, in the usual sense used for logical interpretations, would only allow the reduction of the emptiness problem.

Finally, true to our credo, throughout our paper we treated solutions as first-class citizens. Hence, we also solved decision problems about solutions in $\copil$: the solvability problem (does a sequent have a solution) and the termination of proof search problem (does a sequent have an infinite solution). Then, we extracted similar decidability results for $\coLJT$ and $\coLJQ$. For this, the negative and positive translations were, in fact, studied in their more general form, as translations of coinductive proof terms: the maps that achieve the reduction of inhabitation problems emerge as the restrictions of those general translations to (finite) proof terms.

In all, we extended the scope of coinductive proof search, originally developed for $\LJT$ and implicational logic, in the ambitious mode mentioned in the introduction: to a much more expressive logic, $\LJP$, and in a way that encompasses our previous results about $\LJT$ and produces new results about $\LJQ$, because $\LJP$ fully embeds these systems for intuitionistic logic.

\smallskip
\noindent
\textbf{Discussion.}  In the coinductive proof search approach, we privileged a conceptual approach, where the representation of the search space is separated from its analysis. This separation of concerns is reflected in the architecture of our decision procedures, given as the composition of
$\finrepsymb$ with a recursive predicate adequate for the specific problem at hand. This organization is modular, with $\finrepempty{\jay}$, $\finrepempty{\unt{\jay}}$ or $\finrepempty{\upt{\jay}}$ being reused, as we move our attention to a different decision problem; but this organization is not optimized, because knowing the particular predicate we want to compose
$\finrepsymb$ with, in general, should suggest simplifications. Therefore, complexity or optimization issues are not (yet) a concern of coinductive proof search.

One example of the infinity-or-nothing property is the ``monatomic theorem'' \cite{HindleyBook1997} about simple types, telling that in most cases the existence of an inhabitant entails an infinity of inhabitants -- a theorem we generalized in \cite{JESRMLP-APAL2021}. Formulas or sequents with that property have a finite number of inhabitants if and only if they are unprovable and, thus, have a finiteness problem which coincides with the emptiness problem. We have seen, both in $\LJT$ and $\LJQ$, that, as soon as a disjunction is present in the context of a sequent, the sequent has the infinity-or-nothing property. For $\LJQ$, it is obvious that the same is true for a conjunction -- because the left rule repeats the main formula in the premiss (for implications, the situation is more interesting: although the main formula is repeated in the right premiss, the application of the rule has a price, namely the first premiss). Such formulation of left rules is found in \cite{DyckhoffLengrand2007}, where the calculus is used as a typing system for some sort of call-by-value $\lb$-terms. Sharper left rules without the repetition of the main formula are preferable for proof search, but maybe not so for typing of $\lb$-terms. This points to a tension between the concepts of proof and inhabitant.

We now discuss a vague resemblance between our embeddings, of intuitionistic logic into polarized logic, and other embeddings of the same source logic into linear logic \cite{Girard87} or modal logic S4 \cite{TroelstraS00}. Our negative translation of implication is given by the scheme $(\downarrow - )\supset -$ and this reminds of Girard's main translation of implication into linear logic.  Recall Troelstra-Schwichtenberg's textbook \cite{TroelstraS00} also calls Girard's translation a modal translation given by the scheme $(\Box -) \supset -$. As already said in \cref{subsec:LJQ-LJQco}, our positive translation, if restricted to negative formulas, can be simplified: $\upt{(-)}$ is equal to $\Downarrow \ult{(-)}$ and can be considered an abbreviation. In this restricted setting, considering the translation of implication, in each of the recursive definitions of the two translations, we find: in $\ult{(-)}$ the scheme $(\Downarrow -)\supset (\uparrow\Downarrow -)$, while in $\upt{(-)}$ the scheme $\downarrow( - \supset (\uparrow -))$. If we disregard $\uparrow$ in these two schemes: the latter reminds the scheme $\Box(- \supset -)$ of the second modal translation in \cite{TroelstraS00}, and also of the second, ``boring'' embedding in \cite{Girard87}; while the former reminds the scheme $(\Box -) \supset (\Box -)$ of G\"odel's original interpretation into S4.

\smallskip
\noindent
\textbf{On related work.} There is a diverse literature on inhabitation problems for the simply-typed $\lb$-calculus \cite{AlvesBroda2015,BourreauSalvati2011,BrodaDamas2005,HindleyBook1997,TakahashiAkamaHirokawa1996}. For detailed comparisons of coinductive proof search with different methodologies in this literature, we refer to our previous papers \cite{EspiritoSantoMatthesPintoInhabitation,JESRMLP-APAL2021}. In our previous studies about the implicational fragment, we concentrated on $\LJT$, whose notion of proof term corresponds easily with the ordinary $\lb$-terms. As we move to a richer set of connectives and/or switch to $\LJQ$, the notion of proof term becomes richer and assumes different forms which lead to decision problems relative to the notion of inhabitant. Even if the emptiness of a formula relative to $\LJT$ inhabitants is equivalent to emptiness by $\LJQ$ inhabitants (because the two proof systems are equivalent w.~r.~t.~provability), the same is not true about finiteness (consider the sequent $x:a\impl b,y:a\vdash b$, for two distinct atoms $a$, $b$; in $\LJT$ there is one inhabitant, in $\LJQ$ there are infinitely many). The finiteness problem for $\lb$-terms or $\LJT$ proof terms was solved before \cite{TakahashiAkamaHirokawa1996,WellsYakobowski04}, but the one for $\LJQ$ we solved here seems to be new.

The work of \cite{WellsYakobowski04} is the only one we are aware of that deals with
a question of type finiteness for \emph{full} \ipl\ 
(but $\bot$ is not included). That work  considers a cut-free $LJT$-presentation of \ipl\ close to ours, but allowing more proofs,
due to unrestricted RHS in its \emph{contraction} rule (recall our version of $LJT$ imposes an atom or disjunction on the RHS when a formula from the context is selected to the
``focus''). The work \cite{WellsYakobowski04} uses graphs to represent the search space, and such graphs are guaranteed to be finite only in the case where contexts are  sets, in other words, when the \emph{total discharge convention} is assumed. 
The decision of type finiteness is then based on traversal of this finite graph structure and exhaustive checking for the absence of ``cyclic proof structures''.
It should be noted that decision of type finiteness in \cite{WellsYakobowski04} is part of more general algorithms that count the number of inhabitants of a type.
In our case, counting of inhabitants is done by a function defined by structural recursion on finitary forests. 
This worked fine for the implicational fragment of $LJT$ \cite{EspiritoSantoMatthesPintoInhabitation}, and we anticipate no major obstacles in extending the idea to full $LJT$.

The prominence given to the concept of solution, and solving decision problems about solutions is quite unique of coinductive proof search. In the context of $\LJT$ restricted to implication, 
we already obtained in \cite{JESRMLP-FI2019} 
decidability of such decision problems, 
including the problem of termination of proof search.
We now have extended substantially these results to polarized logic $\PIL$, 
and extracted similar results to the systems $\LJT$ and $\LJQ$ of intuitionistic logic, through the analysis of coinductive extensions of the negative and positive translations into $\LJP$.

Similarly to $\LJP$, frameworks like $LJF$ \cite{LiangMillerTCS09}, call-by-push-value \cite{Levy06}, or the $\lambda !$-calculus \cite{BucciarelliKRV23} are platforms for the interpretation of other logical systems. The latter two have been used mostly for the embedding of call-by-name or call-by-value $\lb$-calculi. $\LJP$ can be equipped with cuts and cut-elimination rules, as in \cite{JESENTCS17}, and presumably one can interpret in $\LJP$ the cut-elimination processes of $\LJT$ \cite{HerbelinCSL94} and $\LJQ$ \cite{DyckhoffLengrand2007}, but this task is out of the scope of this paper. 

Much closer to our concerns here is the work on $LJF$ \cite{LiangMillerTCS09}, where we find translations of $\LJT$ and $\LJQ$ to the focused framework. Interestingly,
the embedding of $\LJQ$ into $LJF$ also shows a left mode and a right mode (${(\_)^l}$ and ${(\_)}^r$), similar to what happens with our positive embedding into $\LJP$ (with $\ult{(\_)}$ and $\upt{(\_)}$). But, at the same time, the embedding into $LJF$ shows some differences, already at the translation of formulas. This is expected, since the classes of negative/positive formulas of $LJF$ are less restrictive than those of $\LJP$. In the latter, there are operators for polarity shift, while in $LJF$ such changes occur silently, \emph{e.\,g.,~}two negative formulas linked by a positive connective form a positive formula. In addition, $LJF$ has a positive conjunction, while $LJP$ doesn't.
Polarity shifts may sometimes feel like a nuisance, but, on the other hand, they provide naturally the ``delay'' operators which in $LJF$ require some encoding. 
Our study of the translations of $\LJT$ and $\LJQ$ goes farther, showing them to be full embeddings and effecting reductions of decision problems, and this is possible very much due to our focus on the proof terms level (as opposed to mere provability). 

\smallskip
\noindent
\textbf{On ongoing and future work.} It is high time to try coinductive proof search with fragments of first-order logic. Nevertheless our current work is still concerned with propositional logic, but with tighter proof systems having higher avoidance of the infinity-or-nothing phenomenon and entailing sharper notions of finite formula. In the optimal situation, such systems only allow \emph{canonical} inhabitants \cite{SchererRemy15} and give the correct characterization of formulas with a unique inhabitant. 
Our past experience with uniqueness of inhabitants and ``coherence theorems'' \cite{EspiritoSantoMatthesPintoInhabitation} was restricted to implication/simple types; we can build on the present paper to address these questions for richer languages.


%% file: app-analysis-PIPL.tex
\subsection{On well-definedness of infinitary representation in Section~\ref{sec:proof-search-PIPL}}
\label{sec:appendix-definednessofS}

This section is dedicated to the proof of Lemma~\ref{lem:welldefinedness-S}.

It remains to check the parity condition. As mentioned in the main text,
this comes from the observation that all the ``intermediary''
corecursive calls to $\solfunction{\jay'}$ in the calculation of $\solfunction{\jay}$ lower the ``weight'' of the logical sequent.
We will now define that weight and give the precice statement about which corecursive calls lower the weight (instead of making a formal definition of ``intermediary'').

\begin{definition}[weight]
  Weight of a formula: $w(\falsity,a^+):=0$, $w(a^-):=1$, and for composite formulas, add the weights of the components and add the following for the extra symbols: $w(\downarrow,\wedge):=0$, $w(\vee):=1$, $w(\uparrow):=2$, $w(\impl):=3$. Then $w(N)\geq1$ and $w(P)\geq0$.

  Weight of context $\Gamma$: the sum of the weights of all the formulas associated with the variables.

  Weight of logical sequent: $w(\Gamma\vdash A):=w(\Gamma)+w(A)$, $w(\Gamma\Longrightarrow N):=w(\Gamma)+w(N)-1\geq0$. $w(\Gamma\vdash[P]):=w(\Gamma)+w(P)$, $w(\Gamma|P\Longrightarrow A):=w(\Gamma)+w(P)+w(A)+1$, $w(\Gamma[N]\vdash R):=w(\Gamma)+w(N)+w(R)$. Then for all $\jay$, $w(\jay)\geq0$.
\end{definition}

For the analysis of the recursive call structure of $\solletter$, we consider two minor modifications of \cref{def:sol}: Firstly, in the case of an argument $\Gamma\vdash P$, the first summand is not $\ret{\solfunction{\Gamma\vdash[P]}}$, but $\retsymb$ with the following argument: depending on the form of $P$ (as classified by the first four defining rules of $\solletter$), it is the definiens of the respective case of $\solfunction{\Gamma\vdash[P]}$.
In other words, we do not consider a different outcome of $\solfunction\jay$ for any logical sequent $\jay$, but we ``inline'' the four defining rules for $\solfunction{\Gamma\vdash[P]}$, which only amounts to ``short-circuit'' the recursive call structure in that the call under $\retsymb$ for $\solfunction{\Gamma\vdash\downshift N}$ is to $\solfunction{\Gamma\Longrightarrow N}$ and for $\solfunction{\Gamma\vdash P_1\vee P_2}$, it is to $\solfunction{\Gamma\vdash [P_i]}$. The inlining itself is not done recursively, hence the four defining rules for $\solfunction{\Gamma\vdash[P]}$ have to stay in place.
Secondly, we do a modification in the same spirit as follows: In the fifth defining rule of $\solletter$, we replace $\solfunction{\Gamma\vdash a^-}$ by its definiens (of the penultimate defining rule of $\solletter$).

\begin{lemma}\label{lem:weight-S}
  Every direct corecursive call in the definition of $\solfunction\jay$ to some $\solfunction{\jay'}$ which is not an argument to $\coretsymb$ lowers the \emph{weight} of the logical sequent.
\end{lemma}
\begin{proof}
We distinguish two cases: (1) neither $\jay$ nor $\jay'$ are R-stable sequents and (2) otherwise.
We have to show the following inequalities for (1):

\noindent Concerning the rules for stable sequents, there is only the rule introducing $\dlvsymb$, and the inequality is $$w(\Gamma \vdash  C) > w(\Gamma \Longrightarrow  C)\enspace:$$
this is why $\cdot\Longrightarrow \cdot$ has to weigh less.

\noindent Concerning the rules for co-terms (of sort $p$), we distinguish cases for $L$ and even show $$w(\Gamma | a^+ \Longrightarrow  A) > w(\Gamma, x:a^+ \vdash  A)\enspace.$$ This is why $\cdot|\cdot\Longrightarrow \cdot$ has to weigh more (and variable names must not enter the weight of contexts $\Gamma$) -- we only need to prove this when $A$ is a $C$, but it holds for all $A$, and we also even show $$w(\Gamma | \downshift  N \Longrightarrow  A) > w(\Gamma, x:N \vdash  A)\enspace:$$ $w(\downarrow)=0$ suffices -- we only need to prove this when $A$ is a $C$, but it holds for all $A$.

\noindent The final rule for sort $p$ is captured by
$$w(\Gamma | P_1 \vee  P_2 \Longrightarrow  A) > w(\Gamma | P_i \Longrightarrow A)\enspace:$$ trivial since $w(\vee)>0$.

\noindent The first twelve defining rules of $\solletter$ are dealt with a bit more briefly:

$w(\Gamma \vdash [ \downshift  N ]) > w(\Gamma \Longrightarrow  N)$: $w(\downarrow)=0$ suffices.

$w(\Gamma \vdash [ P_1 \vee  P_2 ]) > w(\Gamma \vdash [ P_i] )$: trivial since $w(\vee)>0$.

$w(\Gamma \Longrightarrow  P \impl N) > w(\Gamma | P \Longrightarrow  N)$: since both logical sequent weights are unfavourably modified, the weight of $\impl$ has to be so high.

$w(\Gamma \Longrightarrow  N_1 \wedge  N_2) > w(\Gamma \Longrightarrow  N_i)$: since $w(N_{3-i})\geq1$.

$w(\Gamma[P \impl N] \vdash  R) > w(\Gamma \vdash  [P])$ and $> w(\Gamma[N] \vdash  R)$: both are trivial since $w(\impl)>0$.

$w(\Gamma [\upshift  P] \vdash  R) > w(\Gamma | P \Longrightarrow  R)$: this is why $\uparrow$ has to weigh more (given that $\cdot|\cdot\Longrightarrow \cdot$ weighs more).

$w(\Gamma[N_1 \wedge  N_2] \vdash  R) > w(\Gamma[N_i] \vdash  R)$: since $w(N_{3-i})\geq1$.

\noindent For (2), we consider the following inequalities:

$w(\Gamma \Longrightarrow  a^-) > w(\Gamma \vdash  a^-)$ is not to be shown (and is wrong) since
we applied our second modification on the reading of $\solletter$.

$w(\Gamma \Longrightarrow  \upshift  P) > w(\Gamma \vdash  P)$: this works since $\uparrow$ weighs more (given that $\cdot\Longrightarrow \cdot$ weighs less).

Due to the first modification of the reading of $\solletter$, we show $w(\Gamma\vdash\downshift N) > w(\Gamma\Longrightarrow N)$ and $w(\Gamma\vdash P_1\vee P_2) > w(\Gamma\vdash [P_i])$: value-wise, they are the same inequalities as the fifth and sixth for~(1).
\end{proof}
It is clear that this lemma guarantees the parity condition for all $\solfunction\jay$.

\subsection{On the characterization of predicates on forests in Section~\ref{sec:proof-search-PIPL}}\label{sec:appendix-exfinnofin}

Let $D=(\lopvar,Q,\overline Q,\lopvaralt)$ be dual pair data.
We give a sequence of approximations from above to the coinductive predicate $\cpredsymb:=\cpredd D$ whose intersection characterizes the predicate.
The index $n$ is meant to indicate to which observation depth of $T$ we can guarantee that $\cpred D T$ holds.
For this purpose, we do not take into account the summation operation as giving depth.
We present the notion as a simultaneous inductive definition.
\[
  \begin{array}{c}
    \infer[]{\cpredi 0T}{}\qquad
    \infer{\cpredi{n+1}{f(T_i)_i}}{\overline\lopvar_i\cpredi n{T_i}\lopvarbinop\lopvar_i\overline Q(T_i)}\qquad\infer[]{\cpredi{n+1}{\sum_iT_i}}{\overline\lopvaralt_i\cpredi{n+1}{T_i}}
  \end{array}
\]
As for $\cpredsymb$, we can logically simplify the premiss for the second inference rule for nullary and unary $f$. In fact, $\overline\lopvar_i\cpredi n{T_i}\lopvarbinop\lopvar_i\overline Q(T_i)$ then shrinks down to $\lopvar=\bigvee$ and $\cpredi n{T_1}\lopvarbinop\overline Q(T_1)$, respectively.

A guarantee up to observation depth $0$ does not mean that the root symbol is suitable but the assertion is just void.
Going through a function symbol requires extra depth.
The child has to be fine up to a depth that is one less.
As announced, the summation operation does not provide depth, which is why this simultaneous inductive definition cannot be seen as a definition of $\cpredisymb n$ by recursion over the index $n$.
However, since we have excluded infinite stacking of sums through our parity condition on forests, there is a hidden recursive definition over $n$ of this family of predicates:
instead of applying the third inference rule, one has to unfold the finite and finitely stacked sums whose ultimate members will be dealt with by the second rule that then is on a smaller index, and the constituent conditions are then gathered uniformly through the connective $\overline\lopvaralt$.
In particular, if the three inference rules are viewed just as (properly quantified) fixed-point equations, the previous reasoning shows that there is a unique family $(\cpredisymb n)_{n\geq0}$ satisfying these equations.

By induction on the inductive definition, one can show that $\cpredisymb n$ is antitone in $n$, i.\,e., if $\cpredi {n+1}T$ then $\cpredi nT$.

In continuation of \cref{ex:dualpairsmembership}, we define $\nofinisymb n:=\cpredisymb n$ for $D=(\bigwedge,\bigvee)$, $\inffinisymb n:=\cpredisymb n$ for $D=(\bigwedge,\nofinsymb,\exfinsymb,\bigwedge)$, $\exsolisymb n:=\cpredisymb n$ for $D=(\bigvee,\bigwedge)$, and $\exinfisymb n:=\cpredisymb n$ for $D=(\bigwedge,\nosolsymb,\exsolsymb,\bigwedge)$. We allow ourselves the analogous arrangements that brought us the concise presentation of $\nofinsymb$ and $\inffinsymb$ in \cref{fig:exfin-finfin-rules} and of $\exsolsymb$ and $\exinfsymb$ in \cref{fig:nosol-allfin-rules} and thus arrive at the presentation of the inductive rules in \cref{fig:sliced-rules} that we invite the reader to compare with the unindexed ones in those previous figures.
\begin{figure}[tb]\caption{Predicates $\nofinisymb n$, $\inffinisymb n$, $\exsolisymb n$ and $\exinfisymb n$}\label{fig:sliced-rules}
  \[
    \begin{array}{l@{\qquad}c@{\qquad}r}
        \infer[]{\nofini 0T}{}&
      \infer
      {\nofini{n+1}{f(T_i)_i}}{\nofini n{T_j}}&\infer[]{\nofini{n+1}{\sum_iT_i}}{\bigwedge_i\nofini{n+1}{T_i}}\\[1.5ex]
      \infer[]{\inffini 0T}{}&
      \infer
      {\inffini{n+1}{f(T_i)_i}}{\inffini n{T_j}&\bigwedge_i\exfin{T_i}}&\infer
      {\inffini{n+1}{\sum_iT_i}}{\inffini{n+1}{T_j}}\\[1.5ex]
              \infer[]{\exsoli 0T}{}&
      \infer
      {\exsoli{n+1}{f(T_i)_i}}{\bigwedge_i\exsoli n{T_i}}&\infer[]{\exsoli{n+1}{\sum_iT_i}}{\exsoli{n+1}{T_j}}\\[1.5ex]
      \infer[]{\exinfi 0T}{}&
      \infer
      {\exinfi{n+1}{f(T_i)_i}}{\exinfi n{T_j}&\bigwedge_i\exsol{T_i}}&\infer
      {\exinfi{n+1}{\sum_iT_i}}{\exinfi{n+1}{T_j}}
    \end{array}
  \]
\end{figure}

\begin{lemma}[Inductive characterization of predicate $\cpredd D$]\label{lem:cprediok}
	Given dual pair data $D$ and forest $T$. Then, $\cpred DT$ iff\/ $\cpredi n T$ for all $n$ (for $\cpredisymb n$ defined with respect to $D$).
\end{lemma}
\begin{proof}
  Let $D=(\lopvar,Q,\overline Q,\lopvaralt)$.
  From left to right, this is by induction on $n$. One decomposes
  (thanks to priority~$1$) the sums until one reaches finitely many
  expressions $f(T_i)_i$ to which the induction hypothesis
  applies (depending on the value of $\lopvaralt$, the reasoning is for one of those expressions or for all of them).
  From right to left, 
  one proves coinductively $R\subseteq\cpredd D$, for \mbox{$R:=\{T:\forall\,n\geq0,\,\cpredi n T\}$}.
  This amounts to proving
  \begin{enumerate}
  \item \label{item:cprediokf} $R(f(T_i)_i)$ implies $\overline\lopvar_iR(T_i)\lopvarbinop\lopvar_i\overline Q(T_i)$, and
  \item \label{item:cpredioksum} $R(\sum_iT_i)$ implies $\overline\lopvaralt_iR(T_i)$.
  \end{enumerate}
  Concerning (\ref{item:cprediokf}), assume $R(f(T_i)_i)$. We first consider $\lopvar=\bigwedge$. In particular, $\cpredi 1{f(T_i)_i}$, hence by inversion $\lopvar_i\overline Q(T_i)$ and the existence of a child $T_i$.
  The proof is then indirect: if for all $i$ we had $T_i\not\in R$, then, for each $i$, there would be an $n_i$ s.\,t.\ $\neg\cpredi{n_i}{T_i}$ (hence $n_i>0$), and letting $m$ be the maximum of these $n_i$'s, $\neg\cpredi m{T_i}$ by antitonicity; hence we would have $\neg\cpredi{m+1}{f(T_i)_i}$, contradicting $R(f(T_i)_i)$. 
  The case $\lopvar=\bigvee$ is easier: we are fine if $\bigvee_i\overline Q(T_i)$, so we assume it does not hold. We have to show for all $i$ that $R(T_i)$. This is again indirect, and it is easier than the previous case since no maximum has to be built.
  
  Concerning (\ref{item:cpredioksum}), this is by cases on $\lopvaralt$. If $\lopvaralt=\bigwedge$, we reason as in the first case of (\ref{item:cprediokf}). Since there is a child $T_i$, the maximum $m$ is necessarily not $0$, hence of the form $m'+1$. This would then imply $\neg \cpredi{m'+1}{\sum_i T_i}$, contradicting the hypothesis (giving, for all $n$, $\cpredi{n}{\sum_i T_i}$). The case $\lopvaralt=\bigvee$ is even simpler than the second case of (\ref{item:cprediokf}).
\end{proof}
An immediate consequence of the preceding lemma is that if $\cpredisymb n$ is closed under decontraction
for each $n\geq0$, this also holds of $\cpredd D$.

\begin{lemma}[Closedness under decontraction of $\cpredisymb n$]
  \label{lem:cpredi-decontraction}
  If the predicate $\overline Q$ is closed under decontraction, then for all $n\geq0$, $\cpredisymb n$ is closed under decontraction.
\end{lemma}
\begin{proof}
  Follows by induction on the inductive definition of $\cpredisymb n$ -- we profit from not counting sums as providing depth.
  This is particularly easy to argue since the non-trivial cases of the definition of decontraction in \cref{fig:decontr} concern the nullary variables $z$ and the unary function symbols $\coretR x\cdot R$.\end{proof}
In particular, $\nofinisymb n$ is closed under decontraction (recall in this case $\overline Q=\truepred$, trivially closed under decontraction).
By (the remark after) \cref{lem:cprediok}, this also holds of $\nofinsymb$.
For its complement $\exfinsymb$, we can prove closure under decontraction directly and easily by induction on the inductive definition of $\exfinsymb$.
Hence, as a second instance of the previous lemma, also $\inffinisymb n$ is closed under decontraction (recall here $\overline Q=\exfinsymb$).
The third instance is $\exsolisymb n$ (in this case $\overline Q=\falsepred$, trivially closed under decontraction).
By \cref{lem:cprediok}, also $\exsolsymb$ is closed under decontraction.
The fourth instance is then $\exinfisymb n$ (with $\overline Q=\exsolsymb$).

In \cref{sec:proof-search-PIPL}, we announced that we will prove that $\nofinsymb$ and $\nofinmemsymb$ hold of the same forests, and similarly for $\inffinsymb$ and $\inffinmemsymb$.
We can do this in a more informative way by defining approximations $\nofinmemi n T$ and $\inffinmemi n T$ to $\nofinmem T$ and $\inffinmem T$ in terms of $\finext T$, where for each $n$ individually, $\nofinmemisymb n = \nofinisymb n$ and $\inffinmemisymb n = \inffinisymb n$.

Define the height of $\PIL$ terms by $\height{f(T_i)_i}:=1+\max_i \height{T_i}$. Thus, $\height T$ is always a positive number (not $0$).
Define $\exfinmemi n T:\Leftrightarrow \exists T'\in\finext{T}, \height {T'}\leq n$ and $\nofinmemi n T:\Leftrightarrow\neg\exfinmemi n T$.
Obviously, we get that $\exfinmem T$ holds iff $\exfinmemi n T$ holds for some $n\geq0$. Hence, $\nofinmemsymb=\cap_{n\geq0}\nofinmemisymb n$.

\begin{definition}[Slices of extensional predicate $\finfinmemsymb$]\label{def:finfinmemi}
  We require that  $\finfinmemi 0 T$ never holds and $\finfinmemi n T:\Leftrightarrow \forall T'\in\finext{T}, \height {T'}< n$ for $n>0$.
\end{definition}
In particular, $\finfinmemi 1 T$ iff $\nofinmem T$.
We now simply define $\inffinmemi n T$ as $\neg\finfinmemi n T$.

\begin{lemma}[Characterization of slices of the extensional predicates $\nofinmemsymb$ / $\inffinmemsymb$]\label{lem:charmemi}
  Let $T$ be any forest.
  \begin{enumerate}
  \item \label{item:charmemi1} For all $n\geq 0$, $\nofinmemi n T$ iff $\nofini n T$.
  \item \label{item:charmemi2} $\nofinmem T$ iff $\nofin T$. Equivalently, $\exfinmem T$ iff $\exfin T$.
  \item \label{item:charmemi3} For all $n\geq 0$, $\inffinmemi n T$ iff\/ $\inffini n T$.
  \item \label{item:charmemi4} $\inffinmem T$ iff\/ $\inffin T$. Equivalently, $\finfinmem T$ iff\/ $\finfin T$.
  \end{enumerate}
\end{lemma}
\begin{proof}
  The right-to-left direction for (\ref{item:charmemi1}) is just by induction over the inductive generation of the predicate on the right-hand side.
  The first equivalence in (\ref{item:charmemi2}) is an immediate consequence of (\ref{item:charmemi1}), the above observation $\nofinmemsymb=\cap_{n\geq0}\nofinmemisymb n$ and $\nofinsymb=\cap_{n\geq0}\nofinisymb n$, obtained as instance of \cref{lem:cprediok}.
  The second equivalence in (\ref{item:charmemi2}) is by duality. But notice that its right-to-left direction hinges on the opposite direction in (\ref{item:charmemi1}).
  This is the direction that is used in the right-to-left direction for (\ref{item:charmemi3}), and we develop that latter in more detail.
  It is done by induction over $\inffini n T$, and the first and third rules are easy to deal with.
  Assume $\inffini{n+1}{f(T_i)_i}$ coming from $\inffini{n}{T_j}$ and $\exfin{T_i}$ for all $i$. We have to show $\inffinmemi{n+1}{f(T_i)_i}$. By IH we have $\inffinmemi{n}{T_j}$.
  In case $n=0$, our goal is equivalent to $\exfinmem{f(T_i)_i}$ by the remark after \cref{def:finfinmemi}.
  The right-to-left direction in the second statement in (\ref{item:charmemi2}) allows us to proceed with $\exfin{f(T_i)_i}$, and this is guaranteed by our assumptions.
  If $n>0$, then $\inffinmemi{n}{T_j}$ brings us $T_j'\in\finext{T_j}$ with $\height{T_j'}\geq n$.
  For $i\neq j$, use $\exfin{T_i}$. The same part of (\ref{item:charmemi2}) gives $\exfinmem{T_i}$, hence a $T_i'\in\finext{T_i}$.
  Set $T':=f(T_i')_i\in\finext{f(T_i)_i}$. Then, $\height{T'}\geq 1+\height{T_j'}\geq 1+n$, which proves our goal.

  For the left-to-right direction, we only discuss (\ref{item:charmemi1}), since (\ref{item:charmemi3}) is slightly simpler, except that it also uses the left-to-right direction in the second statement in (\ref{item:charmemi2}).
  The proof is by induction on $n$. For $n=0$, the conclusion is trivial.
  Inside the step from $n$ to $n+1$, an auxiliary observation is: for all terms $T$ of the form $f(T_i)_i$, $\nofinmemi{n+1}T$ implies $\nofini{n+1}T$.
  We show its contraposition: assume $\neg\nofini{n+1}T$, hence for all $i$, we have $\neg \nofini{n}{T_i}$.
  By the IH we obtain $\neg\nofinmemi{n}{T_i}$, hence there is $T_i'\in\finext{T_i}$ with $\height{T'_i}\leq n$.
  Set $T':=f(T'_i)_i\in\finext T$. $\height{T'}=1+\max_i \height{T'_i}\leq 1+n$, which gives $\neg\nofinmemi{n+1}T$.
  For the general case, we have to decompose sums in the given forest $T$ recursively from the outside, which is possible since summation has priority $1$ in the coinductive grammar, until a (possibly deeply but finitely nested) finite sum of terms of the above form is reached. From $\nofinmemi{n+1}T$ we get by definition of $\finextsymb$ that each such summand satisfies $\nofinmemisymb{n+1}$, hence by the auxiliary observation also $\nofinisymb{n+1}$. Since we have this for all the summands, we can rebuild an evidence for $\nofini{n+1}T$.

  The proof of (\ref{item:charmemi4}) requires the following observation: $\finfinmem T$ holds iff $\finfinmemi n T$ holds for some $n\geq0$.
  From left to right, set $n:=1+\max_{T'\in\finext{T}}\height{T'}$.
  From right to left, observe that there can only be finitely many finite members below a certain height in a given forest (again, to argue for this, sums have to be decomposed finitely until one hits a height-increasing function symbol). Put differently, $\inffinmemsymb=\cap_{n\geq0}\inffinmemisymb n$.
  Together with $\inffinsymb=\cap_{n\geq0}\inffinisymb n$, obtained as instance of \cref{lem:cprediok}, (\ref{item:charmemi3}) yields that $\inffinmemsymb=\inffinsymb$ and $\finfinmemsymb=\finfinsymb$, as sets of forests.
\end{proof}

We cannot hope for a similar informative analysis of $\exsolisymb n$ in terms of a predicate on forests that builds on membership, not even for the case $n=1$.
The simple reason is that we can find forests $T_1$ and $T_2$ of sort $v$ for which $\ext{T_1}=\ext{T_2}$ and $\neg\exsoli 1{T_1}$ but $\exsoli 1{T_2}$:
Let $T_1:=\oo^v$ and $T_2:=\inj i P{T_1}$ for some $i$ and $P$.
Obviously, $\ext {T_1}$ and $\ext {T_2}$ are empty.
And $T_1$ and $T_2$ are distinguished by $\exsolisymb 1$.
There is a possibility of characterizing $\exsolisymb n$ through a sliced membership predicate, but the information gain seems too small to pursue this here.

\begin{definition}[Slices of extensional predicate $\exinfmemsymb$]\label{def:exinfmemi}
  We require that  $\exinfmemi 0 T$ always holds and $\exinfmemi n T:\Leftrightarrow \exists T'\in\ext{T}, \height {T'}\geq n$ for $n>0$.
  Notice that we have no definition of $\height {T'}$ in case that $T'$ is in $\copil$ and not in $\pil$. We consider $\height {T'}\geq n$ as a two-place predicate with arguments $T'$ and $n$, with the expected interpretation for $T'$ in $\pil$, and otherwise, we just assume it holds (for any $n$).\footnote{Equivalenty, one could define $\height {T'}$ to be $\omega$ in this case and then use the extension of the order on the natural numbers to $\omega$.}
\end{definition}

\begin{lemma}[Characterization of the extensional predicates $\exsolmemsymb$ and $\exinfmemsymb$]\label{lem:charmemsol}
  Let $T$ be any forest.
  \begin{enumerate}
  \item \label{item:charmemsol1}
    $\exsolmem T$ iff $\exsol T$. Equivalently, $\nosolmem T$ iff $\nosol T$.
  \item \label{item:charmemsol2}
    For all $n\geq 0$, $\exinfmemi n T$ iff $\exinfi n T$.
  \item \label{item:charmemsol3}
    $\exinfmem T$ iff $\exinf T$ iff for all $n\geq0$, $\exinfmemi n T$. Hence, $\allfinmem T$ iff $\allfin T$.
  \end{enumerate}
\end{lemma}
\begin{proof}
  Part~(\ref{item:charmemsol1}). The direction from left to right is shown by a simple coinductive proof: $\exsolmemsymb$ is backward closed relative to the coinductive definition of $\exsolsymb$.
  From right to left, we use a corecursive extraction procedure $\exfsymb$ from forests satisfying $\exsolsymb$ into expressions of $\copil$.
  Given a forest $T$, we decompose sums recursively from the outside until only expressions of the form $f(T_i)_i$ are reached as summands, and by the assumption $\exsol T$, one such summand fulfills $\bigwedge_i\exsol{T_i}$.
  We then set $\exf T:=f(\exf{T_i})_i$, thus corecursively applying the extraction procedure to $T_i$ for all $i$.
  We think of this rather as an extraction procedure than a set-theoretic function definition because the choices of the summands need not be unique.
  Since the constructor $f$ guards the definition of $\exf T$, this properly defines an expression of $\copil$.
  And a coinductive proof immediately shows $\exf T\in\ext T$, thus we have $\exsolmem T$.
  (By the remark after \cref{eq:mem-rules}, $\exf T$ is even a co-proof term.)
  The second equivalence is by duality.

  Part~(\ref{item:charmemsol2}). The right-to-left direction resembles the same direction of \cref{lem:charmemi}.\ref{item:charmemi3}. It is done by induction over $\exinfi n T$, and the first and third rules are easy to deal with.
  Assume $\exinfi{n+1}{f(T_i)_i}$ coming from $\exinfi{n}{T_j}$ and $\exsol{T_i}$ for all $i$. We have to show $\exinfmemi{n+1}{f(T_i)_i}$. By IH we have $\exinfmemi{n}{T_j}$.
  In case $n=0$, our goal is equivalent to $\exsolmem{f(T_i)_i}$ by definition of $\exinfmemisymb 1$.
  Part~(\ref{item:charmemsol1}) allows us to proceed with $\exsol{f(T_i)_i}$, and this is guaranteed by our assumptions.
  If $n>0$, then $\exinfmemi{n}{T_j}$ brings us $T_j'\in\ext{T_j}$ with $\height{T_j'}\geq n$.
  For $i\neq j$, use $\exsol{T_i}$. Part~(\ref{item:charmemsol1}) gives $\exsolmem{T_i}$, hence a $T_i'\in\ext{T_i}$.
  Set $T':=f(T_i')_i\in\ext{f(T_i)_i}$.
  Then, $\height{T'}\geq 1+n$ (this reasoning is also correct when $T'$ is not in $\pil$), which proves our goal.

  The left-to-right direction is again similar to \cref{lem:charmemi}.\ref{item:charmemi3}, but that proof is not given in detail, which we do here.
  The proof is by induction on $n$.
  For $n=0$, the conclusion is trivial.
  We do the step from $n$ to $n+1$, first decomposing sums in the given forest $T$, until we get from $\exinfmemi{n+1}T$ -- hence a $T'\in\ext T$ with $\height{T'}\geq n+1$ -- a (possibly nested) summand $f(T_i)_i$ of $T$ still with $T'\in\ext{f(T_i)_i}$, hence $\exinfmemi{n+1}{f(T_i)_i}$.
  $T'=f(T'_i)_i$ with $T'_i\in\ext{T_i}$ for all $i$, and since $\height{T'}\geq n+1$, there is a $j$ such that $\height{T_j'}\geq n$ (this reasoning is also correct when $T'$ is not in $\pil$).
  For $n>0$ this means $\exinfmemi n{T_j}$, but the latter trivially also holds for $n=0$.
  By IH, we obtain $\exinfi n{T_j}$.
  For all $i$, we have $\exsolmem{T_i}$, hence $\exsol{T_i}$ by part~(\ref{item:charmemsol1}).
  Thus, $\exinfi{n+1}{f(T_i)_i}$, and we can build the evidence for $\exinfi{n+1}{T}$ by going into the sum(s).
  
  Part~(\ref{item:charmemsol3}).
  The direction from the first to the third formulation is trivial by our reading of ``$\height {T'}\geq n$''.
  The direction from the third to the second formulation follows from part~(\ref{item:charmemsol2}) and \cref{lem:cprediok}.
  The missing implication from $\exinf T$ to $\exinfmem T$ requires a refinement of our construction of part~(\ref{item:charmemsol1}). We define a corecursive extraction procedure $\exfvarsymb$ from forests satisfying $\exinfsymb$ into expressions of $\copil$.
  Given a forest $T$, as for $\exfsymb$, we decompose sums until summands are of the form $f(T_i)_i$, and by the assumption $\exinf T$, one such summand fulfills $\exinf{T_{i_0}}$ for some $i_0$ and $\bigwedge_i\exsol{T_i}$.
  We define $\exfvar T$ as $f(\tilde T_i)_i$, with the corecursively obtained $\tilde T_{i_0}:=\exfvar {T_{i_0}}$ and for $i\neq i_0$ the setting $\tilde T_i:=\exf{T_i}$, using part~(\ref{item:charmemsol1}).
  The nature of this definition is as for $\exfsymb$, and it is guarded by the constructor $f$, hence properly defines an expression of $\copil$.
  As for $\exfsymb$, a coinductive proof immediately shows $\exfvar T\in\ext T$.
  Moreover, coinduction shows that $\exfvar T$ satisfies the coinductive predicate of not belonging to $\pil$, thus we even have $\exinfmem T$.
\end{proof}


\subsection{On termination of finitary representation in \cref{sec:finrep}}
\label{sec:appendix-terminationofF}

Definition~\ref{def:finrep} contains recursive equations that are not justified by calls to the same function for ``smaller'' sequents, in particular not for the rules governing R-stable sequents as first argument. We mentioned
that the proof of termination of an analogous function for implicational logic \cite[Lemma 24]{JESRMLP-APAL2021} can be adapted to establish also termination of $\finrep{\jay}{\Xi}$ for any valid arguments. Here, we substantiate this claim.

The difficulty comes from the rich syntax of $\PIL$, so that the ``true'' recursive structure of $\finrep\rho\Xi$---for R-stable sequents that spawn the formal fixed points---gets hidden through intermediary recursive calls with the other forms of logical sequents. However, we will now argue that all those can be seen as plainly auxiliary since they just decrease the ``weight'' of the problem to be solved.

\begin{lemma}
  Every direct recursive call in the definition of $\finrep\jay\Xi$ to some $\finrep{\jay'}{\Xi'}$ for neither $\jay$ nor $\jay'$ R-stable sequents lowers the \emph{weight} of the first argument.
\end{lemma}
\begin{proof}
  This requires to check the very same
  inequations as in the proof case (1) of \cref{lem:weight-S}.
\end{proof}
The message of the lemma is that the proof search through all
the other forms of logical sequents (including the form $\Gamma\vdash C$) is
by itself terminating. Of course, this was to be expected. Otherwise,
we could not have ``solved'' them by a recursive definition in
$\finrepsymb$ where only R-stable sequents ask to be hypothetically solved
through fixed-point variables.

The present argument comes from an analysis that is deeply connected
to $\PIL$, it has nothing to do with an abstract approach of defining
(infinitary or finitary) forests. As seen directly in the
definition of $\finrepsymb$, only by cycling finitely through the
$\dlv\cdot$ construction is the context $\Gamma$ extended in the
arguments $\jay$ to $\finrepsymb$. And the context of the last
fixed-point variable in $\Xi$ grows in lockstep.

It is trivial to observe that all the formula material of the
right-hand sides lies in the same subformula-closed sets (see~\cite{JESRMLP-APAL2021})  as the
left-hand sides (in other words, the logical sequents in the recursive calls
are taken from the same formula material, and there is no
reconstruction whatsoever).

Therefore, the previous proof for the implicational case \cite[Lemma 24]{JESRMLP-APAL2021}
can be carried over without substantial
changes. What counts are recursive calls with first argument an R-stable
sequent for the calculation when the first argument is an R-stable
sequent. In the implicational case, these ``big'' steps were enforced
by the grammar for finitary forests (and the logical sequents $\Gamma\vdash R$
had even only atomic $R$ there, but this change is rather irrelevant
for the proof (instead of counting atoms, one has to count $R$ formulas for
getting the measure, but this does not affect finiteness of it).
The preparatory steps in the proof of \cite[Lemma 24]{JESRMLP-APAL2021} 
are also easily adapted, where the $\Gamma$ part of the first argument to $\finrepsymb$
takes the role of the context $\Gamma$ in that proof.

\subsection{On forest transformation for inessential extensions
in \cref{sec:analysis-PIPL}
}\label{sec:appendix-decontraction}
\begin{definition}[Decontraction]\label{def:co-cont-forests}
	Let $\Gamma\leq\Gamma'$. For a forest $T$ of $\copils$,
	the forest $[\Gamma'/\Gamma]T$ of $\copils$ is defined  by
	corecursion in Fig.~\ref{fig:decontr},
	where, for $w\in dom(\Gamma)$,
	$$D_w:=\{w\}\cup\{w': (w':\Gamma(w))\in(\Gamma'\setminus\Gamma)\}\enspace.$$ 
	In other words, the occurrences of variables (in the syntactic way they
	are introduced in the forests) are duplicated for all other variables
	of the same type that $\Gamma'$ has in addition.
	Moreover, if $\rho=(\Gamma\vdash R)$ and $\rho'=({\Gamma'}\vdash R)$, then $[\rho'/\rho]T$ is defined to be $[\Gamma'/\Gamma]T$. The operation $[\rho'/\rho]\_$ is called \emph{decontraction}.
\end{definition}
\begin{figure}[tb]\caption{Corecursive equations for definition of decontraction}\label{fig:decontr}
	\[
	\begin{array}{lcll}
		{[}\Gamma'/\Gamma]f(T_1,\ldots,T_k)&=&f([\Gamma'/\Gamma]T_i,\ldots,[\Gamma'/\Gamma]T_k)&\textrm{for $f$ neither $z$ nor $\coretR x\cdot R$}\\[1.5ex]
		{[}\Gamma'/\Gamma]\sum_i T_i&=&\sum_i {[}\Gamma'/\Gamma]T_i\\[1.5ex]
		{[}\Gamma'/\Gamma]z&=&z&\textrm{if $z\notin dom(\Gamma)$}\\
		{[}\Gamma'/\Gamma]z&=&\kern-1em\s{z'\in D_z}{z'}&\textrm{if $z\in dom(\Gamma)$}\\[1.5ex]
		{[}\Gamma'/\Gamma]\coretR x s R&=&\coretR{x}{{[}\Gamma'/\Gamma]s}R&\textrm{if $x\not\in dom(\Gamma)$}\\
		{[}\Gamma'/\Gamma]\coretR x s R&=&\kern-1em\s{x'\in D_x}{\coretR{x'}{[\Gamma'/\Gamma]s}R}&\textrm{if $x\in dom(\Gamma)$}
	\end{array}
	\]
\end{figure}

\begin{lemma}[Solution spaces and decontraction]\label{lem:solextension}
	Let $\rho\leq\rho'$. Then $\solfunction{\rho'}=[\rho'/\rho]\solfunction\rho$.
\end{lemma}
\begin{proof}
	Analogous to the proof for implicational logic \cite{JESRMLP-APAL2021}. Obviously, the decontraction operation for forests
	has to be used to define decontraction operations for all forms of logical sequents (analogously to the R-stable sequents, where only $\Gamma$ varies).
	Then, the coinductive proof is done simultaneously for all forms of logical sequents.
\end{proof}

\subsection{Completing the proofs of \cref{prop:genfinchar}.\ref{prop:genfinchar.2} with the indexed predicates}\label{sec:appendix-mainpropgencoindanalysis}

We want to prove \cref{prop:genfinchar}.\ref{prop:genfinchar.2}.
Thus, assume given finitary dual pair data $D^+=(D,\Pi,\overline\Pi,P)$ with dual pair data $D=(\lopvar,Q,\overline Q,\lopvaralt)$ so that $\overline Q$ is closed under decontraction.
As before, we write $\cpredisymb n$ for the approximations to $\cpredd D$, with $D$ understood.

For $T\in\pils$, we write $\bigass nT$ for the following
assumption: For every free occurrence of some $X^\rho$ in $T$ (those
$X^\rho$ are found in $\FPV(T)$) such that $\neg P(\rho)$, there
is an $n_0$ with $\cpredi {n_0}{\solfunction\rho}$ and
$d+n_0\geq n$ for $d$ the depth of the occurrence in $T$ as
defined earlier, where sums and generations of fixed points do not
contribute to depth.

Notice that, trivially $n'\leq n$ and $\bigass {n}T$ imply
$\bigass {n'}T$.

\begin{lemma}[Ramification of \cref{prop:genfinchar}.\ref{prop:genfinchar.2}]\label{lem:genramif}
  Let $T\in\pils$ be well-bound, proper and guarded and such
  that $\cpredf{D^+}T$ holds. Then, for all $n\geq0$, $\bigass nT$
  implies $\cpredi n{\interps T}$.
\end{lemma}

\begin{proof}
  By induction on the finitary forests $T$ (which can also be seen as a proof by induction on predicate
  $\cpredfd{D^+}$).

  Case $T=X^\rho$.  Then $\interps T=\solfunction\rho$. Assume
  $n\geq0$ such that $\bigass nT$. By inversion on $\cpredfd{D^+}$, we have $\neg P(\rho)$,
  hence, since $X^\rho\in\FPV(T)$ at depth $0$ in $T$, $\bigass nT$ gives
  $n_0\geq n$ with $\cpredi {n_0}{\solfunction\rho}$. Since
  $\cpredisymb m$ is antitone in $m$, we also have
  $\cpredi n{\interps T}$.

  Case $T=\gfp X^\rho.T_1$. $\cpredf{D^+}{T}$ comes from $\cpredf{D^+}{T_1}$. Let
  $N:=\interps T=\interps{T_1}$.  As $T$ is proper,
  $N=\solfunction{\rho}$. We do the proof by a side induction on
  $n$. The case $n=0$ is trivial. So assume $n=n'+1$ and $\bigass nT$ and that we
  already know that $\bigass{n'}T$ implies
  $\cpredi {n'}{\solfunction\rho}$.
  We have to show $\cpredi n{\solfunction\rho}$, i.\,e., $\cpredi n{\interps{T_1}}$.
  We use the main induction hypothesis on $T_1$ with the same index $n$.
  Hence, it suffices to show $\bigass n{T_1}$.
  Consider any free occurrence of some $Y^{\rho'}$ in $T_1$ such
  that $\neg P(\rho')$. We have to show that there is an $n_0$ with
  $\cpredi {n_0}{\solfunction{\rho'}}$ and $d+n_0\geq n$ for $d$ the
  depth of the occurrence in $T_1$.

  First sub-case: the considered occurrence is also a free occurrence
  in $T$. Since we disregard fixed-point constructions for depth, $d$
  is also the depth in $T$. Because of $\bigass nT$, we get an $n_0$
  as desired.

  Second sub-case: the remaining case is with $Y=X$ and, since $T$ is
  well-bound, $\rho\leq\rho'$. As remarked before, $\bigass nT$
  gives us $\bigass {n'}T$. The side induction hypothesis therefore
  yields $\cpredi {n'}{\solfunction\rho}$. By closure of
  $\cpredisymb {n'}$ under decontraction (thanks to \cref{lem:cpredi-decontraction}, where we use our general assumption that $\overline Q$ is closed under decontraction), we get
  $\cpredi {n'}{[\rho'/\rho]\solfunction\rho}$, but that latter
  forest is $\solfunction{\rho'}$ by Lemma~\ref{lem:solextension}.
  By guardedness of $T$, this occurrence of $X^{\rho'}$ has depth
  $d\geq1$ in $T_1$. Hence, $d+n'\geq 1+n'=n$.

  Case $T=f(T_1,\ldots,T_k)$ with a proper function symbol $f$.
  Assume $n\geq0$ such that
  $\bigass nT$. We have to show that $\cpredi n{\interps{T}}$. This is
  trivial for $n=0$. Thus, assume $n=n'+1$.
  We first argue that, for every $i$, $\cpredf{D^+}{T_i}$ implies
  $\cpredi {n'}{\interps{T_i}}$. Fix some $i$. We use the induction hypothesis on
  $T_i$ (even with this smaller index $n'$). Therefore, we are left to
  show $\bigass {n'}{T_i}$. Consider any free occurrence of some
  $X^{\rho}$ in $T_i$ such that $\neg P(\rho)$, of depth $d$ in
  $T_i$. This occurrence is then also a free occurrence in $T$ of
  depth $d+1$ in $T$. From $\bigass nT$, we get an $n_0$ with
  $\cpredi {n_0}{\solfunction\rho}$ and $d+1+n_0\geq n$, hence with
  $d+n_0\geq n'$, hence $n_0$ is as required for showing
  $\bigass {n'}{T_i}$. Now, $\cpredf{D^+}T$ comes from
  $\overline\lopvar_i\cpredf {D^+}{T_i}\lopvarbinop\lopvar_i\overline\Pi(T_i)$.
  By definition of finitary dual pairs, for every $i$, $\overline\Pi(T_i)$ iff $\overline Q(\interps{T_i})$.
  Together with our previous observation, $\overline\lopvar_i\cpredf {D^+}{T_i}\lopvarbinop\lopvar_i\overline\Pi(T_i)$ thus implies
  $\overline\lopvar_i\cpredi {n'}{\interps{T_i}}\lopvarbinop\lopvar_i\overline Q(\interps{T_i})$, whence $\cpredi {n'+1}{f(\interps{T_i})_i}$.

  Case $T=\sum_i T_i$. Assume $n\geq0$ such that $\bigass nT$. We have to show
  that $\cpredi n{\interps{T}}$. This is trivial for $n=0$. Thus,
  assume $n=n'+1$.
  We first argue that, for every $i$, $\cpredf{D^+}{T_i}$ implies
  $\cpredi {n}{\interps{T_i}}$. Fix some $i$.
  We use the induction hypothesis on $T_i$
  (with the same index $n$). Therefore, we are left to show
  $\bigass {n}{T_i}$. Consider any free occurrence of some
  $X^{\rho}$ in $T_i$ such that $\neg P(\rho)$, of depth $d$ in
  $T_i$. This occurrence is then also a free occurrence in $T$ of
  depth $d$ in $T$. From $\bigass nT$, we get an $n_0$ with
  $\cpredi {n_0}{\solfunction\rho}$ and $d+n_0\geq n$, hence $n_0$
  is as required for showing $\bigass {n}{T_i}$.
  Now,  $\cpredf{D^+}T$ comes from $\overline\lopvaralt_i\cpredf {D^+}{T_i}$,
  and, by the previous observation, this implies $\overline\lopvaralt_i\cpredi {n'+1}{\interps{T_i}}$,
  whence $\cpredi {n'+1}{\sum_i\interps{T_i}}$.
  (Of course, it is
  important that sums do not count for depth in finitary terms if they
  do not count for the index of the approximations to
  $\cpredd D$. Therefore, this proof case is so simple.)
 \end{proof}

We return to \cref{prop:genfinchar}.\ref{prop:genfinchar.2}:
\begin{proof}
  Let $T\in\pils$ be well-bound, proper and
  guarded, assume $\cpredf{D^+} T$ and that for all
  $X^\rho\in\FPV(T)$, $\ipred D{\solfunction\rho}$ implies
  $P(\rho)$.  We have to show $\cpred D{\interps T}$. By
  \cref{lem:cprediok} it suffices to show
  $\cpredi n{\interps T}$ for all $n$. Let $n\geq0$. By the just
  proven refinement, it suffices to show $\bigass nT$.  Consider any
  free occurrence of some $X^{\rho}$ in $T$ such that
  $\neg P(\rho)$, of depth $d$ in $T$.  By contraposition of the
  assumption on $\FPV(T)$ and by the complementarity of the dual pair of predicates, we have
  $\cpred D{\solfunction\rho}$, hence by
  Lemma~\ref{lem:cprediok} $\cpredi n{\solfunction\rho}$, and
  $d+n\geq n$, as required for $\bigass nT$.
\end{proof}


%% file: appendix-negative-embedding.tex
\subsection{On the  negative embeddings in \cref{subsec:LJT-LJTco}}
\label{app:appendix-negative-embedding}

The characterization of the 
co-proof terms in the $\untsymb$-fragment of $\copil$ is 
done through the unary predicates $\ntermsymb$, $\nexpsymb$ and $\nspinesymb$ on  co-proof terms of $\copil$ (of sorts $t$, $e$ and $s$, respectively), whose  simultaneous coinductive definition is given in \cref{fig:imageLJTco}.  In this definition,  type annotations are tacitly assumed to range over formulas in the $\untsymb$-fragment. 
 We call $\untsymb$-terms (resp.\ $\untesymb$-expressions, resp.\ $\untsymb$-spines) those co-proof terms $t$ (resp.\ $e$, resp.\ $s$) with $\nterm t$ (resp.\ $\nexp e$, resp.\ $\nexp s$).
\begin{figure}[tb]\caption{Predicates $\ntermsymb$, $\nexpsymb$ and $\nspinesymb$}\label{fig:imageLJTco}
\[
    \begin{array}{c}
\infer=[]{\nterm{\lb(x^N.e)}}{\nexp{e}&\textrm{$e$ atomic}}\quad
\infer=[]{\nterm{\lb(x^N.\dlv t)}}{\nterm{t}}\quad  \infer=[]{\nterm{ \pairljt{t_1}{t_2}}}{\nterm{t_1}&\nterm{t_2}}\\[1ex]
\infer=[]{\nterm{\ea{e}}}{\nexp{e}&\textrm{$e$ atomic}}\quad  \infer=[]{\nterm{\ep{e}}}{\nexp{e}&\textrm{$e$ not atomic}}
\\[1ex]
\infer=[]{\nexp{\coretR xsR}}{\nspine{s}}\quad  \infer=[]{\nexp{\ret{\inj i {\downshift N}{\thunk t}}}}{\nterm{t}}\\[1ex]
\infer=[]{\nspine{\nil}}{}\quad  \infer=[]{\nspine{\cothunk{\abortt R}}}{}\quad  \infer=[]{\nspine{\cothunk{[x_1^{N_1}.e_1,x_2^{N_2}.e_2]}}}{\nexp{e_1}&\nexp{e_2}}\\[1ex]
\infer=[]{\nspine{\thunk{t}::s}}{\nterm{t}&\nspine{s}}\quad  \infer=[]{\nspine{i::s}}{\nspine{s}}\\
\end{array}
\]  
\end{figure}
Notice that it is crucial that these subclasses are designated from within the co-proof terms -- this is not a new generative process. While, from the perspective of \cref{fig:imageLJTco} alone, the fixed point $t$ of the equation $t=\lb(x^N.\dlv t)$ seems to qualify as $\untsymb$-term, this $t$ is no co-proof term of $\coLJT$, for lack of cycling through expressions of the form $\focl ysR$.

The \emph{forgetful} map $\fgt{\_}$, from $\untsymb$-terms (resp.~$\untesymb$-expressions, $\untsymb$-spines) to co-proof terms of sort $t$ (resp.~$s$, $e$) of $\coLJT$
is given corecursively in \cref{fig:forgetful-map}. 
One immediately sees that corecursive calls always occur as arguments to co-proof term constructors of $\coLJT$, and also that
a constructor with priority 2 in the source originates a constructor with priority 2 in the target.  A consequence of the latter is that infinite co-proof terms are mapped to infinite co-proof terms.
An easy induction on $\PIL$-proof terms shows that the restriction of this forgetful map 
to $\untsymb$-terms, $\untesymb$-expressions and $\untsymb$-spines in $\PIL$  defines a forgetful map into $\LJT$

\begin{figure}\caption{Forgetful map $\fgt{\_}$ into $\coLJT/\LJT$}\label{fig:forgetful-map}
\[
\begin{array}{rcllrcl}
\fgt{\lb(x^N.\dlv t)}&=&\lb x^{\fgt N}.\fgt t&&\fgt{\nil}&=&\nil\\
\fgt{\lb(x^N.e)}&=&\lb x^{\fgt N}.\fgt e&\textrm{for $e$ atomic}
&\fgt{\cothunk{\abortt R}}&=&\abortt{\fgt R}\\
\fgt{\langle t_1,t_2\rangle}&=&\langle\fgt{t_1},\fgt{t_2}\rangle&&\fgt{\cothunk{[x_1^{N_1}.e_1,x_2^{N_2}.e_2]}}&=&[x_1^{\fgt{N_1}}.\fgt{e_1},x_2^{\fgt{N_2}}.\fgt{e_2}]\\
  \fgt{\ea e}&=&\fgt e&\textrm{for $e$ atomic}
                                                         &\fgt{\thunk{t}::s}&=&\fgt{t}::\fgt{s}\\
  \fgt{\ep e}&=&\fgt e&\textrm{$e$ not atomic}
                                                         &\fgt{i::s}&=&i::\fgt{s}\\[1.5ex]
\fgt{\coretR x s R}&=& \focl x{\fgt s}{\fgt R}&&\fgt{\ret{\inj i{\downshift N}{\thunk t}}}&=&\inj i{\fgt N}{\fgt t}
\end{array}
\]
\end{figure}
\begin{lemma}[Preservation of validity by the forgetful map into $\coLJT$] \label{lem:soundness-forgetful-map-to-coLJT}
For $t,e,s$ in $\copil$:
\begin{enumerate}
\item If\/ $\Gamma\Longrightarrow t:N$ is valid in $\copil$ and $\nterm t$, then ${\fgt{\Gamma}\Longrightarrow\fgt{t}:\fgt{N}}$ is valid in $\coLJT$.
\item If\/ $\Gamma\vdash e:R$ is valid in $\copil$ and $\nexp e$, then ${\fgt{\Gamma}\vdash\fgt{e}:\fgt{R}}$ is valid in $\coLJT$.
\item If\/ $\Gamma[s:N]\vdash R$ is valid in $\copil$ and $\nspine s$, then $\fgt{\Gamma}[\fgt s:\fgt N]\vdash\fgt{R}$  is valid in $\coLJT$.
\end{enumerate}
\end{lemma}
\begin{proof}
The three items are proved simultaneously by coinduction on the typing relation of $\coLJT$.
\end{proof}
\begin{corollary}[Preservation of derivability by the forgetful map to $\LJT$] \label{cor:soundness-forgetful-map-to-LJT}
\Cref{lem:soundness-forgetful-map-to-coLJT} holds with all upper ``co'' indices removed.
\end{corollary}
\begin{proof}
We argue about item 1 (the other items are analogous).  From  the assumption and \cref{lem:conservativeness-typing} follows validity of $\Gamma\Longrightarrow t:A$ in $\copil$.  Hence,  ${\fgt{\Gamma}\Longrightarrow\fgt{t}:\fgt{A}}$ is valid in $\coLJT$ by \cref{lem:soundness-forgetful-map-to-coLJT}. We have already observed that $\fgt{t}$ is an $\LJT$-term when $t$ is both a term of $\PIL$ and a $\ntermsymb$. Therefore, by  \cref{lem:conservativeness-typing-LJT}, ${\fgt{\Gamma}\Longrightarrow\fgt{t}:\fgt{A}}$ is valid in $\LJT$.
\end{proof}


%% file: appendix-positive-embedding.tex
\subsection{On the positive embeddings in \cref{subsec:LJQ-LJQco}}
\label{app:appendix-positive-embedding}

The characterization of the 
co-proof terms in the $\uptsymb$-fragment of $\copil$ is 
done through the unary predicates $\pexpsymb$, $\vvalsymb$ on  co-proof terms of $\copil$ (of sorts $e$ and $v$, respectively), whose  simultaneous coinductive definition is given in \cref{fig:imageLJQco}.  In this definition type annotations are tacitly assumed to range over formulas in the $\uptsymb$-fragment. 

\begin{figure}[tb]\caption{Predicates $\pexpsymb$ and $\vvalsymb$}\label{fig:imageLJQco}
\[
    \begin{array}{c}
\infer=[]{\pexp{\ret v}}{\vval v}\quad\infer=[]{\pexp{\elimimpforqt xv{y^L}{e}P}}{\vval{v}& \pexp{e}}\quad\infer=[]{\pexp{\elimandforqt x{y^L}eiP}}{\pexp{e}}\quad\infer=[]{\pexp{\elimorforqt x{y_1^{L_1}}{e_1}{y_2^{L_2}}{e_2}P}}{\pexp{e_1}& \pexp{e_2}}\quad\infer=[]{\pexp{\elimabsforqt x{{P}}}}{}\\[1.5ex]
\infer=[]{\vval x}{}\quad\infer=[]{\vval {\lbforqut{x^{L}}{e}}}{ \pexp{e}}\quad\infer=[]{\vval {\pairforqut{e_1}{e_2}}}{ \pexp{e_1}&\pexp{e_2}}\quad\infer=[]{\vval {\inj i {P}{ v} }}{ \vval{v}}\quad
    \end{array}
\]  
\end{figure}

The \emph{forgetful} map  $\fgtQ{\_}$, from $\pexpsymb$-expressions (resp.~$\vvalsymb$-values) to co-proof terms of sort $t$ (resp.~$v$) of $\coLJQ$,
is given corecursively in \cref{fig:coLJQ-forgetful-map}. 
One immediately sees that 
for every constructor with priority 2 in the source, a constructor with priority 2 appears in the target (as a consequence, infinite co-proof terms are mapped to infinite co-proof terms of $\coLJQ$), and  that corecursive calls always occur as arguments to co-proof term constructors of $\coLJQ$, and are thus legitimate. 
Note also, the restriction of this map to $\pexpsymb$-expressions in $\PIL$ and to  $\vvalsymb$-values in $\PIL$ defines a forgetful map into $\LJQ$ (an easy induction on $\PIL$-co-proof terms confirms this).
\begin{figure}[tb]\caption{Forgetful map  $\fgtQ{\_}$ into $\coLJQ/\LJQ$}\label{fig:coLJQ-forgetful-map}
$$
\begin{array}{c}
\begin{array}{rclcrcl}

\fgtQ{\ret{{v}}}&=&\crc{\fgtQ{v}}&&\fgtQ x & = &x \\
\fgtQ{\elimimpforqt x{ v}{y^{ L}}{ e}P}&=&\impleftljqt x{\fgtQ v}y{\fgt L}{\fgtQ e}{\fgt P}& &\fgtQ{\lbforqut{x^{L}}{e}}&=&  \lb{x^{\fgt L}}.{\fgtQ e} \\
\fgtQ{\elimandforqt x{y^{L}}{e}iP} &=& \conjleftljqt xy{\fgt{L}}{\fgtQ e}i{\fgt P}&&\fgtQ{\pairforqut{e_1}{e_2}}&=&  \langle \fgtQ{{e_1}}, \fgtQ {{e_2}} \rangle \\
\fgtQ{\elimorforqt  x{{y_1}^{L_1}}{e_1}{{y_2}^{L_2}}{e_2}{P}  } 
&=&\disjleftljqt x{y_1}{\fgt{L_1}}{\fgtQ{e_1}}{y_2}{\fgt{L_2}}{\fgtQ{e_2}}{\fgt P}
&&\fgtQ{\inj i P v} &=& {\inj i {\fgt P}{\fgtQ v}} \\
\fgtQ{\elimabsforqt x{{P}}} &=&\abortljqt x{\fgt{{P}}}\\

\end{array}
\end{array}
$$
\end{figure}
\begin{lemma}[Preservation of validity by the forgetful map to $\coLJQ$] \label{lem:soundness-forgetful-map-to-coLJQ}
For $e,v$ in $\copil$:
\begin{enumerate}
\item If\/ $\Gamma\vdash e:A$ is valid in $\copil$ and $\pexp e$, then ${\fgt{\Gamma}\Longrightarrow\fgtQ{e}:\fgt{A}}$ is valid in $\coLJQ$.
\item If\/ $\Gamma\vdash[v:P]$ is valid in $\copil$ and $\vval v$, then ${\fgt{\Gamma}\vdash[\fgtQ{v}:\fgt{P}]}$  is valid in $\coLJQ$.
\end{enumerate}
\end{lemma}
\begin{proof}
The two items are proved simultaneously by coinduction on the typing relation of $\coLJQ$.
\end{proof}
\begin{corollary}[Preservation of derivability by the forgetful map to $\LJQ$] \label{cor:soundness-forgetful-map-to-LJQ}
\Cref{lem:soundness-forgetful-map-to-coLJQ} holds with all upper ``co'' indices removed.
\end{corollary}
\begin{proof}
We argue about item 1 (item 2 is analogous).  From  the assumption and \cref{lem:conservativeness-typing} follows the validity of $\Gamma\vdash e:A$ in $\copil$.  Hence,  ${\fgt{\Gamma}\Longrightarrow\fgtQ{e}:\fgt{A}}$ is valid in $\coLJQ$ by \cref{lem:soundness-forgetful-map-to-coLJQ}. We have already observed that $\fgtQ{e}$ is an $\LJQ$-expression when $e$ is both a stable-expression of $\PIL$ and a $\pexpsymb$-expression. Therefore, by  \cref{lem:conservativeness-typing-LJQ}, ${\fgt{\Gamma}\Longrightarrow\fgtQ{e}:\fgt{A}}$ is valid in $\LJQ$.
\end{proof}
